\documentclass[twocolumn]{aastex631}

\usepackage{amssymb}
\usepackage{bold-extra}
\AtBeginDocument{
}

\newcommand{\n}{\nodata}

\defcitealias{MOJAVE_XVIII}{MOJAVE~XVIII}

\def\gtrsim{\mathrel{\hbox{\rlap{\hbox{\lower4pt\hbox{$\sim$}}}\hbox{$>$}}}}
\def\lesssim{\mathrel{\hbox{\rlap{\hbox{\lower4pt\hbox{$\sim$}}}\hbox{$<$}}}}
\def\gtrsim{\mathrel{\hbox{\rlap{\hbox{\lower4pt\hbox{$\sim$}}}\hbox{$>$}}}}
\def\lesssim{\mathrel{\hbox{\rlap{\hbox{\lower4pt\hbox{$\sim$}}}\hbox{$<$}}}}

\received{2021 August 9}
\accepted{2021 September 9}
\revised{2021 September 8}
\submitjournal{Astrophysical Journal}
\shortauthors{Homan et al.}
\shorttitle{MOJAVE. XIX. Brightness Temperatures of Blazar Jets}

\begin{document}

\title{MOJAVE XIX: Brightness Temperatures and Intrinsic Properties of Blazar Jets}

\author[0000-0002-4431-0890]{D. C. Homan}
\affiliation{Department of Physics and Astronomy, Denison University, Granville, OH 43023, USA
  \email{homand@denison.edu}
}

\author[0000-0002-8033-5972]{M. H. Cohen}
\affiliation{Department of Astronomy, California Institute of Technology, Pasadena, CA 91125, USA}

\author[0000-0002-2024-8199]{T. Hovatta}
\affiliation{Finnish Centre for Astronomy with ESO, FINCA, University of Turku, Finland}
\affiliation{Aalto University Mets\"ahovi Radio Observatory, Mets\"ahovintie 114,
FI-02540 Kylm\"al\"a, Finland}

\author[0000-0002-0093-4917]{K. I. Kellermann}
\affiliation{National Radio Astronomy Observatory, 520 Edgemont Road,
Charlottesville, VA 22903, USA}

\author[0000-0001-9303-3263]{Y. Y. Kovalev}
\affiliation{Lebedev Physical Institute of the Russian Academy of Sciences, Leninsky prospekt 53, 119991 Moscow, Russia}
\affiliation{Moscow Institute of Physics and Technology, Institutsky per. 9, Dolgoprudny, Moscow region, 141700, Russia}
\affiliation{Max-Planck-Institut f\"ur Radioastronomie, Auf dem H\"ugel 69,
53121 Bonn, Germany}

\author[0000-0003-1315-3412]{M. L. Lister}
\affiliation{Department of Physics and Astronomy, Purdue University, 525 Northwestern Avenue,
West Lafayette, IN 47907, USA
%\email{mlister@purdue.edu}
}

\author[0000-0002-0739-700X]{A. V. Popkov}
\affiliation{Moscow Institute of Physics and Technology, Institutsky per. 9, Dolgoprudny, Moscow region, 141700, Russia}
\affiliation{Lebedev Physical Institute of the Russian Academy of Sciences, Leninsky prospekt 53, 119991 Moscow, Russia}

\author[0000-0002-9702-2307]{A. B. Pushkarev}
\affiliation{Crimean Astrophysical Observatory, 298409 Nauchny, Crimea, Russia}
\affiliation{Lebedev Physical Institute of the Russian Academy of Sciences, Leninsky prospekt 53, 119991 Moscow, Russia}
\affiliation{Moscow Institute of Physics and Technology, Institutsky per. 9, Dolgoprudny, Moscow region, 141700, Russia}

\author[0000-0001-9503-4892]{E. Ros}
\affiliation{Max-Planck-Institut f\"ur Radioastronomie, Auf dem H\"ugel 69,
53121 Bonn, Germany}

\author[0000-0001-6214-1085]{T. Savolainen}
\affiliation{Aalto University Department of Electronics and
  Nanoengineering, PL 15500, FI-00076 Aalto, Finland}
\affiliation{Aalto University Mets\"ahovi Radio Observatory, Mets\"ahovintie 114,
FI-02540 Kylm\"al\"a, Finland}
\affiliation{Max-Planck-Institut f\"ur Radioastronomie, Auf dem H\"ugel 69,
53121 Bonn, Germany}

\begin{abstract}
%250 word limit -- currently at 240
We present multi-epoch, parsec-scale core brightness temperature observations 
of 447 AGN jets from the MOJAVE and 2cm Survey programs at 15 GHz from 1994 to 2019. The 
brightness temperature of each jet over time is characterized by its median 
value and variability.
We find that the range of median brightness temperatures for AGN jets in our 
sample is much larger than the
variations within individual jets, consistent with Doppler boosting being the 
primary difference between the brightness temperatures of jets in their 
median state. We combine the
observed median brightness temperatures with apparent jet speed measurements to
find the typical intrinsic Gaussian brightness temperature of 
$4.1(\pm 0.6)\times10^{10}$~K, suggesting that jet cores are at or 
below equipartition between particle and magnetic field energy in their 
median state.  We use this value to derive estimates for the Doppler factor 
for every source
in our sample. For the 309 jets with both apparent speed and brightness
temperature data, we estimate their Lorentz factors and viewing
angles to the line of sight.  
Within the BL\,Lac optical class, we find
that high-synchrotron-peaked (HSP) BL\,Lacs have smaller Doppler factors,
lower Lorentz factors, and larger angles to the line of sight than 
intermediate and low-synchrotron-peaked (LSP) BL\,Lacs.  We confirm that 
AGN jets with larger Doppler factors measured in their parsec-scale radio cores are 
more likely to be detected in $\gamma$\,rays, and we find a strong correlation
between $\gamma$-ray luminosity and Doppler factor for the detected sources. 
\end{abstract}

\keywords{
Active galaxies ---
Galaxy jets ---
Radio galaxies ---
Quasars ---
BL Lacertae objects ---
Surveys
}

\section{Introduction}
\label{s:intro}

Extra-galactic jets from Active Galactic Nuclei (AGN) flow
outward from the central super-massive black hole (SMBH)/accretion
disk system at nearly the speed of light, and for observers
at a small angle to the jet direction, emission from the 
approaching jet is Doppler boosted and variable, 
creating some of the most spectacular displays in the 
Universe.  The relativistic charged particles and magnetic 
fields that comprise the jets
create broadband synchrotron and inverse-Compton emission that
together span the observable spectrum from radio to TeV 
$\gamma$-rays, and the jets may serve as a source of 
high-energy neutrino emission as well 
\citep[e.g., ][]{2018Sci...361.1378I, PhysRevLett.124.051103,2020AdSpR..65..745K,2020ApJ...894..101P,2021ApJ...908..157P,2021A&A...650A..83H}.  

Unfortunately, the extreme nature of these jets also 
complicates our study of their intrinsic properties and
physical processes.  In addition to Doppler boosting
of the intrinsic emission, the flow of the jets toward us at
nearly the speed of light leads to a compression of
the apparent timescale, creating observed ``superluminal'' 
motions \citep[e.g., ][]{1971ApJ...170..207C} in the jets with  
$\beta_\mathrm{obs} = \beta\sin\theta/(1-\beta\cos\theta)$, where
$\beta$ is the intrinsic speed and $\theta$ is the angle
the jet axis makes with the line of sight.  To untangle
these effects, we need to measure both the observed 
speed of the jet, and its Doppler factor, 
$\delta = 1/[\Gamma(1-\beta\cos\theta)]$, where 
$\Gamma = 1/\sqrt{1-\beta^2}$ is the Lorentz factor of the flow; 
however, Doppler factors are extraordinarily difficult to
measure in synchrotron jets as they lack sharp spectral features
of a known wavelength.

\citet{Readhead94} suggested using the apparent brightness
temperatures of jet cores measured at radio wavelengths, 
along with an assumption of 
equipartition between magnetic field and particle energy
in the emission region to estimate jet Doppler factors.  The radio jet
core in Very Long Baseline Interferometry (VLBI) images is the apparent base
of the jet where the transition from optically thin to optically thick
emission occurs.
In the frame of the host galaxy, the Doppler boosted 
observed brightness temperature in the direction of the
observer is given by $T_\mathrm{b,obs} = \delta T_\mathrm{b,int}$, where 
$T_\mathrm{b,int}$ is the intrinsic, un-boosted brightness 
temperature of the region\footnote{Note that variability brightness
temperatures include two additional powers of $\delta$ due to the estimation
of the angular size by the variability timescale \citep[e.g., ][]
{1999ApJ...521..493L}}.  The assumption of 
equipartition between field and particle energy has been 
used by a number of authors to estimate Doppler factors
from either VLBI data \citep[e.g., ][]{1996ApJ...461..600G,2001ApJ...549L..55T} 
or integrated flux density variability 
\citep[e.g., ][]{1999ApJ...521..493L, 2009A&A...494..527H, 2017MNRAS.466.4625L}.  

\citet{H06} showed that it was
possible to estimate a global value for $T_\mathrm{b,int}$ directly
from VLBI apparent motion and brightness temperature data
without the need to assume equipartition or any other ratio of 
particle to magnetic field energy, and recently 
\citet{2018ApJ...866..137L}
used Doppler factor distributions from population models 
to constrain $T_\mathrm{b,int}$ independent of the assumption
of equipartition.  We also note that the VLBI-based flux-density 
variability approach of \citet{2005AJ....130.1418J} can estimate the
Doppler factor of a moving jet feature from its angular size 
and variability timescale without any assumptions about its brightness
temperature.

In this paper we present multi-epoch, parsec-scale
core brightness temperature observations of 447 AGN
jets from the MOJAVE program \citep[e.g., ][]{2005AJ....130.1389L, 
2018ApJS..234...12L}, and we combine those 
observations with apparent speed 
measurements in 309 of our jets by \citet[][hereafter \citetalias{MOJAVE_XVIII}]{MOJAVE_XVIII}.  We use our multi-epoch Very Long Baseline
Array (VLBA) 
observations from the entire available
span of the MOJAVE and 2cm Survey programs, from 1994 to 2019, 
to characterize the brightness temperature of each jet core over time 
by its median value and variability, and by comparing the jets to one 
another in their median state, we strengthen our confidence that a 
single representative value of $T_\mathrm{b,int}$ can apply broadly 
across our sample. 
Rather than assume equipartition, we follow \citet{H06} 
and combine our median brightness temperature observations
with apparent speed measurements to estimate the global 
value for $T_\mathrm{b,int}$.  As a result of this analysis we 
obtain estimates of the Doppler factor for almost every source
in our sample, and for the 309 jets where we have apparent speed 
measurements, we also estimate their Lorentz factors and 
jet viewing angles to the line of sight.   We compare these
intrinsic properties between sources as a function of their optical
class, spectral energy distribution (SED) peak frequency,
and $\gamma$-ray properties, and we discuss the implications
of our measurement of $T_\mathrm{b,int}$ for the energy balance between
particles and magnetic fields in jet cores.

The paper is organized as follows.  In \autoref{s:data} 
we describe our data analysis, including both our methods
for measuring brightness temperatures and for combining those
measurements with apparent jet speeds to find $T_\mathrm{b,int}$ and
estimate the intrinsic properties of the jets.  In \autoref{s:discuss}
we present and discuss our results, and we summarize our conclusions
in \autoref{s:conclude}.  We assume a $\Lambda$CDM cosmology 
with $H_0=71$ km s$^{-1}$ Mpc$^{-1}$, $\Omega_\Lambda = 0.73$, 
and $\Omega_M=0.27$ \citep{2009ApJS..180..330K} throughout the paper.

\section{Data Analysis}
\label{s:data}

Our sample consists of the 447 AGN recently studied by the MOJAVE 
program for kinematics in \citetalias{MOJAVE_XVIII}, of which 206 are 
members of the MOJAVE\,1.5\,Jy quarter-century (QC) flux-density limited \href{https://www.physics.purdue.edu/astro/MOJAVE/sample.html}{sample} 
selected on the basis of parsec-scale jet emission 
\citep[e.g., ][]{LHH19}. Our whole sample of 447 AGN includes
sources that are outside the 1.5 Jy QC sample added over the years for a variety
of reasons including their high energy emission and membership in other AGN monitoring programs, but all have a minimum 15 GHz correlated flux density
larger than $\sim 50$ mJy and J2000 declinations $> -30^\circ$ as
described in \citetalias{MOJAVE_XVIII}.
\autoref{t:sources} lists the sources in our 
sample along with several of their properties. For each source 
we measure its core brightness temperature as described in 
\autoref{s:measureTb} in all the 15~GHz VLBA epochs analyzed 
by our program through August 6, 2019, and in 
\autoref{s:betaT_analysis} we describe our method that combines the 
brightness temperature observations with 
apparent speeds from \citetalias{MOJAVE_XVIII} to estimate Doppler factors 
($\delta$) Lorentz factors ($\Gamma$) and viewing angles to the line 
of sight ($\theta$) for sources that have the necessary information.

\begin{deluxetable*}{lllcccccr} 
\tablecolumns{9} 
\tabletypesize{\scriptsize} 
%\rotate
\tablewidth{0pt}  
\tablecaption{\label{t:sources} Source Properties}  
\tablehead{
\colhead{Source} & \colhead{Alias} &  \colhead {$z$} & \colhead{Class} & \colhead{MOJ 1.5} &
\colhead{Spectrum} & \colhead{$\nu_\mathrm{peak,obs}$} & 
\colhead{$L_{\gamma}$} & \colhead {References} \\
\colhead{} &   \colhead {} &  \colhead {} & \colhead{} & \colhead{} &
\colhead{} & \colhead{(log$_{10}$ Hz)} & \colhead{(log$_{10}$ ergs/s)} & \colhead {} \\
\colhead{(1)} & \colhead{(2)} & \colhead{(3)} & \colhead{(4)} &  \colhead{(5)} &
\colhead{(6)} & \colhead{(7)} & \colhead{(8)} & \colhead{(9)} 
}
\startdata 
0003$+$380 & \objectname{S4 0003+38} & $0.229$ & Q & N & LSP & $13.14$ & $45.12$ & \cite{1994AAS..103..349S},1 \\ 
0003$-$066 & \objectname{NRAO 005} & $0.3467$ & B & Y & LSP & $12.92$ & $44.81$ & \cite{2005PASA...22..277J},2 \\ 
0006$+$061 & \objectname{TXS 0006+061} & \ldots & B & N & LSP & $13.44$ & \ldots & \cite{2012AA...538A..26R},1 \\ 
0007$+$106 & \objectname{III Zw 2} & $0.0893$ & G & Y & LSP & $13.30$ & \ldots & \cite{1970ApJ...160..405S},3 \\ 
0010$+$405 & \objectname{4C +40.01} & $0.256$ & Q & N & LSP & $12.79$ & $44.59$ & \cite{1992ApJS...81....1T},2 \\ 
0011$+$189 & \objectname{RGB J0013+191} & $0.477$ & B & N & LSP & $13.67$ & $45.41$ & \cite{2013ApJ...764..135S},2 \\ 
0012$+$610 & \objectname{4C +60.01} & \ldots & U & N & LSP & $13.11$ & \ldots & \ldots,1 \\ 
0014$+$813 & \objectname{S5 0014+813} & $3.382$ & Q & N & LSP & $12.50$ & \ldots & \cite{1987AZh....64..262V},3 \\ 
0015$-$054 & \objectname{PMN J0017-0512} & $0.226$ & Q & N & LSP & $13.60$ & $45.27$ & \cite{2012ApJ...748...49S},1 \\ 
0016$+$731 & \objectname{S5 0016+73} & $1.781$ & Q & Y & LSP & $12.32$ & $47.91$ & \cite{1986AJ.....91..494L},2 \\ 
\ldots \\
\enddata
\tablecomments{
The complete version of this table appears in the online journal.
Columns are as follows: 
(1) Source name in B1950.0 coordinates;
(2) Alias;
(3) Redshift;
(4) Optical Class (Q=quasar, B=BL\,Lac, G=radio galaxy, 
N=narrow-line Seyfert 1, U=unknown);
(5) Member of the MOJAVE 1.5 Jy QC Sample (Y = yes, N = no);
(6) SED Class (LSP/ISP/HSP = Low/Intermediate/High Synchrotron Peaked);
(7) SED Peak in Observer Frame;
(8) $\gamma$-ray luminosity, computed as described in \autoref{s:Tb-trends};
(9) References for Redshift/Optical Classification, SED property references are as follows:
1 = \cite{2015ApJ...810...14A},
2 = \cite{4LAC},
3 = ASDCfit, \cite{2011arXiv1103.0749S},
4 = \cite{2011ApJ...740...98M},
5 = \cite{2015MNRAS.450.3568X},
6 = \cite{2017AA...598A..17C},
7 = \cite{2008AA...488..867N},
8 = \cite{2017ApJS..232...18A},
9 = \cite{2011ApJ...743..171A},
10 = \cite{2009ApJ...707L.142A},
11 = \cite{2006AA...445..441N},
12 = \cite{3HSP},
13 = \cite{2009ApJ...707...55A}, and
14 = \cite{2015AA...578A..69H}
}
\end{deluxetable*}

\subsection{Measuring Core Brightness Temperatures}
\label{s:measureTb}

We measure the brightness temperature in the core region
in each epoch by fitting a single elliptical Gaussian in 
the $(u,v)$-plane. The core region is isolated by first 
starting with our final \textsc{clean} image of the jet and using
the Caltech VLBI program, \textsc{Difmap} \citep{difmap97, difmap11}, 
to delete the \textsc{clean} components around the core location in 
an area equal in size to the full-width half-maximum dimensions
of the naturally weighted beam.  In some cases, this area may 
be enlarged somewhat if doing so reduces the final $\chi^2$ of 
the fitted Gaussian. The central location for the area from 
which the \textsc{clean} components are deleted is either the pixel 
closest to the core location as used in our kinematics fits 
\citepalias{MOJAVE_XVIII} or the nearest local maximum if a local maximum 
can be found within half a beam-width of the kinematics core 
location.
The deleted \textsc{clean} components are replaced with a single 
elliptical Gaussian which is fit 
in the $(u,v)$-plane. The result is a hybrid Gaussian/\textsc{clean} component 
model, with the Gaussian properties representing the core region (near 
optical depth equals unity) and with \textsc{clean} components modeling 
the remainder of the source structure.

\begin{figure*}
\centering
\includegraphics[scale=0.34,angle=-90]{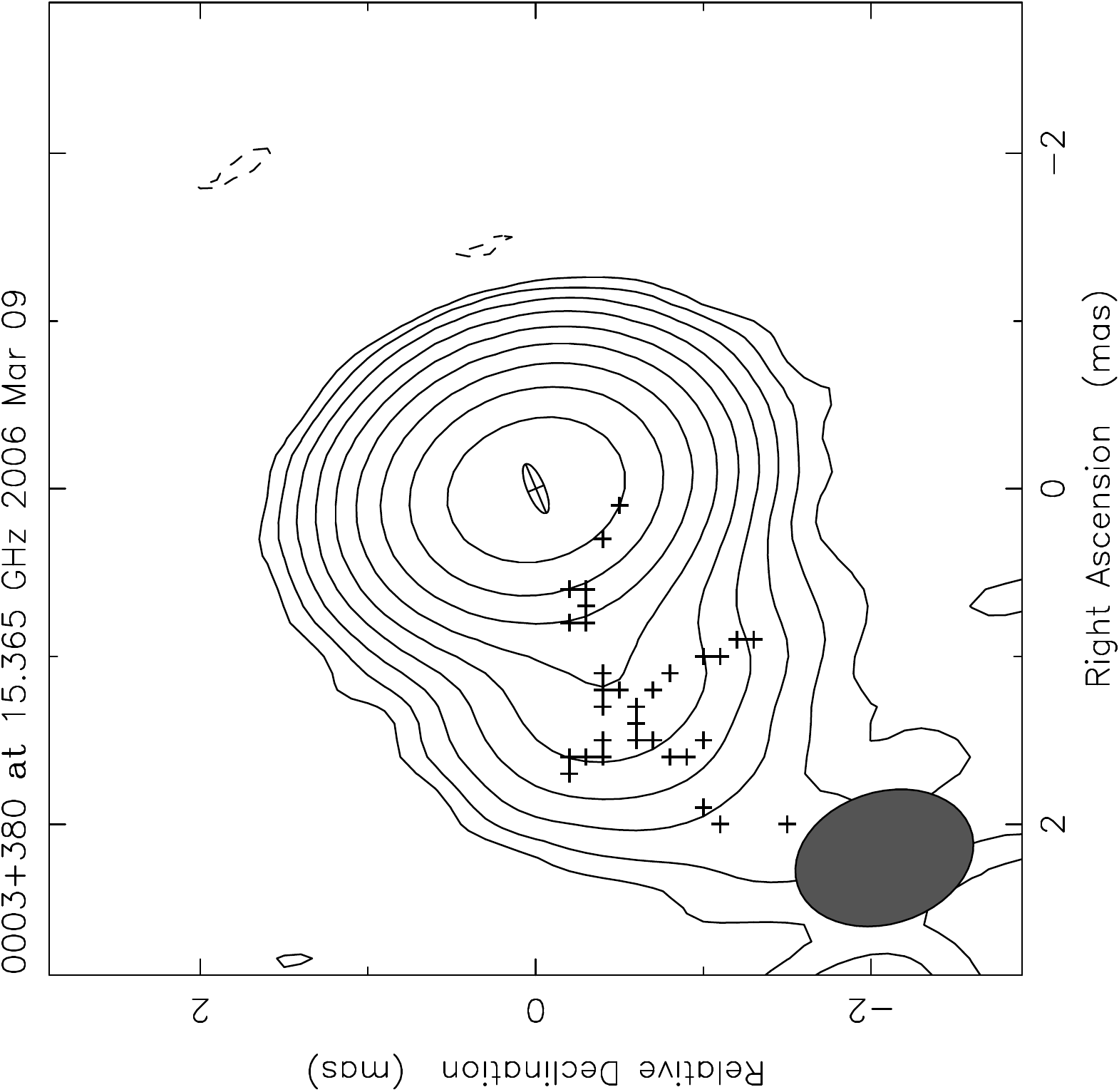}
\includegraphics[scale=0.34,angle=-90]{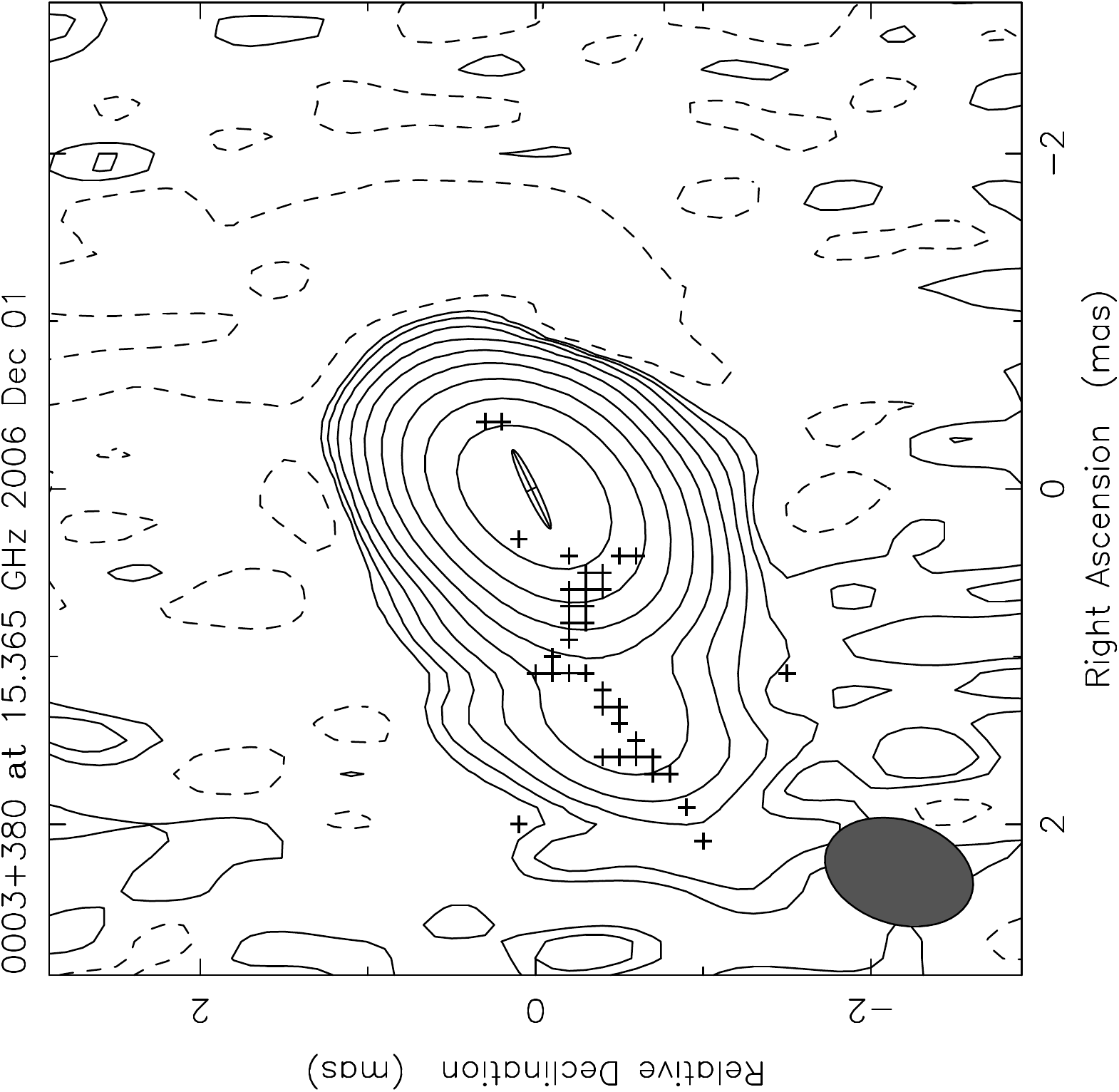}
\includegraphics[scale=0.33,angle=-90]{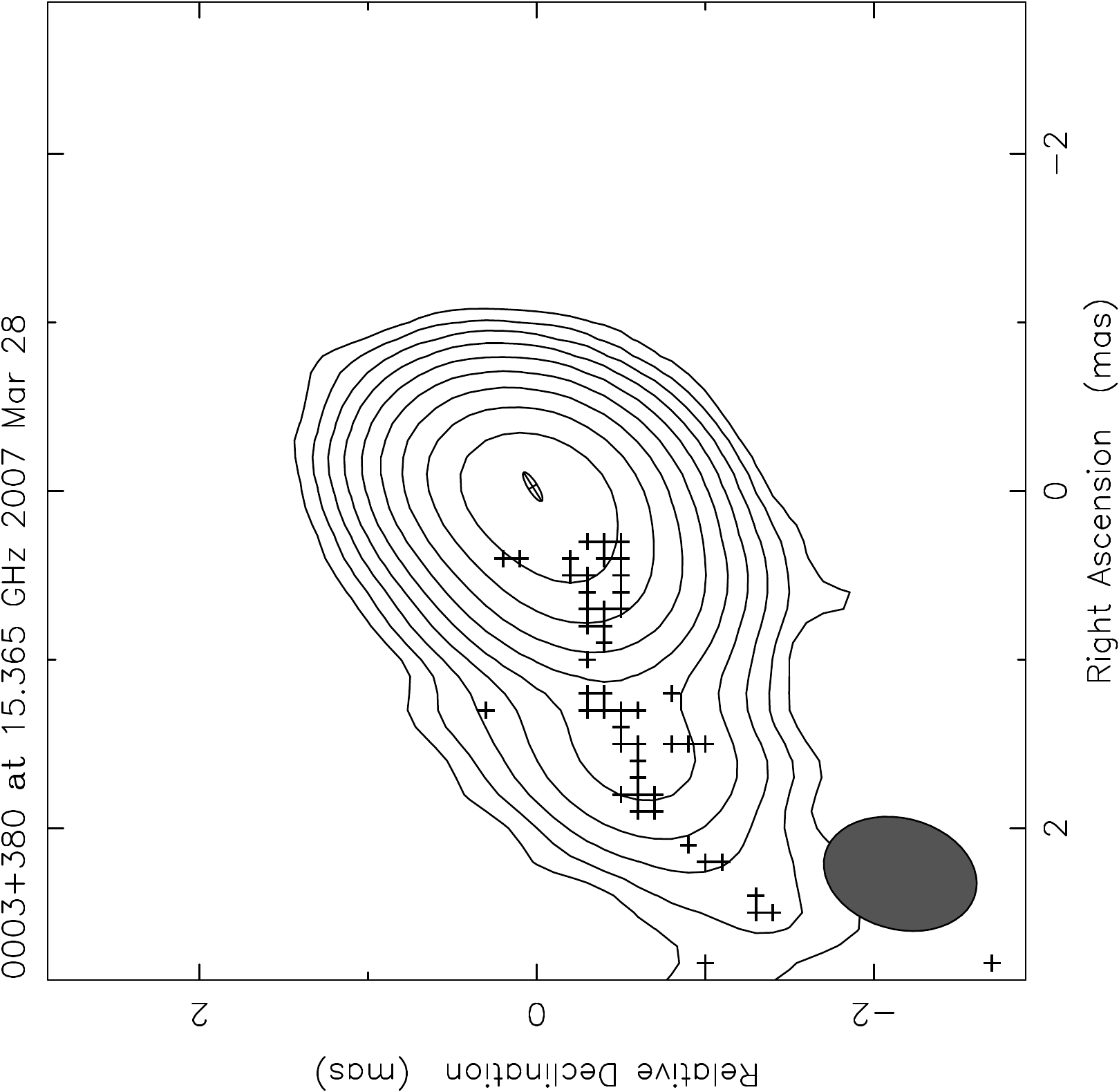}
\includegraphics[scale=0.34,angle=-90]{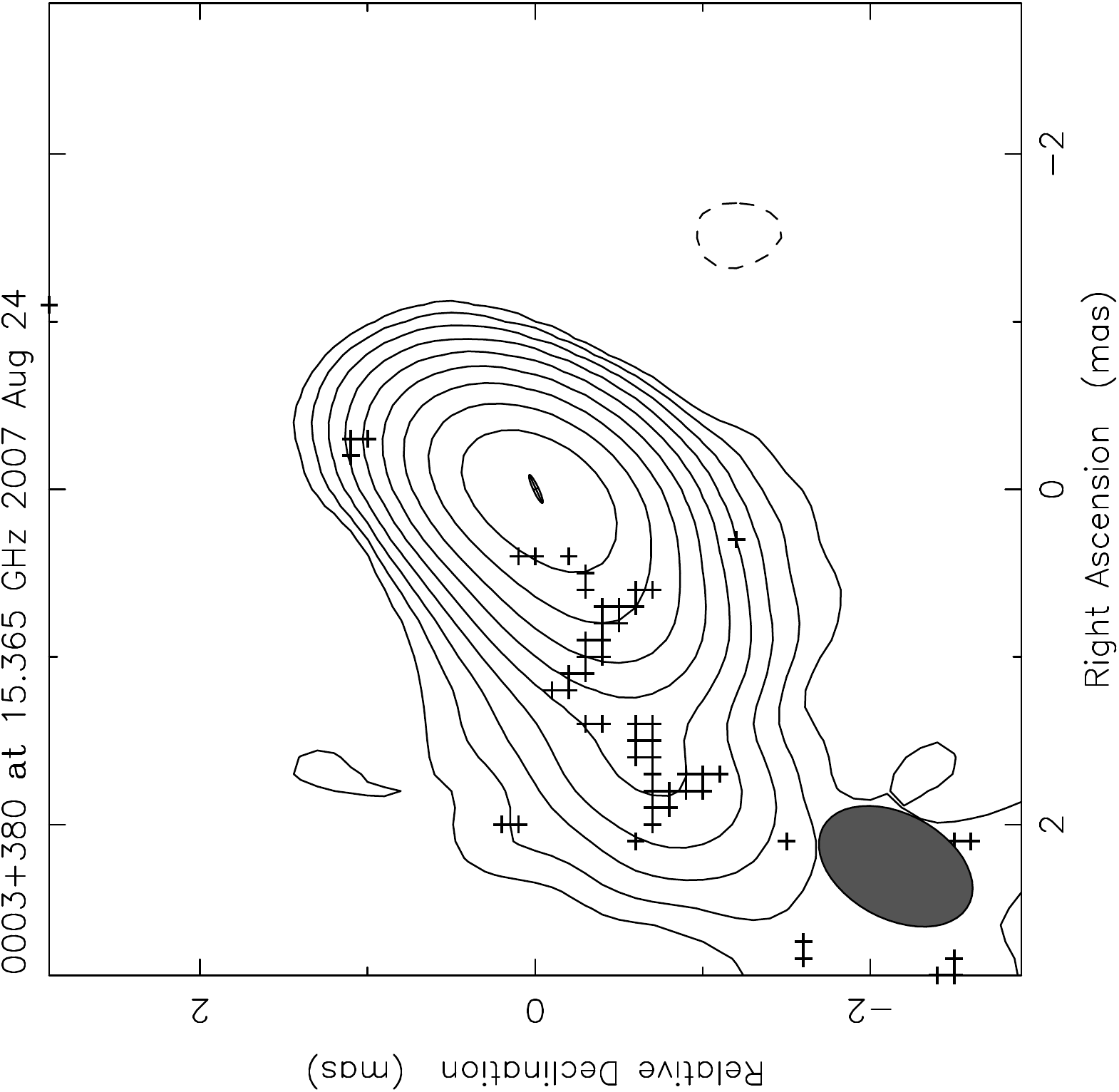}
\includegraphics[scale=0.33,angle=-90]{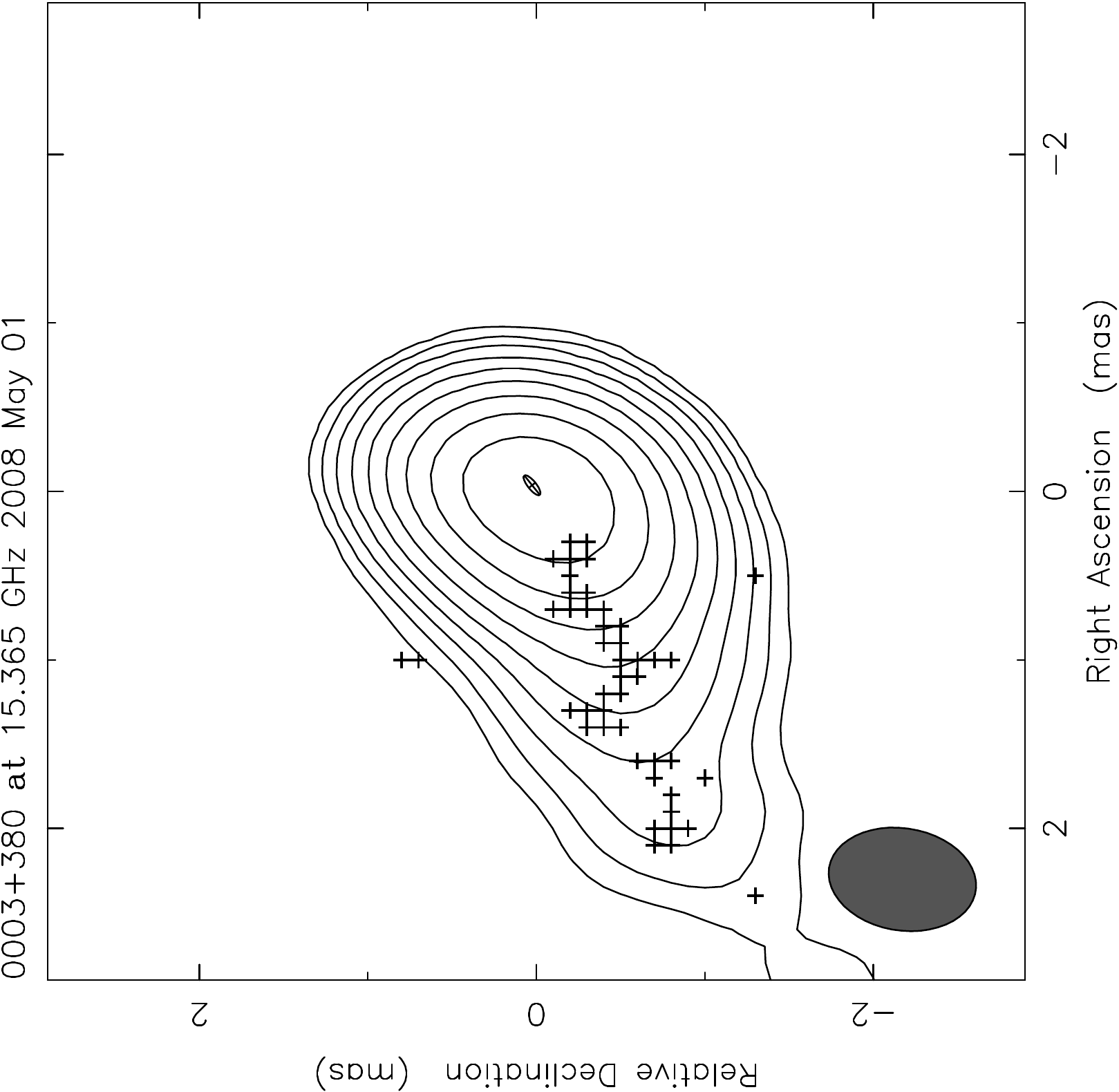}
\includegraphics[scale=0.33,angle=-90]{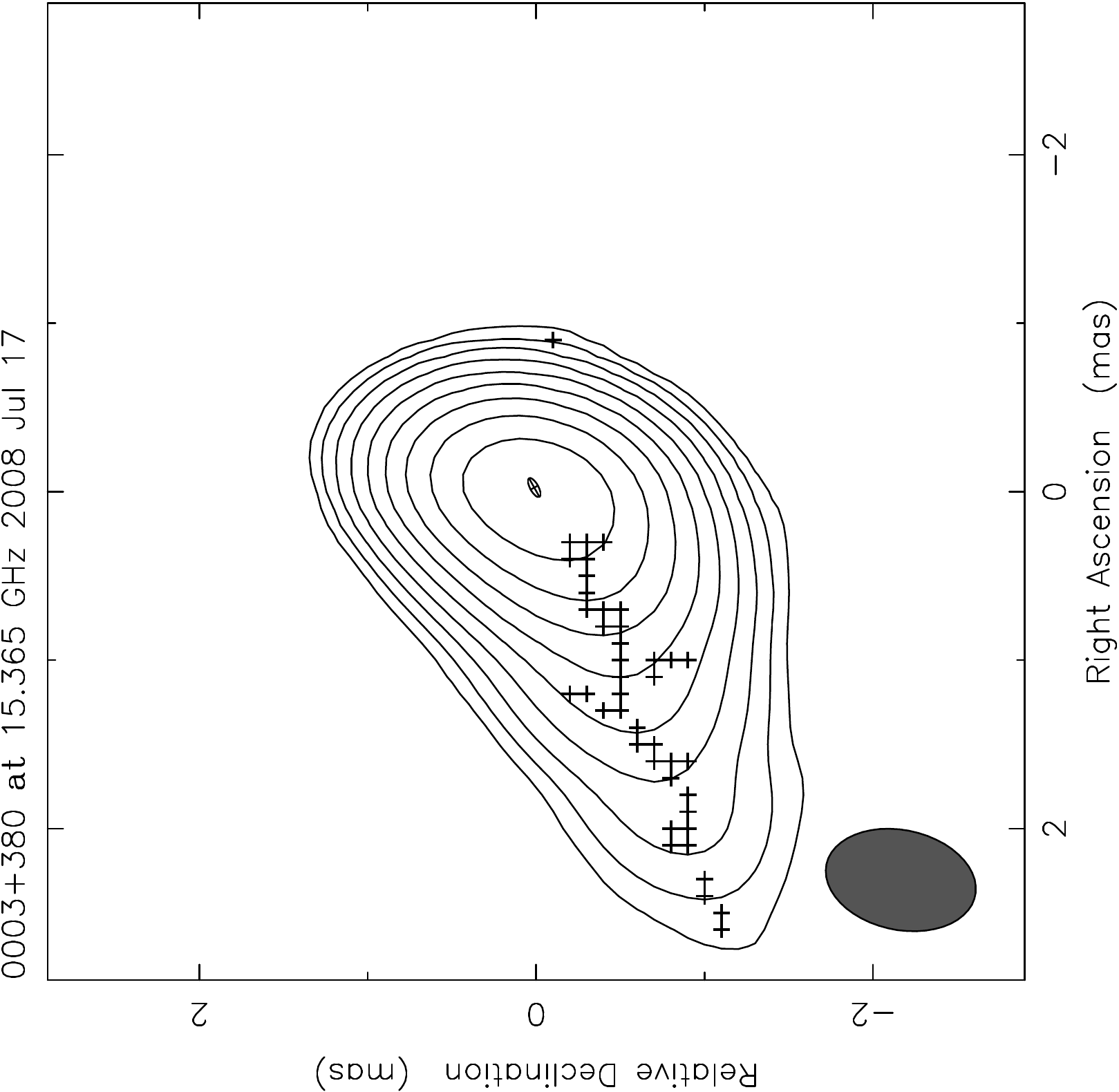}
\figcaption{\label{f:fig1}
Naturally weighted images illustrating the modeling of 
the core region of 0003+380 in our first six epochs.  Contours begin at 
0.2\% and increase in factors of two until 51.2\% of the peak 
intensity of 0.543, 0.363, 0427, 0.417, 0.601, 0.545\,Jy/beam 
in each epoch respectively.  The full-width half-maximum (FWHM) dimensions of the restoring 
beam are illustrated by the filled ellipse in the lower left corner 
of each image.  As described in the text, \textsc{clean} components (crosses) 
from the core region are replaced by a single Gaussian component 
(ellipse). The increased noise in the second epoch is due to a newly emerging feature that is too close to the core to be resolved by this procedure, as described in \autoref{s:data}.
}
\end{figure*}

\autoref{f:fig1} illustrates this technique by showing the inner jet of 
the source 0003$+$380 over its first six epochs.  Because the 
entire core region is modeled by a single Gaussian, this approach will
average over any substructure, and will occasionally lead to noisier
than average fits, such as in the second epoch illustrated in 
\autoref{f:fig1}.  In this epoch, a newly emerging feature in the jet is not
sufficiently distinct from the core region to be modeled by the \textsc{clean}
components directly.  In these cases, it is tempting to fit a
second Gaussian component, and indeed we experimented with a 
multi-Gaussian approach.  However, it is difficult to define robust
criteria under which two Gaussians should replace a single Gaussian 
while still producing a reliable brightness temperature measurement of 
the core region.  By sticking to a single Gaussian in all cases 
we ensure consistency across epochs and between sources while 
allowing that there will be times where the emergence of a new feature 
may enlarge the core region and possibly reduce the measured brightness
temperature.  We report measured brightness temperatures in the 
frame of the host galaxy as the peak brightness temperature of 
the fitted Gaussian \citep[e.g., ][]{K05}

\begin{equation}
\label{e:Tb}
T_b = 1.22\times10^{12}\frac{S_G(1+z)}{\Omega_\mathrm{maj}\Omega_\mathrm{min}\nu^2_\mathrm{obs}} \,\mathrm{K}\,,
\end{equation}

\noindent where $z$ is the source redshift, $S_G$ is the 
integrated flux density of the fitted Gaussian in Jy,
$\Omega_\mathrm{maj,min}$ are the full-width half-maximum (FWHM) 
dimensions of the Gaussian in milliarcseconds, and $\nu_\mathrm{obs}$ 
is the observing frequency in GHz.  The result is in the rest frame of 
the host galaxy. \autoref{t:comp_fits} lists the properties of the 
brightness temperature fit in every epoch for each source.  
Upper limits on our measured angular
sizes were determined in one of two ways: either (1) following \citet{K05}
where the signal to noise ratio $SNR = S_G/\sigma_\mathrm{rms}$,
or (2) by enlarging the angular size of the fitted Gaussian 
until the normalized $\chi^2$ of the fit increased by 1.0.
Unresolved features have their upper limit size reported as the 
larger of methods (1) and (2) in \autoref{t:comp_fits}.

%\documentclass[10pt,preprint]{aastex}

%\begin{document}

%\section{Full Table 2, Starts Next Page} 

\begin{deluxetable*}{llrrrrrrrrrrrrr} 
\tablenum{2}
\tablecolumns{15} 
\tabletypesize{\scriptsize} 
\rotate
\tablewidth{0pt}  
\tablecaption{\label{t:comp_fits} Brightness Temperature Fitting Results}  
\tablehead{
\colhead{} &   \colhead {} & \colhead {$\nu_\mathrm{obs}$} & \colhead{$B_\mathrm{maj}$} & \colhead{$B_{min}$} & 
\colhead{$B_\mathrm{PA}$} & \colhead {$C_X$} &  \colhead{$C_Y$} & \colhead{$C_{fact}$} &
\colhead{$S_\mathrm{G}$} & \colhead{$\Omega_\mathrm{maj}$} &
\colhead{$\Omega_\mathrm{min}$} & \colhead {$\Omega_\mathrm{PA}$} &   \colhead{$\sigma_\mathrm{rms}$} & \colhead{$T_\mathrm{b}$} \\
\colhead{Source} &   \colhead {Epoch} & \colhead{(GHz)} & \colhead{(mas)} & \colhead{(mas)} &
\colhead{(deg)} &   \colhead {(mas)} & \colhead{(mas)} & \colhead{} &
\colhead{(Jy)} & \colhead{(mas)} &
\colhead{(mas)} & \colhead {(deg)} &   \colhead {(mJy/bm)} & \colhead{(log$_{10}$ K)} \\
\colhead{(1)} & \colhead{(2)} & \colhead{(3)} & \colhead{(4)} &  \colhead{(5)} &
\colhead{(6)} & \colhead{(7)} & \colhead{(8)} & \colhead{(9)} &  \colhead{(10)} &
\colhead{(11)} & \colhead{(12)} & \colhead{(13)} & \colhead{(14)} & \colhead {(15)}  
}
\startdata 
% Created on Wed Jul  7 14:21:02 EDT 2021 for 0003+380 by make_sourceTb_plot_tab.pl of 2021_07_07
% Columns are described in corresponding sample_Tb_table.dat 
0003$+$380  & 2006 Mar 09 & $15.37$ & $1.01$ & $0.73$ & $ 17.6$ & $ 0.0$ & $ 0.0$ & $1.00$ & $ 0.586$ & $  0.317$ & $  0.103$ & $-67.1$ & $  1.90$ & $ 11.057$ \\ 
          & 2006 Dec 01 & $15.37$ & $0.85$ & $0.58$ & $-17.4$ & $ 0.0$ & $ 0.0$ & $1.00$ & $ 0.433$ & $  0.520$ & $  0.067$ & $-64.3$ & $  4.00$ & $ 10.893$ \\ 
          & 2007 Mar 28 & $15.37$ & $0.86$ & $0.61$ & $-14.9$ & $ 0.0$ & $ 0.0$ & $1.00$ & $ 0.399$ & $  0.200$ & $  0.056$ & $-57.8$ & $  1.45$ & $ 11.356$ \\ 
          & 2007 Aug 24 & $15.37$ & $0.92$ & $0.58$ & $-28.1$ & $ 0.0$ & $ 0.0$ & $1.00$ & $ 0.408$ & $  0.185$ & $< 0.036$ & $-65.1$ & $  1.17$ & $>11.594$ \\ 
          & 2008 May 01 & $15.37$ & $0.82$ & $0.57$ & $ -9.1$ & $ 0.0$ & $ 0.0$ & $1.00$ & $ 0.545$ & $  0.146$ & $  0.049$ & $-51.2$ & $ 0.159$ & $ 11.682$ \\ 
          & 2008 Jul 17 & $15.37$ & $0.84$ & $0.55$ & $-11.9$ & $ 0.0$ & $ 0.0$ & $1.00$ & $ 0.511$ & $  0.126$ & $  0.049$ & $-61.1$ & $ 0.203$ & $ 11.721$ \\ 
          & 2009 Mar 25 & $15.36$ & $0.85$ & $0.62$ & $-12.3$ & $ 0.0$ & $ 0.0$ & $1.00$ & $ 0.346$ & $  0.284$ & $< 0.060$ & $-63.6$ & $  2.31$ & $>11.107$ \\ 
          & 2010 Jul 12 & $15.36$ & $0.89$ & $0.54$ & $-12.3$ & $ 0.0$ & $ 0.0$ & $1.00$ & $ 0.378$ & $  0.325$ & $  0.067$ & $-71.4$ & $  2.50$ & $ 11.041$ \\ 
          & 2011 Jun 06 & $15.36$ & $0.91$ & $0.54$ & $-10.2$ & $ 0.0$ & $ 0.0$ & $1.00$ & $ 0.472$ & $  0.138$ & $  0.061$ & $-61.0$ & $ 0.371$ & $ 11.553$ \\ 
          & 2013 Aug 12 & $15.36$ & $0.84$ & $0.53$ & $ -4.1$ & $ 0.0$ & $ 0.0$ & $1.00$ & $ 0.581$ & $  0.159$ & $  0.045$ & $-61.6$ & $ 0.846$ & $ 11.713$ \\ 
\ldots \\
% Created on Wed Jul  7 14:21:12 EDT 2021 for 0118-272 by make_sourceTb_plot_tab.pl of 2021_07_07
% Columns are described in corresponding sample_Tb_table.dat 
0118$-$272  & 2009 Dec 26 & $15.36$ & $1.43$ & $0.50$ & $ -6.7$ & $-0.1$ & $ 0.1$ & $1.00$ & $ 0.195$ & $  0.186$ & $  0.097$ & $-33.9$ & $ 0.156$ & $>10.746$\tablenotemark{a} \\ 
          & 2010 Sep 17 & $15.36$ & $1.52$ & $0.49$ & $ -6.9$ & $ 0.0$ & $ 0.0$ & $1.00$ & $ 0.190$ & $  0.181$ & $  0.076$ & $-27.9$ & $ 0.154$ & $>10.856$\tablenotemark{a} \\ 
          & 2011 Jul 15 & $15.36$ & $1.46$ & $0.52$ & $ -7.7$ & $ 0.0$ & $ 0.0$ & $1.00$ & $ 0.179$ & $  0.155$ & $  0.056$ & $-35.9$ & $ 0.216$ & $>11.029$\tablenotemark{a} \\ 
          & 2012 Jul 12 & $15.36$ & $1.61$ & $0.48$ & $-10.4$ & $ 0.0$ & $ 0.0$ & $1.00$ & $ 0.192$ & $  0.243$ & $  0.083$ & $-23.9$ & $ 0.325$ & $>10.693$\tablenotemark{a} \\ 
          & 2013 Jul 08 & $15.36$ & $1.42$ & $0.50$ & $ -3.3$ & $ 0.0$ & $ 0.0$ & $1.00$ & $ 0.263$ & $  0.183$ & $  0.048$ & $-21.4$ & $ 0.543$ & $>11.188$\tablenotemark{a} \\ 
\ldots \\
\enddata
\tablenotetext{a}{Lower limit value ($z=0$) only on account of unknown source redshift.}
\tablecomments{
The complete version of this table appears in the online journal.
Columns are as follows: 
(1) Source name in B1950 coordinates; 
(2) Epoch;
(3) Central Observing Frequency;
(4)$-$(6) Dimensions of naturally weighed beam;
(7)$-$(8) Center location of removed clean component area;
(9) Factor times beam dimensions used for removing clean components;
(10) Flux density of fitted Gaussian;
(11)$-$(13) Dimensions of fitted Gaussian and its position angle;
(14) RMS residual noise in an region twice the beam dimensions centered at ($C_X$,$C_Y$); 
(15) Peak brightness temperature of the fitted Gaussian in rest frame of host galaxy;   
}
\end{deluxetable*} 

%\end{document}

To test the validity of our approach, we generated a set of optically thin,
homogeneous spherical models, each with 1.0\,Jy of flux density but a range
of diameters: 0.010, 0.025, 0.050, 0.100, 0.250, 0.500, 1.000, 
and 2.000\,milli-arcseconds.  This range of size encompasses 
completely unresolved structure all the way through objects with 
significant structure beyond the one-beam area around the center 
where the Gaussian will be fit.  We used the National Radio Astronomy
Observatory's AIPS package \citep{2003ASSL..285..109G} \textsc{uvmod} task to substitute 
these models and thermal noise into the $(u,v)$-coverage of 
several epochs of two different sources: 0415$+$379 and 1510$-$089.
The goal here was to see how this approach to measuring brightness
temperature might depend on $(u,v)$-coverage as it varies over epochs 
or between sources.  Each resulting simulated data set was
first \textsc{\textsc{clean}}'ed in the same fashion as our MOJAVE data and 
then analyzed using the approach described above.  With the 
exception of a small fraction of cases, almost all of the models 
with diameters $<$0.050\,milliarcseconds were unresolved, while 
most of those with diameters 0.050\,milliarcseconds or larger 
were resolved. For each source/diameter combination 
of 0.050\,milliarcseconds or larger, we were able to extract a median 
Gaussian peak brightness temperature across the simulated epochs 
and compare to the expected brightness temperature at the center 
of the sphere for the corresponding case.  We should not expect a 
ratio of 1.0, as a Gaussian is more sharply peaked than a 
sphere, and indeed we found the average ratio was $1.81$.  
This ratio was roughly the same from 0.050 through 2.000 
milliarcseconds with a standard deviation of 0.15 and no 
trend with assumed sphere diameter, indicating that in the large diameter 
cases the remaining \textsc{\textsc{clean}} components that represent the extended 
parts of the structure do not affect the ability of the Gaussian 
to represent the brightness temperature at the center.
Note that in five of our six resolved models, the source template with low 
declination $(u,v)$-coverage had a larger median brightness temperature resulting in 
an average difference of $10\pm4$ \% compared to the high declination template, so 
differing $(u,v)$-coverage between sources may introduce a modest level of 
uncertainty into our measurements.

As an important aside, the ratio of 1.8 between the expected 
central brightness temperature of a homogeneous sphere and the 
measured Gaussian peak brightness temperature illustrates the point 
that brightness temperatures derived from fitted Gaussian parameters 
may be too large in regions that are not peaked as sharply as a 
Gaussian.  It is difficult to know how the brightness distribution
of the inhomogeneous base of a possibly conical or parabolic jet 
will be represented by the single Gaussian fits used in this 
analysis, so some caution should be used in interpreting these 
temperatures directly in terms of the energy balance between
magnetic fields and particles in the jet, discussed in \autoref{s:energy-balance};
however, we note that this constant geometrical factor does not affect
any other aspect of our analysis as it simply divides out of our
estimates of the Doppler factor\footnote{This is because 
$\delta = T_\mathrm{b,obs}/T_\mathrm{b,int}$ and both quantities include the same 
geometrical factor given our method for determining $T_\mathrm{b,int}$ 
described in \autoref{s:measureTb_int}}.

\begin{figure*}
\centering
\includegraphics[scale=0.48]{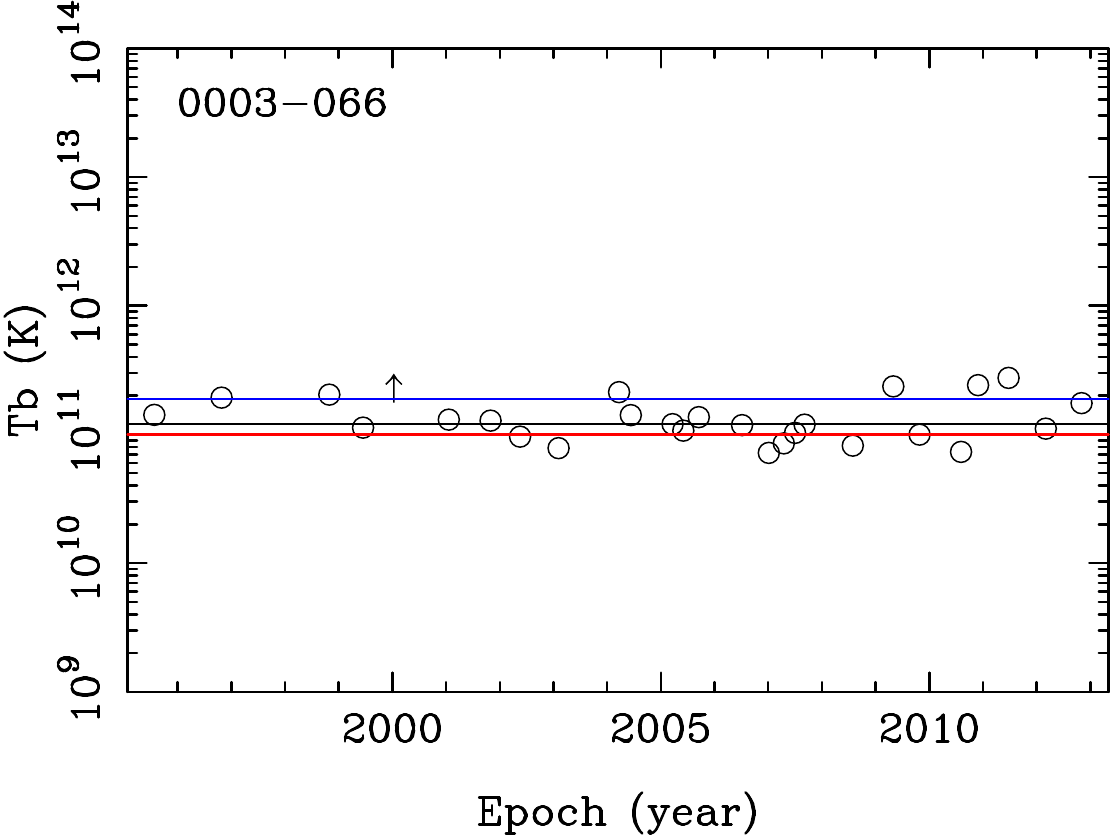}
\includegraphics[scale=0.48]{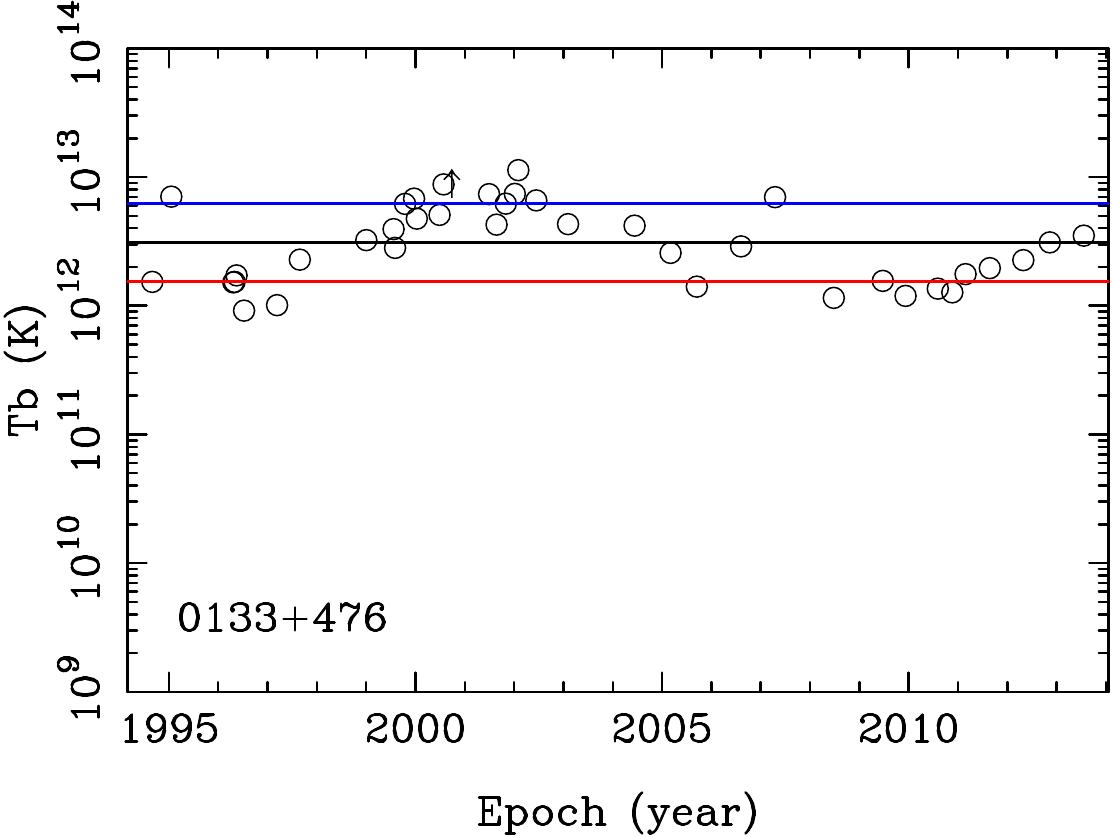}
\includegraphics[scale=0.48]{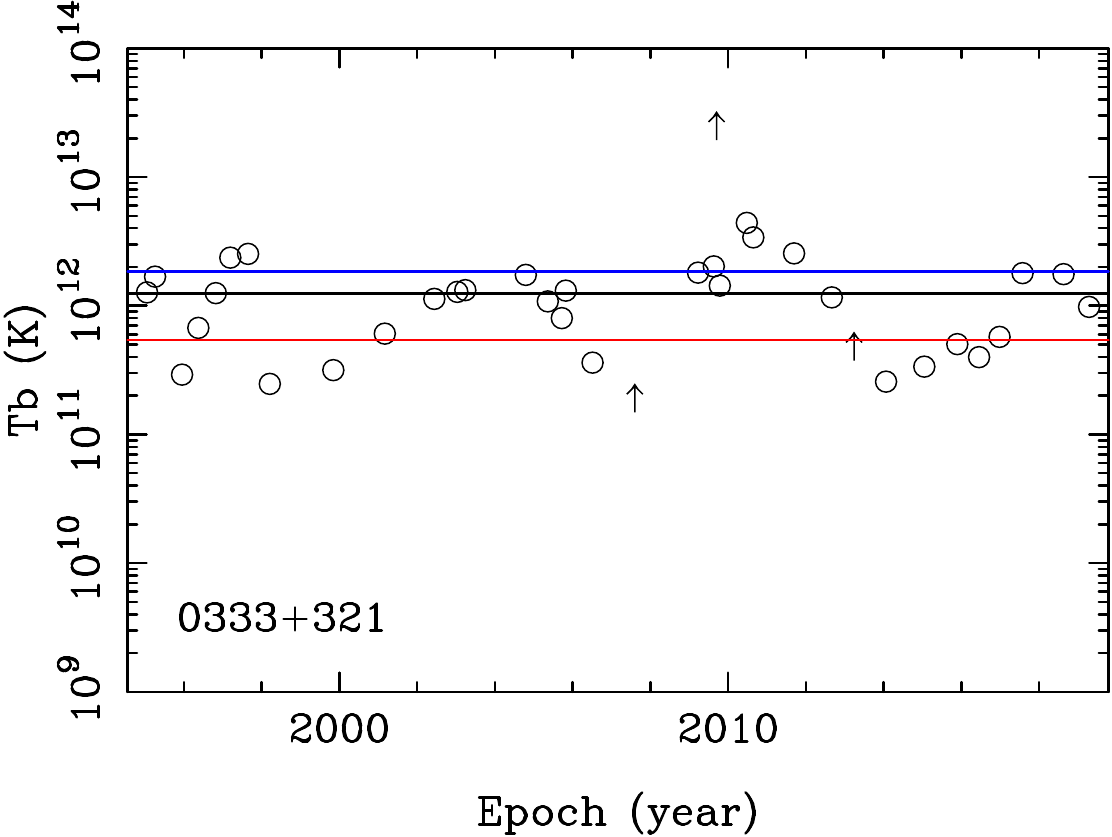}
\includegraphics[scale=0.48]{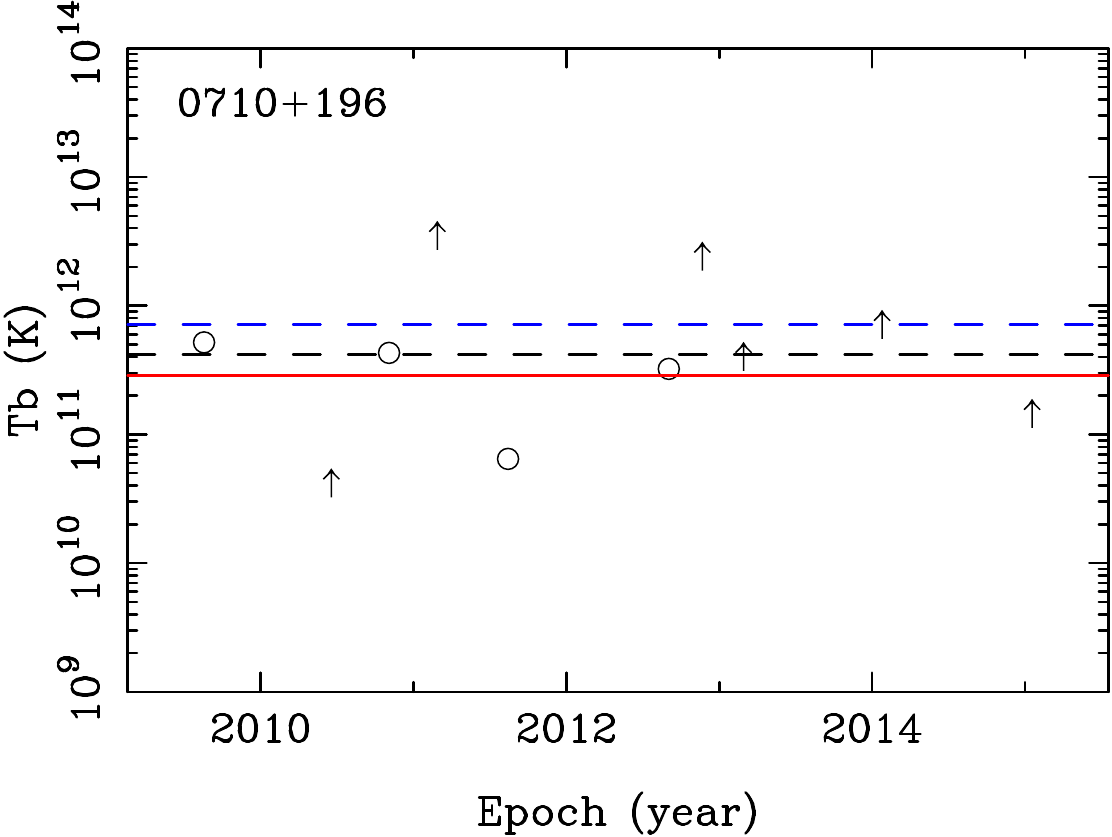}
\includegraphics[scale=0.48]{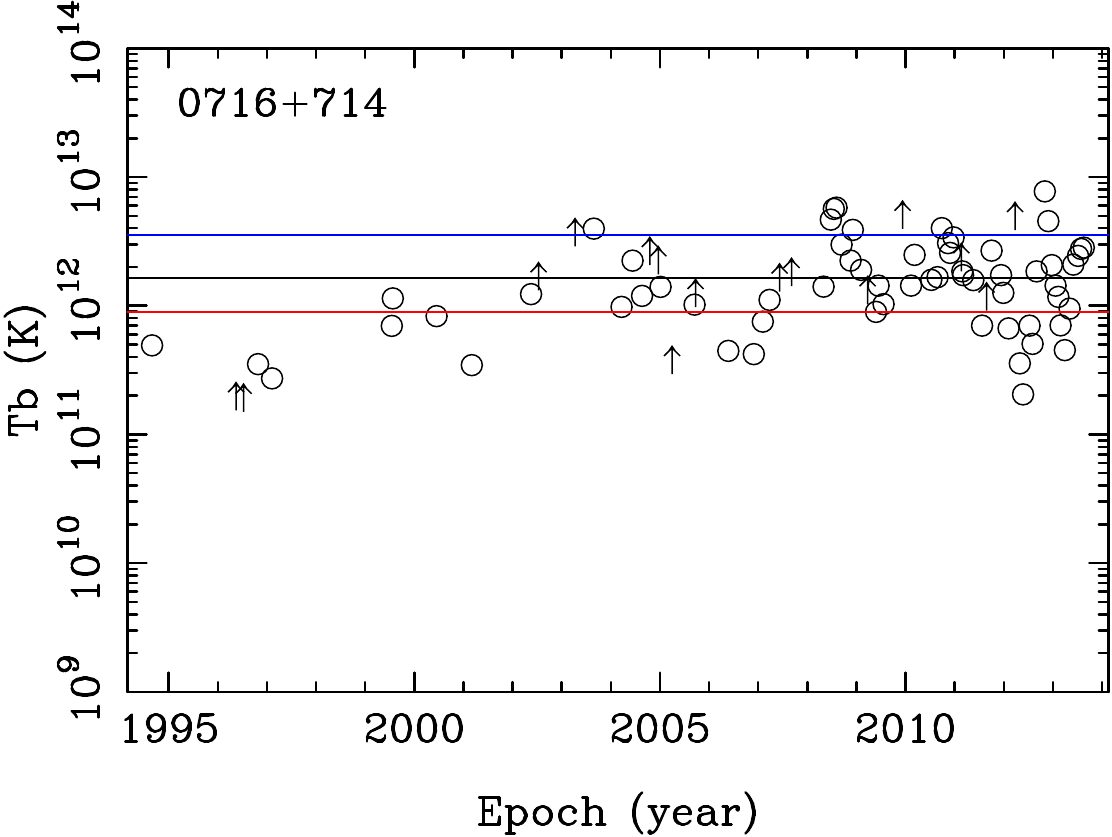}
\includegraphics[scale=0.48]{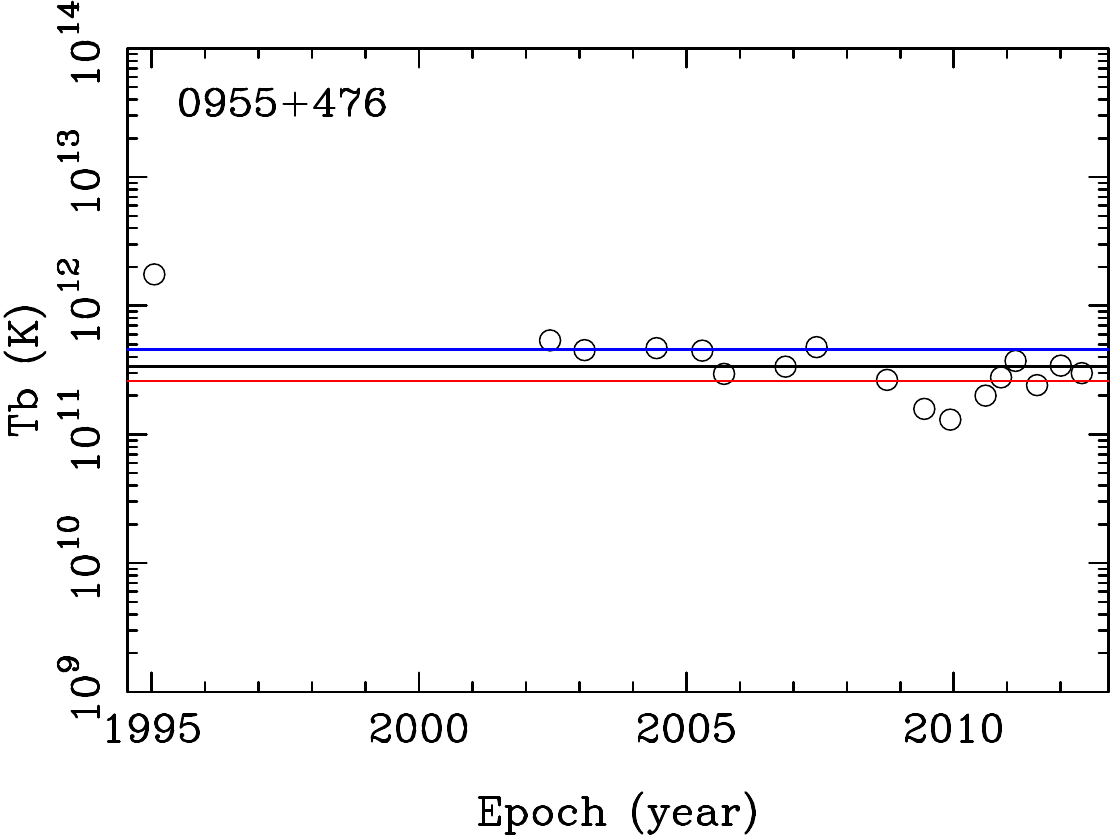}
\includegraphics[scale=0.48]{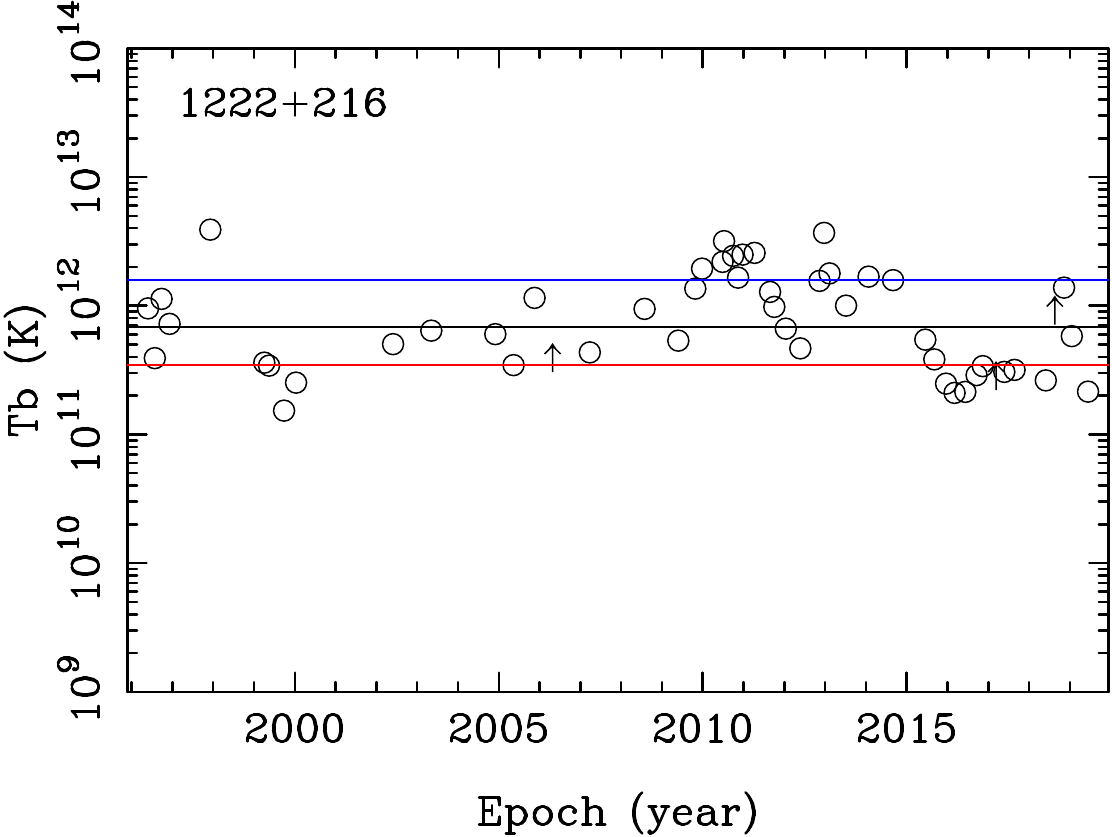}
\includegraphics[scale=0.48]{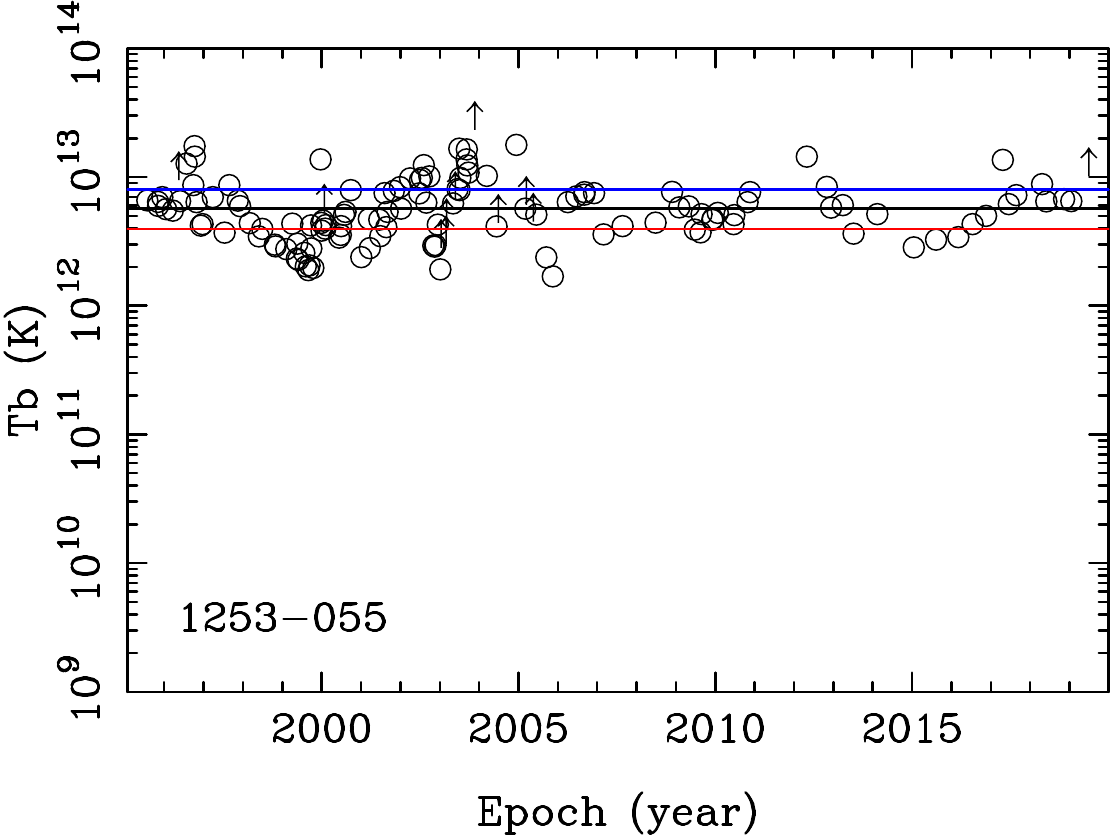}
\includegraphics[scale=0.48]{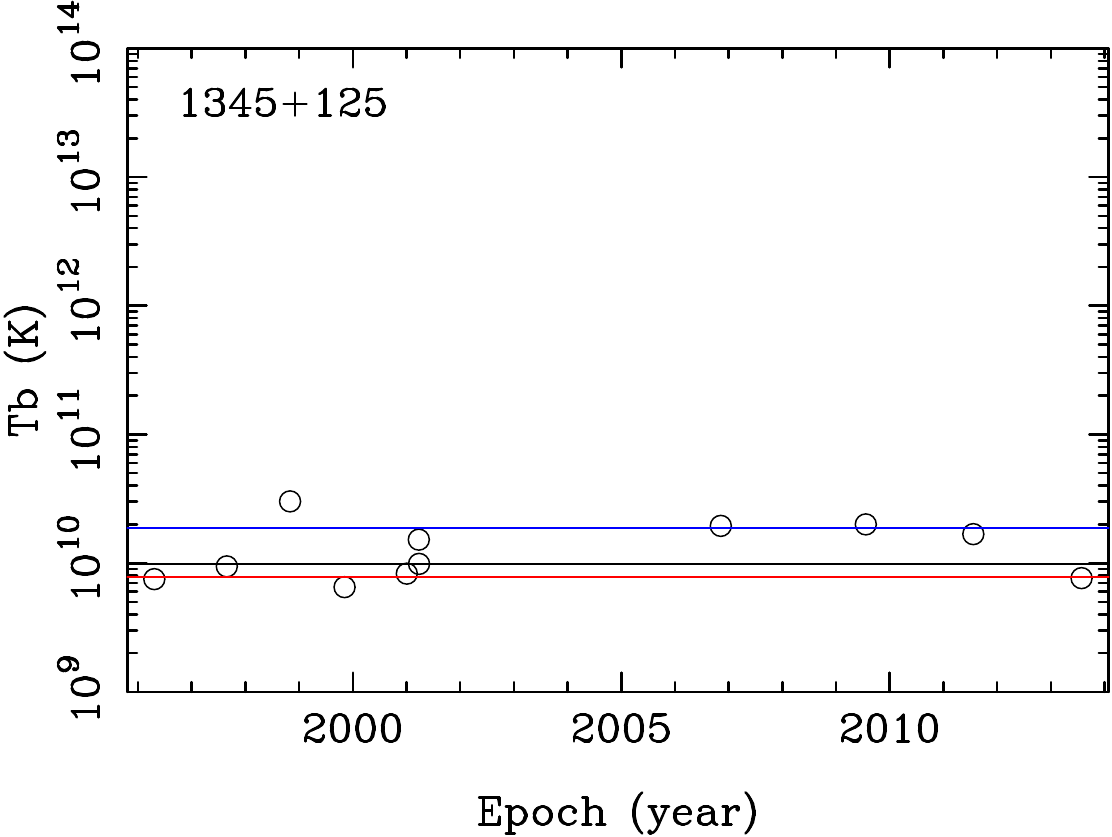}
\includegraphics[scale=0.48]{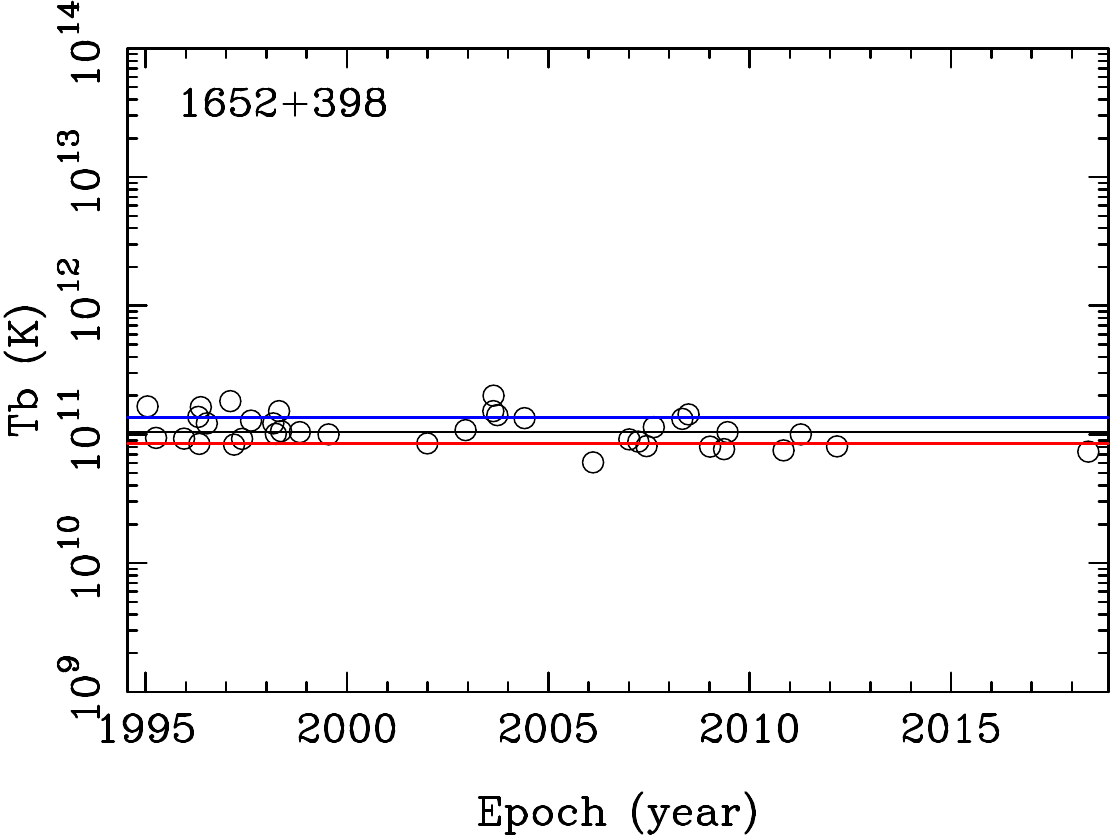}
\includegraphics[scale=0.48]{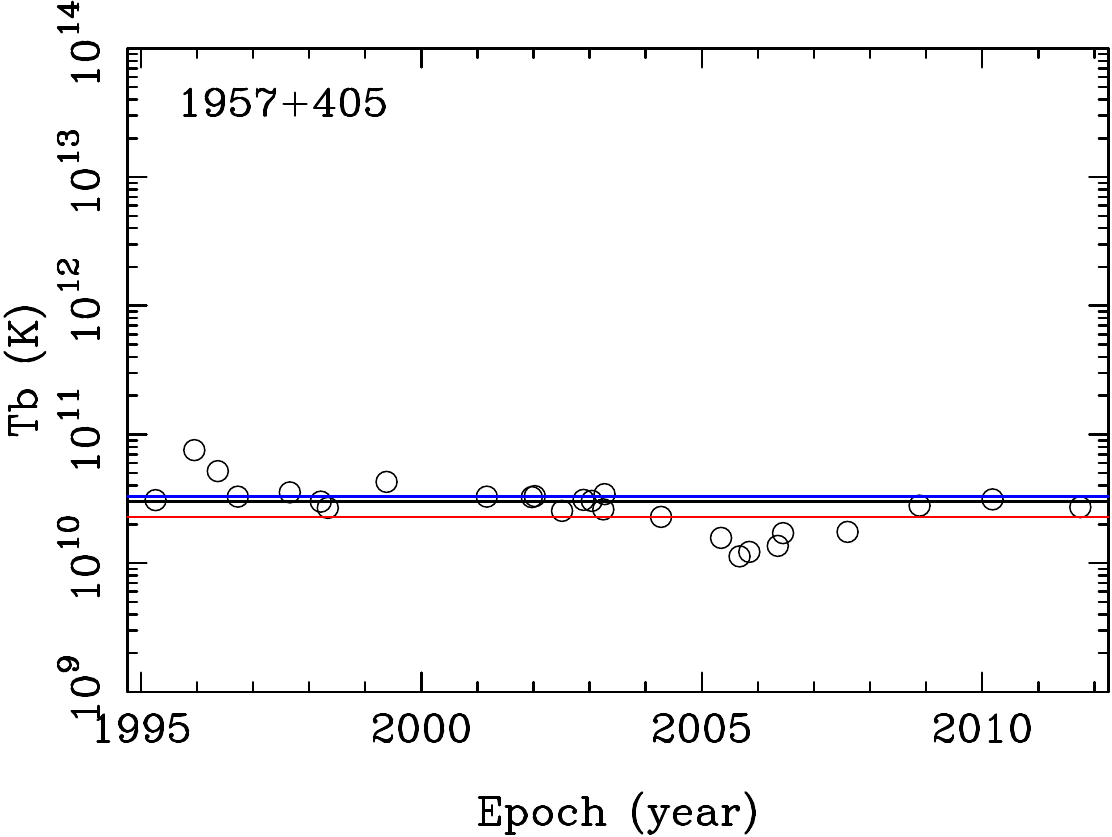}
\includegraphics[scale=0.48]{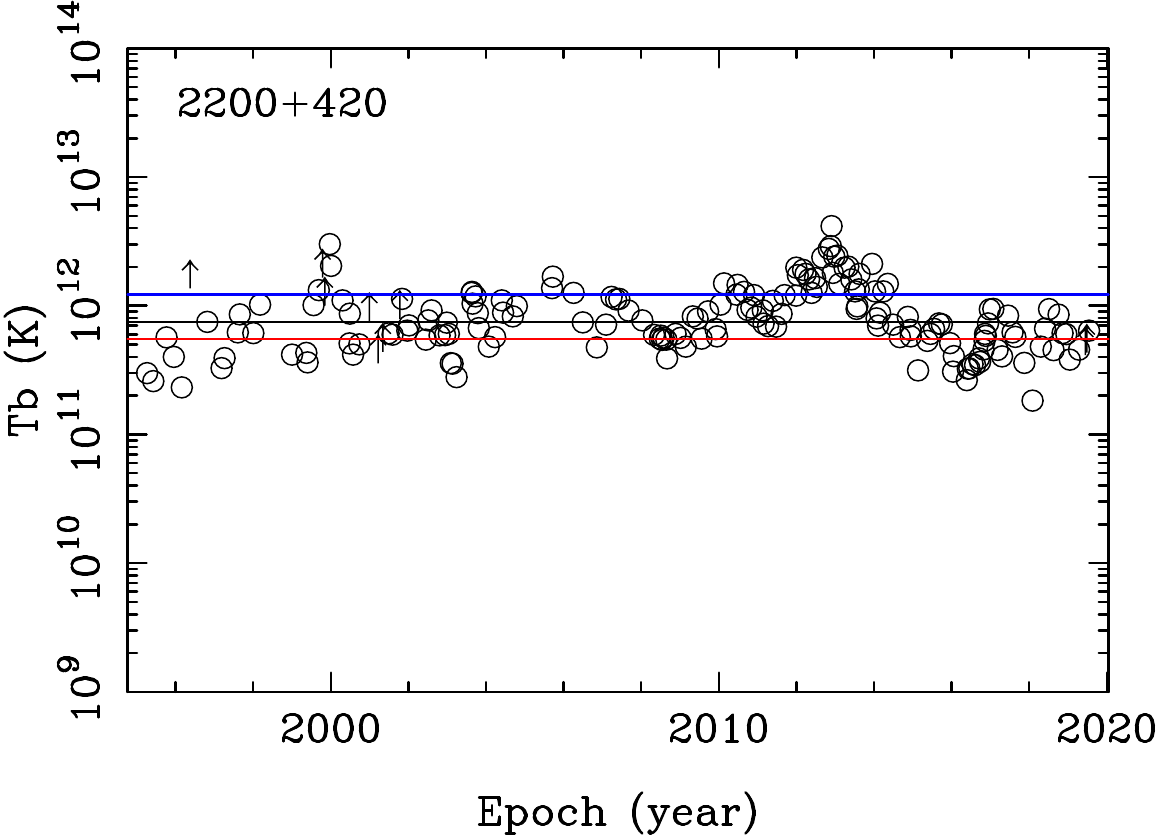}
\figcaption{\label{f:fig2}
  Plots of Brightness Temperature vs.\ Epoch for each source.
  The full set of plots for all 447 sources in our sample appears online.
  Open circles and upward arrows represent measurements and lower limits
  respectively. Estimates of the median value of the distribution are
  shown as black lines; blue and red lines indicate estimates of the
  $75$\% and $25$\% points respectively. Dashed lines are used when
  only a lower limit can be placed on these values. Sources with
  unknown redshifts are plotted with open triangles and dotted lines
  to represent values that otherwise would be considered measurements
  but are too small by an unknown factor of $(1 + z)$.
}
\end{figure*}

\begin{figure*}
\centering
 \includegraphics[width=0.45\textwidth]{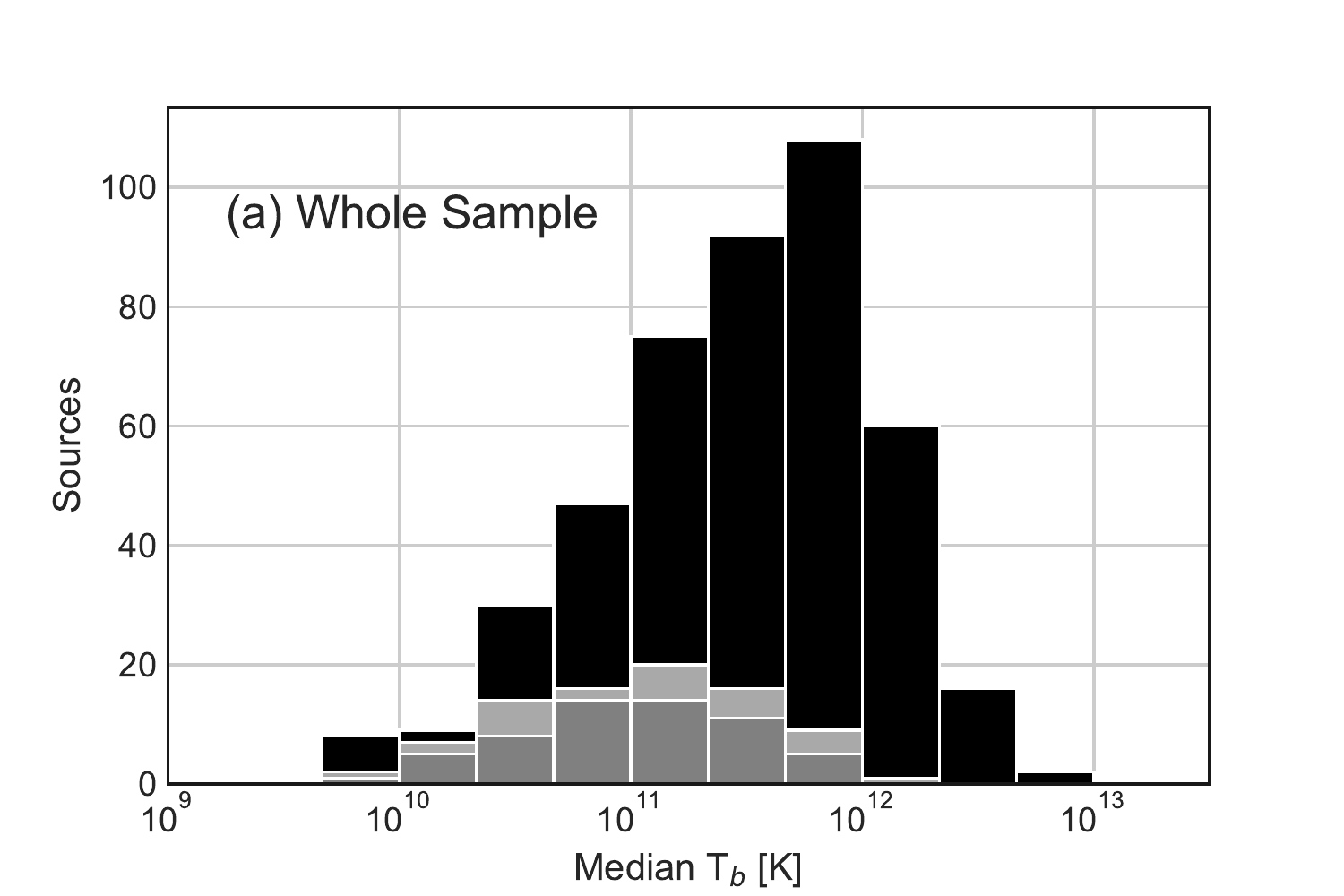}
 \includegraphics[width=0.45\textwidth]{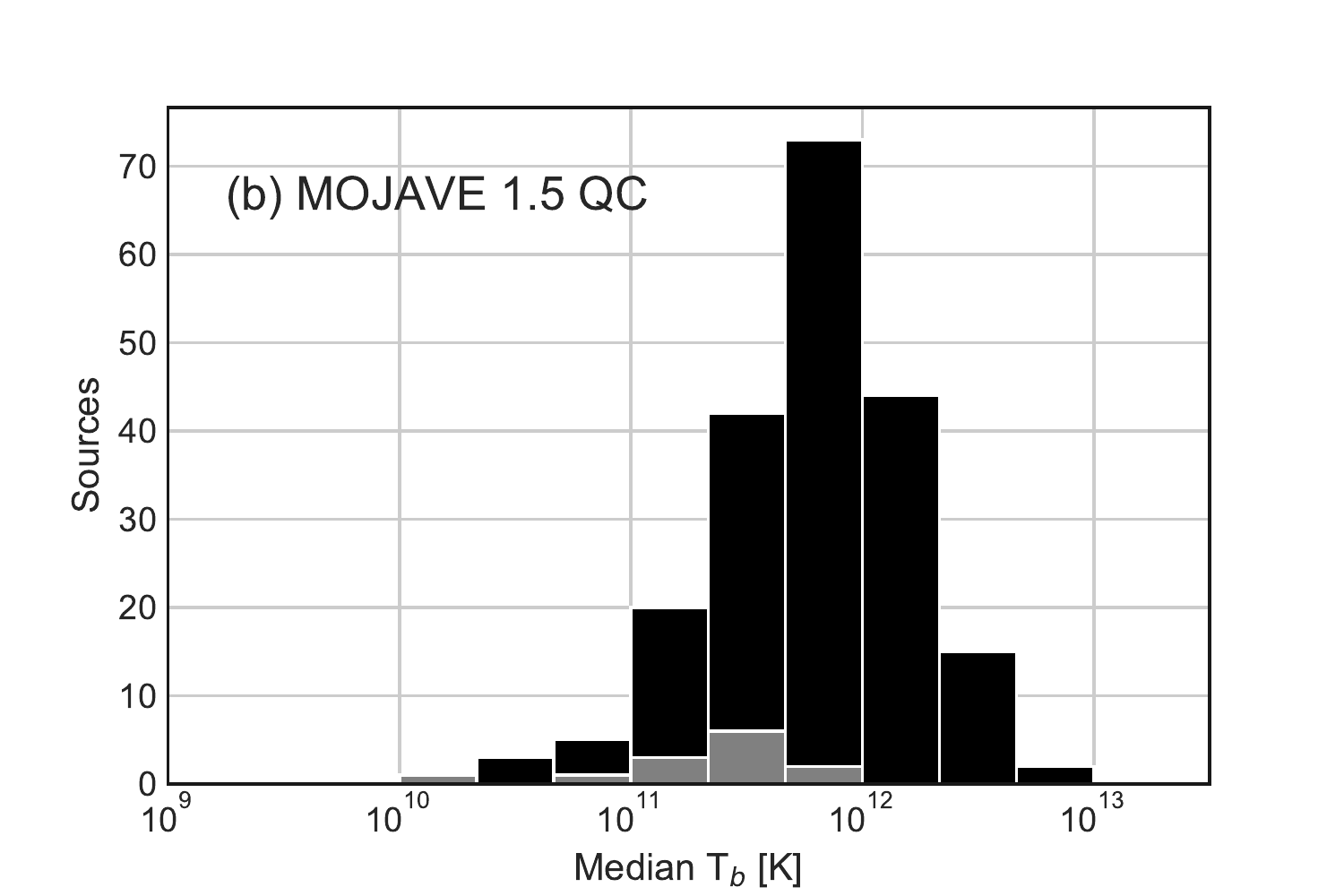}
 \includegraphics[width=0.45\textwidth]{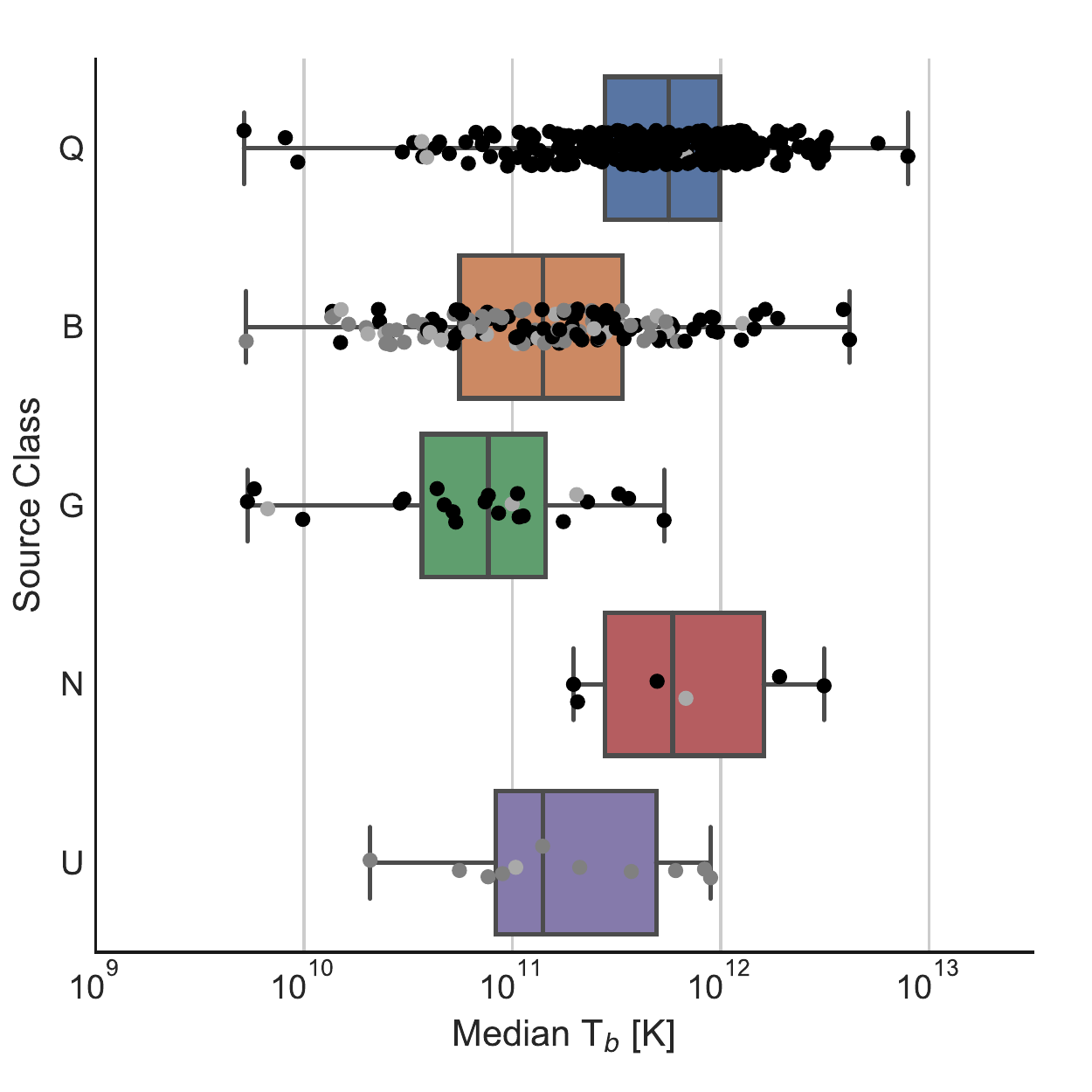}
 \includegraphics[width=0.45\textwidth]{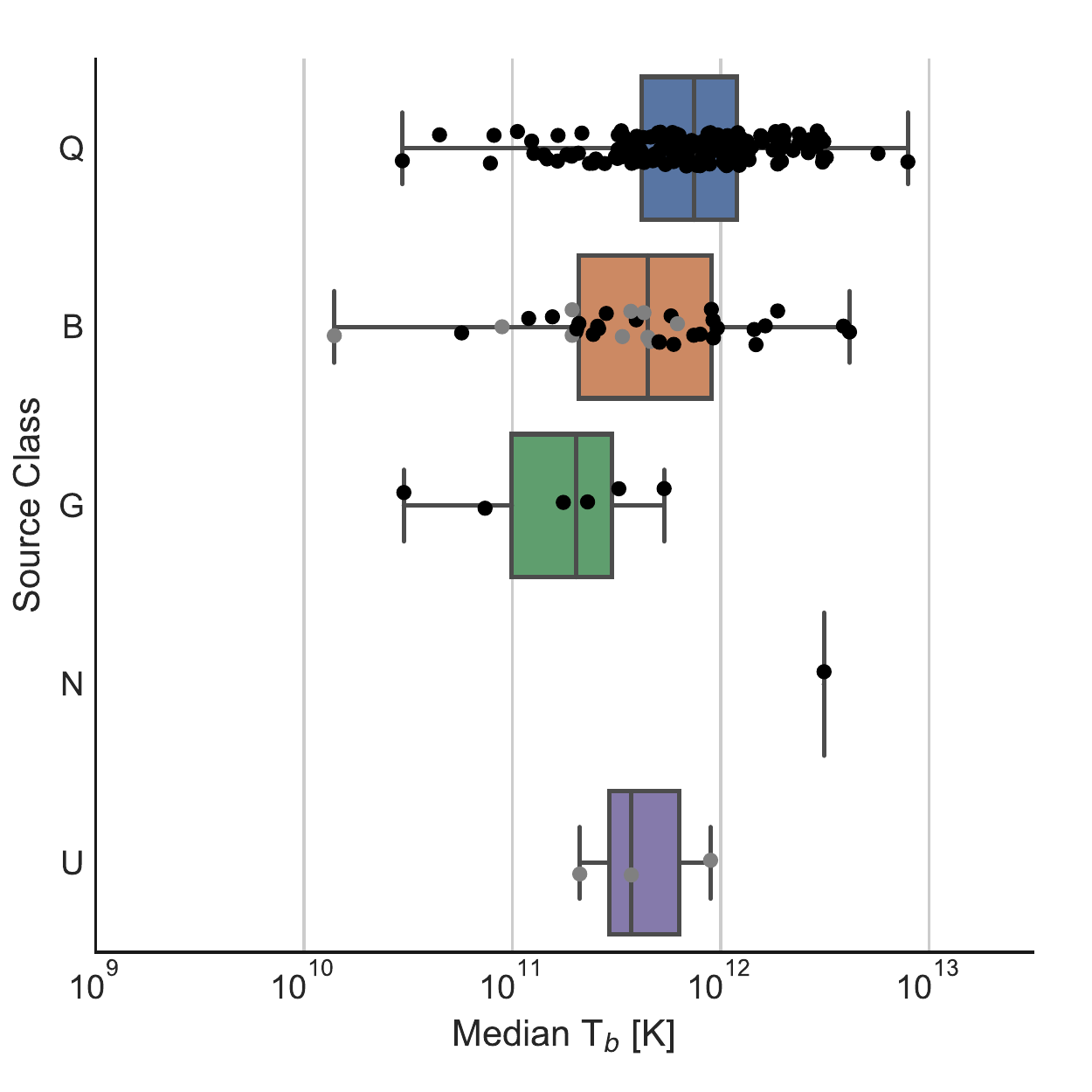}  
\figcaption{\label{f:fig3}
  Distributions of median values of the measured Gaussian peak brightness temperatures
  for each source in the frame of the host galaxy. 
  The upper panels are histograms, and the lower
  panels are combined box and scatter plots that break down the
  distributions by optical class where ``Q'' = quasars, ``B'' = BL\,Lacs,
  ``G'' = radio galaxies, ``N'' = narrow-line Seyfert Is, and ``U'' = unidentified.
  The filled regions of the box plots show the inner-quartile range, 
  while the whiskers show the full extent of the data.  Individual 
  data points are shown as a scatter plot over the box plot to better 
  illustrate the range and density of the data.
  Note that the inner-quartile range in each boxplot is shown without regard to limit status of the individual points; however, the overplotted points are marked as measurements or limits as described below.   In running statistical tests between distributions, we use the log-rank test, as described in the text, to properly account for the limits.
  Gray filling indicates lower limits on the measured brightness 
  temperature, where dark gray is for sources where the lower 
  limit is solely due to the missing redshift.
  Panels on the left are for the entire source sample, while 
  panels on the right contain just the flux-density limited MOJAVE 1.5\,Jy QC sample.
}
\end{figure*}

\autoref{f:fig2} shows plots of our brightness temperature measurements 
over time for each source.  The median value, 25\% value and 
75\% value of the measured distribution for each source are 
indicated by black, blue, and red lines respectively and are 
tabulated in \autoref{t:tb_beta}.  Because some of our brightness temperature 
measurements are lower limits, we determine both the lower bound 
and (where possible) the upper bound on these characteristic points 
in the distribution.  If both lower and upper bounds are  
available, the characteristic point is taken to be their average.  
Lower bounds on the median and other characteristic points are 
determined by treating all limits as measurements. We then 
establish an upper bound on these points by moving all limits 
to the upper end of the distribution.  In some cases, too many 
individual points are limits and determining an upper bound on the
25\%, median, or 75\% point is not possible.  
In these cases the lower bound is listed as a lower limit in \autoref{t:tb_beta} and indicated 
by a dashed line in \autoref{f:fig2}.  Distributions of the median Gaussian 
peak brightness temperature for each source are presented in \autoref{f:fig3} 
and discussed in \autoref{s:Tb-trends}.

\begin{figure*}
\centering
 \includegraphics[width=0.45\textwidth]{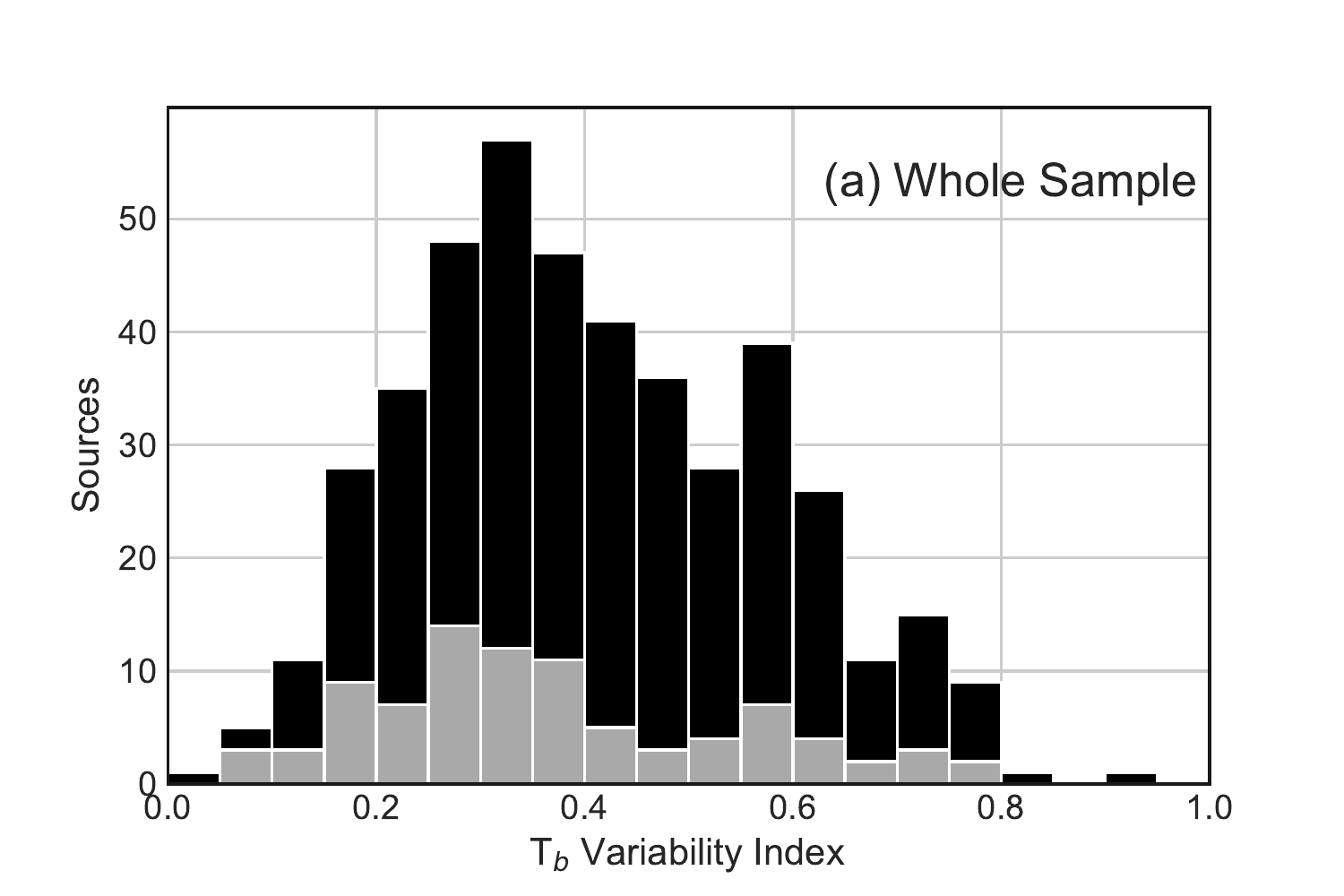}
 \includegraphics[width=0.45\textwidth]{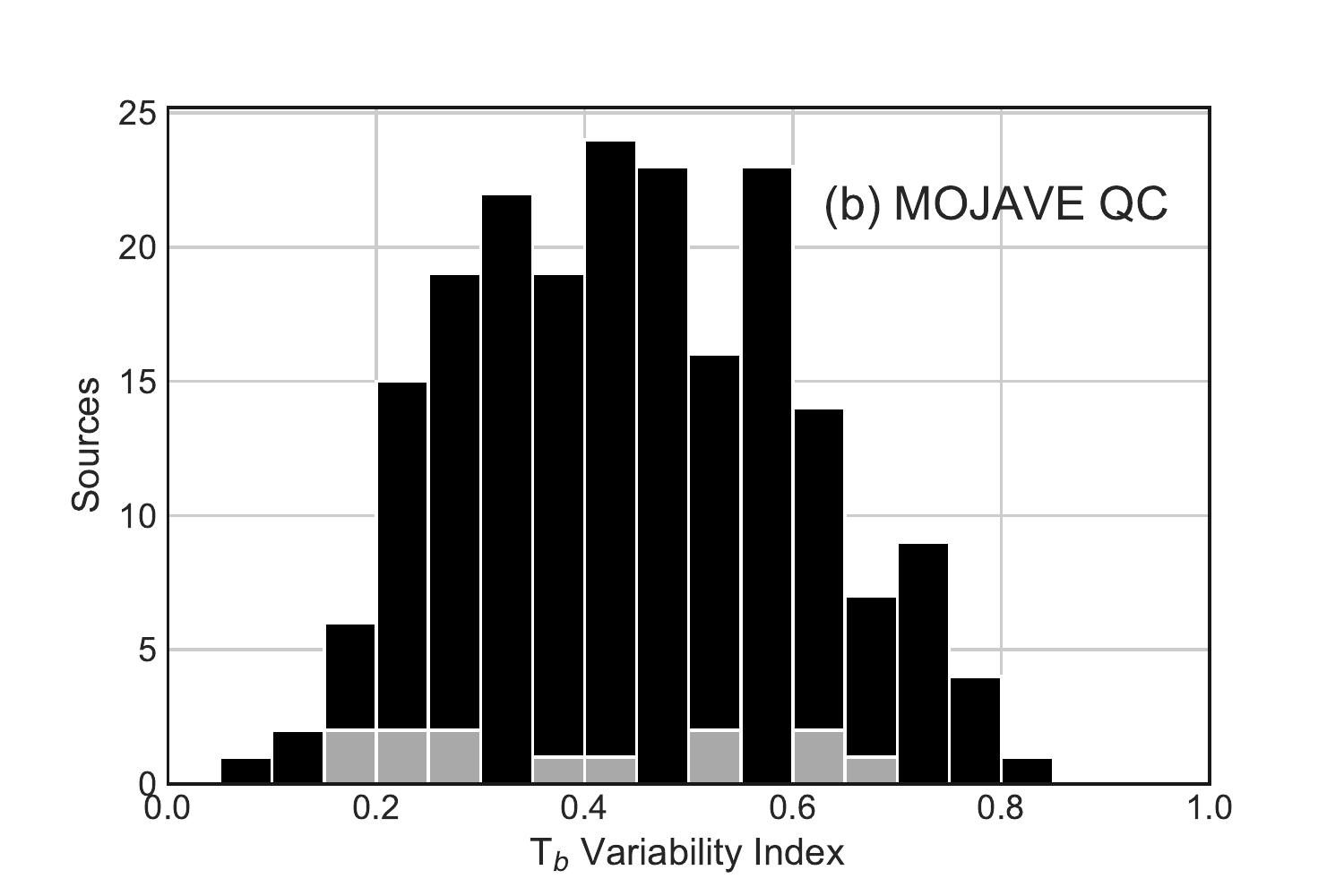}\\  
 \includegraphics[width=0.45\textwidth]{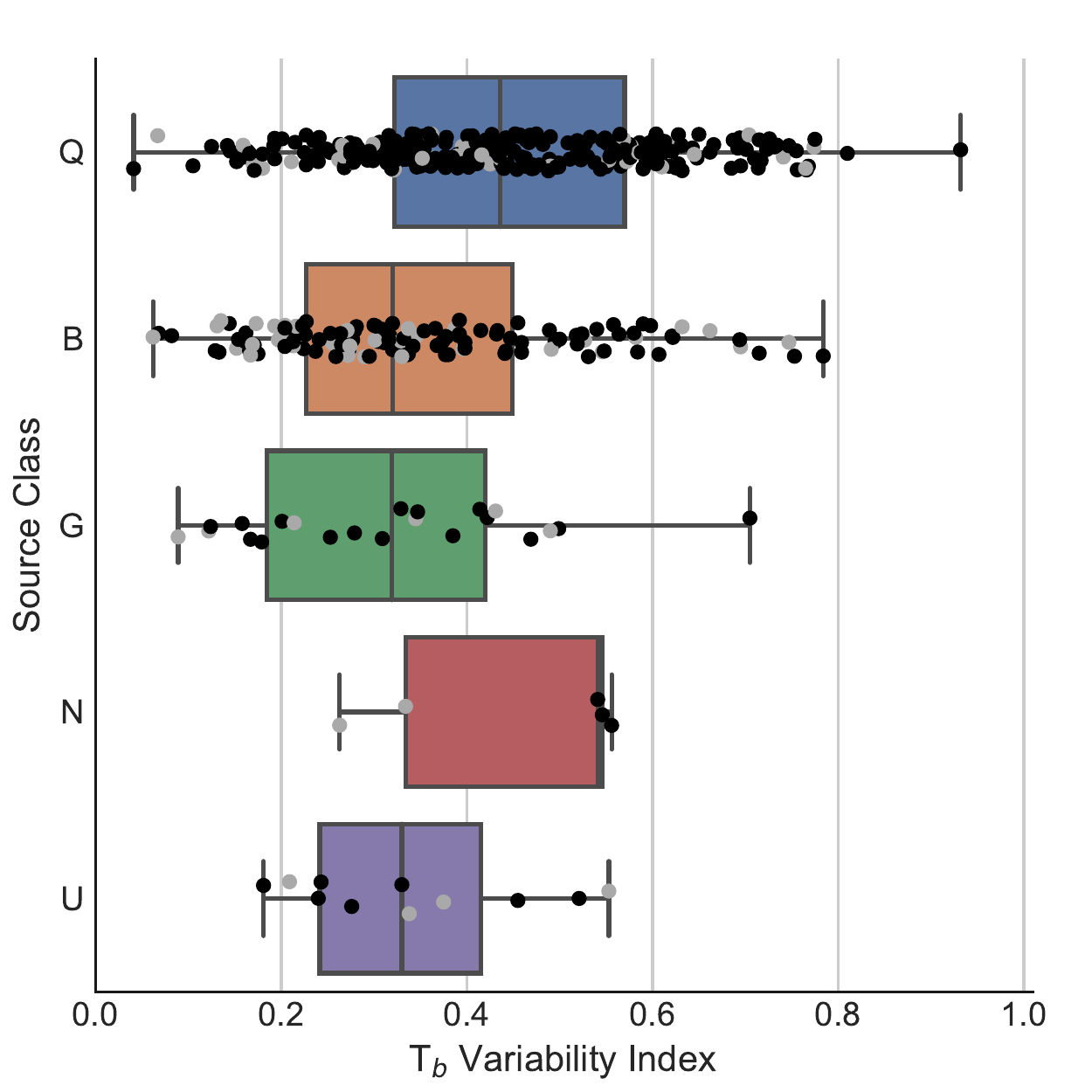}
 \includegraphics[width=0.45\textwidth]{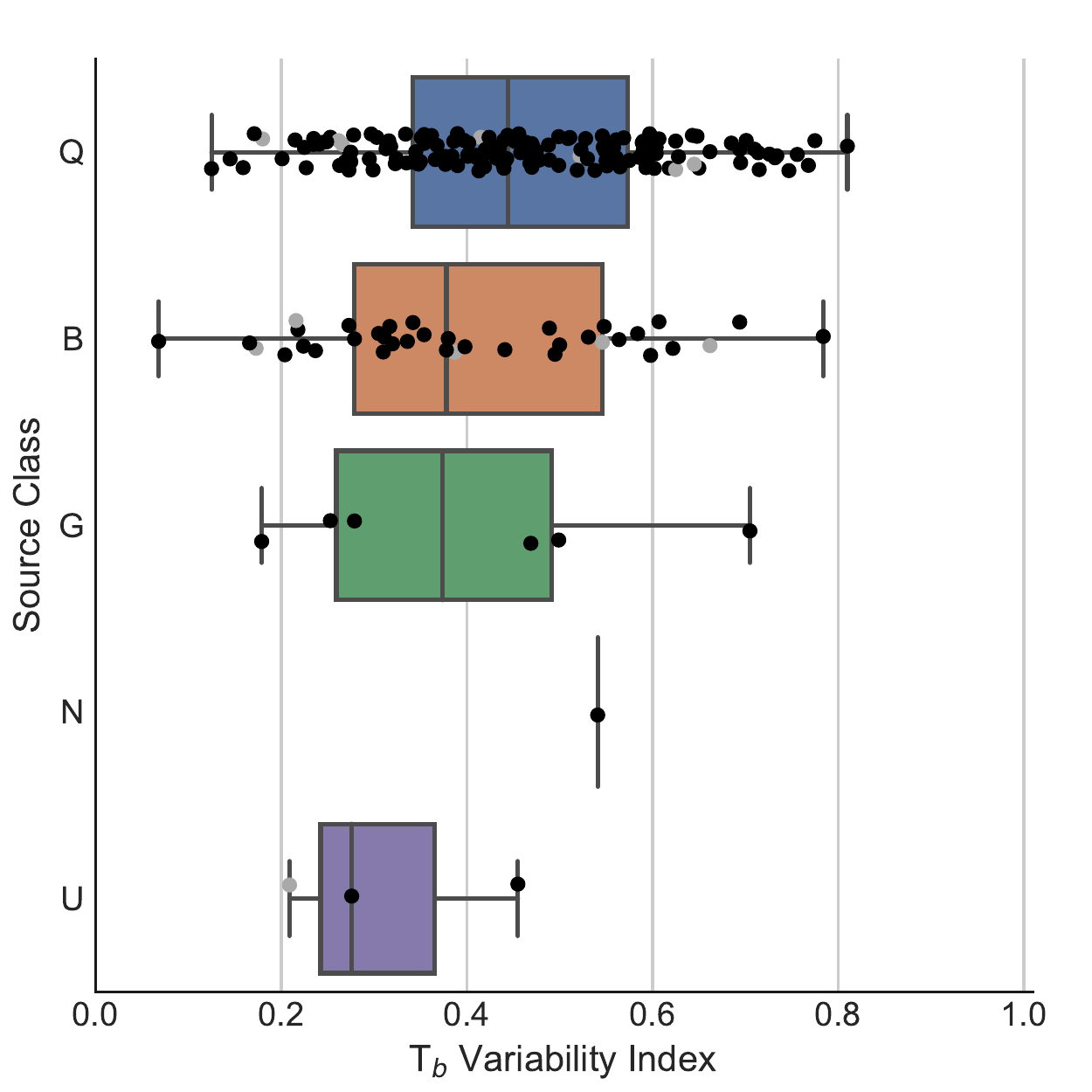}\\  
\figcaption{\label{f:fig4}
  Distributions of the brightness temperature variability index
  for each source. The upper panels are histograms, and the lower
  panels are combined box and scatter plots that break down the
  distributions by optical class where ``Q'' = quasars, ``B'' = BL\,Lacs,
  ``G'' = radio galaxies, ``N'' = narrow-line Seyfert Is, 
  and ``U'' = unidentified.
  The filled regions of the box plots show the inner-quartile range, 
  while the whiskers show the full extent of the data.  Individual 
  data points are shown as a scatter plot over the box plot to better 
  illustrate the range and density of the data.
 Note that the inner-quartile range in each boxplot is shown without regard to limit status of the individual points; however, the overplotted points are marked as measurements or limits as described below.   In running statistical tests between distributions, we use the log-rank test, as described in the text, to properly account for the limits.
 Gray filling indicates lower limits on the variability index.  Panels
  on the left are for the entire source sample, while panels on the right
  contain just the flux-density limited MOJAVE 1.5\,Jy QC sample.
}
\end{figure*}

We use the 25\% and 75\% points in the distribution to also 
define a $T_b$ variability index for each source which is analogous to 
that defined by \citet{AAH92}, 
\begin{equation}
  \label{e:var2575}
  V_{75,25} = \frac{T_\mathrm{b,75}-T_\mathrm{b,25}}{T_\mathrm{b,75}+T_\mathrm{b,25}}
\end{equation}
and these values are tabulated in \autoref{t:tb_beta} with their 
distributions illustrated in \autoref{f:fig4} and discussed in
\autoref{s:var-trends}. We note that several 
brightness temperatures listed in the table are lower limits 
due only to the missing redshift information required for 
\autoref{e:Tb}
%Equation 1 
and are marked accordingly.  These limits are 
computed assuming $z=0$; however, the corresponding variability 
index, $V_{75,25}$, is not a lower limit as the redshift 
dependence cancels out.

\begin{deluxetable}{lrrrrrrrrrrr}
\rotate
\tablenum{3}
\tablecolumns{12} 
\tabletypesize{\scriptsize} 
\tablewidth{0pt}  
\tablecaption{\label{t:tb_beta} Brightness Temperatures and Apparent Speeds}  
\tablehead{
\colhead{Source} &   \colhead {$N$} & \colhead{$T_\mathrm{b,min}$} & \colhead{$T_{b,25}$} & \colhead{$T_\mathrm{b,med}$} &
\colhead{$T_{b,75}$} & \colhead{$T_\mathrm{b,max}$} & \colhead {$V_{75,25}$} &
\colhead{$N_\mathrm{s}$} & \colhead{$\beta_\mathrm{max}$} & \colhead{$\beta_\mathrm{med}$} & \colhead{$\beta_\mathrm{close}$} \\
\colhead{} &   \colhead {} & \colhead{(log$_{10}$ K)} & \colhead{(log$_{10}$ K)} & \colhead{(log$_{10}$ K)} &
\colhead{(log$_{10}$ K)} &\colhead{(log$_{10}$ K)} & \colhead {} &
\colhead {} & \colhead {} & \colhead {} & \colhead {} \\
\colhead{(1)} & \colhead{(2)} & \colhead{(3)} & \colhead{(4)} &  \colhead{(5)} &
\colhead{(6)} & \colhead{(7)} & \colhead{(8)} & \colhead{(9)} &
\colhead{(10)} & \colhead{(11)} & \colhead{(12)}
}
\startdata
0003$+$380 & $ 10$ & $ 10.893$ & $ 11.057$ & $ 11.550$ & $ 11.702$ & $ 11.721$ & $ 0.631$ & $ 3$ & $ 4.61\pm 0.36$ & $ 2.30\pm 0.63$ & $ 0.57\pm 0.15$ \\ 
0003$-$066 & $ 27$ & $ 10.857$ & $ 11.000$ & $ 11.079$ & $ 11.276$ & $ 11.439$ & $ 0.305$ & $ 9$ & $ 7.08\pm 0.21$ & $ 2.48\pm 0.49$ & $1.868\pm0.093$ \\ 
0006$+$061 & $  5$ & $>10.662$\tablenotemark{a} & $>10.917$\tablenotemark{a} & $>11.021$ & $>11.049$ & $>11.064$\tablenotemark{a} & $>0.152$ & \ldots & \ldots & \ldots & \ldots \\ 
0007$+$106 & $ 25$ & $ 10.766$ & $ 11.243$ & $ 11.729$ & $ 12.004$ & $ 12.260$ & $ 0.705$ & $ 2$ & $ 1.58\pm 0.29$ & $ 1.47\pm 0.18$ & $ 1.58\pm 0.29$ \\ 
0010$+$405 & $ 12$ & $>11.201$ & $>11.330$ & $>11.425$ & $>11.534$ & $>11.633$ & \ldots & $ 1$ & $ 6.92\pm 0.64$ & $ 6.92\pm 0.64$ & $ 6.92\pm 0.64$ \\ 
0011$+$189 & $  8$ & $>10.859$ & $>10.992$ & $>11.207$ & $>11.250$ & $>11.593$ & \ldots & $ 1$ & $ 4.54\pm 0.46$ & $ 4.54\pm 0.46$ & $ 4.54\pm 0.46$ \\ 
0012$+$610 & $  6$ & $>10.684$\tablenotemark{a} & $>10.684$\tablenotemark{a} & $>10.747$\tablenotemark{a} & $>10.843$\tablenotemark{a} & $>11.086$\tablenotemark{a} & $ 0.181$ & \ldots & \ldots & \ldots & \ldots \\ 
0014$+$813 & $ 14$ & $ 11.029$ & $ 11.079$ & $ 11.223$ & $ 11.354$ & $ 11.438$ & $ 0.306$ & $ 2$ & $ 9.47\pm 0.91$ & $  9.4\pm  1.2$ & $  9.3\pm  1.5$ \\ 
0015$-$054 & $  8$ & $ 10.436$ & $ 10.908$ & $ 11.246$ & $>11.410$ & $>11.629$ & $>0.521$ & \ldots & \ldots & \ldots & \ldots \\ 
0016$+$731 & $ 16$ & $ 10.676$ & $ 11.584$ & $ 11.902$ & $ 12.384$ & $ 12.801$ & $ 0.726$ & $ 2$ & $ 7.64\pm 0.32$ & $ 5.10\pm 0.22$ & $ 2.57\pm 0.12$ \\ 
\ldots \\
\enddata
\tablenotetext{a}{Lower limit value ($z=0$) only on account of unknown source redshift.}
\tablecomments{
The complete version of this table appears in the online journal.
Columns are as follows: 
(1) Source name in B1950 coordinates; 
(2) Number of Epochs;
(3) Minimum Peak Gaussian Brightness Temperature;
(4) Peak Gaussian Brightness Temperature at 25\% of Distribution;
(5) Median Peak Gaussian Brightness Temperature;
(6) Peak Gaussian Brightness Temperature at 75\% of Distribution;
(7) Maximum Peak Gaussian Brightness Temperature;
(8) Variability Index of Gaussian Brightness Temperature;
(9) Number of robust speeds meeting criteria described in \S{\ref{s:betaT_analysis}};
(10) Fastest apparent speed;
(11) Median apparent speed;
(12) Apparent speed of feature that is closest to the core in its first 
measured epoch;
}
\end{deluxetable}

\subsection{Comparing Brightness Temperatures and Apparent Motions}
\label{s:betaT_analysis}

As described in \autoref{s:intro}, the observed brightness temperature 
in the frame of the host galaxy is the intrinsic brightness temperature 
boosted by the Doppler factor: $T_\mathrm{b,obs} = \delta T_\mathrm{b,int}$.  The unknown 
Doppler factor, $\delta = 1/[\Gamma(1-\beta\cos\theta)]$, depends on
the intrinsic flow speed, $\beta$, and angle to the line of sight, 
$\theta$, in a similar fashion to the observed superluminal motion,
$\beta_\mathrm{obs} = \beta\sin\theta/(1-\beta\cos\theta)$.

Our approach in this section is to compare a characteristic observed 
brightness temperature for each jet to its characteristic observed speed,
following \citet{H06}.  This comparison will allow us to find a typical 
intrinsic brightness temperature, $T_\mathrm{b,int}$, for our sample as a whole.  
We will then take the analysis of \citet{H06} a step further and use 
$T_\mathrm{b,int}$ to estimate the Doppler factor, $\delta$, for each individual 
jet.  Combined with that jet's observed speed, $\beta_\mathrm{obs}$, we determine 
its Lorentz factor, $\Gamma$, and angle to the line of sight, $\theta$.  

\subsubsection{Selecting characteristic values of apparent brightness temperature and kinematics}

\citet{H06} used the 25\% point in the brightness temperature 
distribution of a given source as its characteristic brightness
temperature; however, that choice was driven by the desire to 
avoid too many lower limits in a relatively small set of brightness 
temperature measurements available at the time.
Our new data set is far larger, both in terms of numbers of 
epochs on individual sources and for the number of sources in our 
sample as whole.  Consequently we now simply use the median
brightness temperature of a given source as its characteristic 
brightness temperature.  Only those jets that have a median $T_\mathrm{b}$ 
value, not a limit, are used in the analysis.  Limits are
ambiguous in the statistical comparison and do not allow robust 
estimates of the relativistic properties.  Fortunately only twelve of 
the 321 sources with viable observed speeds have median brightness 
temperature limits, and none of them are part of
the MOJAVE 1.5\,Jy QC flux-density limited sub-sample.

In addition to summarizing the brightness temperature properties of 
each AGN jet, \autoref{t:tb_beta} also includes a summary of the  
distribution of apparent speed of features reported in \citetalias{MOJAVE_XVIII}. For characterizing 
the speed distribution of a given source, we only consider features 
with significant motions, $\geq 3\sigma$, in the approaching jet and 
discard those features identified as 'inward' moving in \citetalias{MOJAVE_XVIII}. For each 
source, \autoref{t:tb_beta} reports the number of measured speeds, $N_s$, which
meet these criteria and lists the maximum apparent speed, median apparent
speed, and speed of the feature that was closest to the VLBI core
in its first measured epoch.  Unlike 
\citetalias{MOJAVE_XVIII}, which required at least five robust features to 
identify a median speed, here we report a fastest, median, and 
closest speed for every jet with at least one motion meeting the 
criteria described above.

In our previous papers we have taken the fastest observed speed 
in a given jet as the most representative of the underlying flow 
\citep[e.g.][]{LCH09,LHH19}; however, the range of speeds in a source 
with many moving features can span a factor of a few, often including 
some very slow features. Jets with at least five features meeting 
our criteria have a median speed that is, on average, about $60\%$ of the 
magnitude of their maximum observed speed. Because the features we 
observe may be propagating shocks 
\citep[e.g., ][]{1985ApJ...298..114M,1989ApJ...341...54H}, 
they may travel at a different
speed than the flow itself and the best observed speed to use in 
representing the flow remains an open question.  To address this 
issue we directly compare three different choices for characterizing the 
observed speed of a jet to the median observed brightness 
temperature of the jet cores for those sources with several 
moving features, $N_s \geq 5$.

\begin{figure}
\centering
\includegraphics[width=0.5\textwidth]{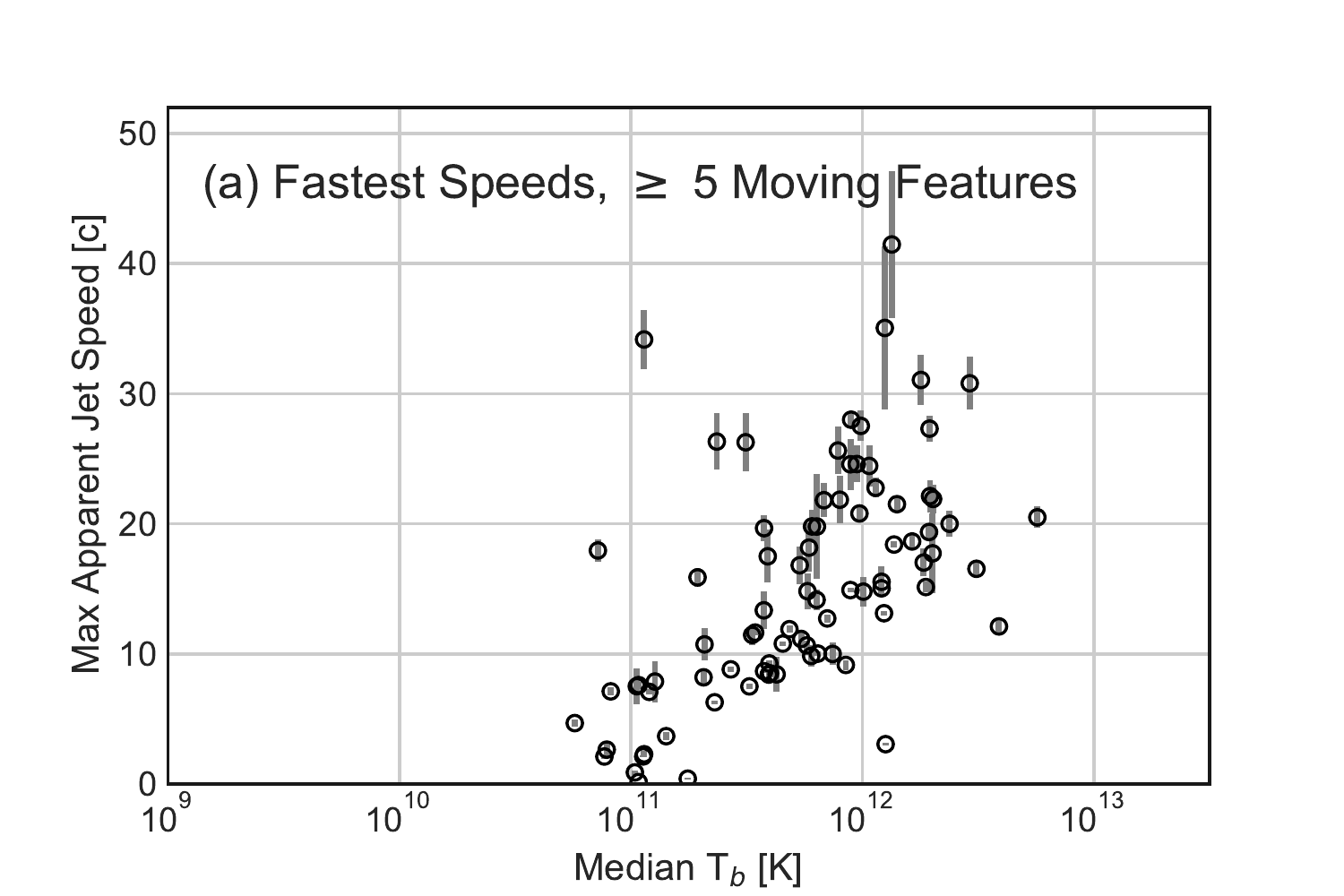}
\includegraphics[width=0.5\textwidth]{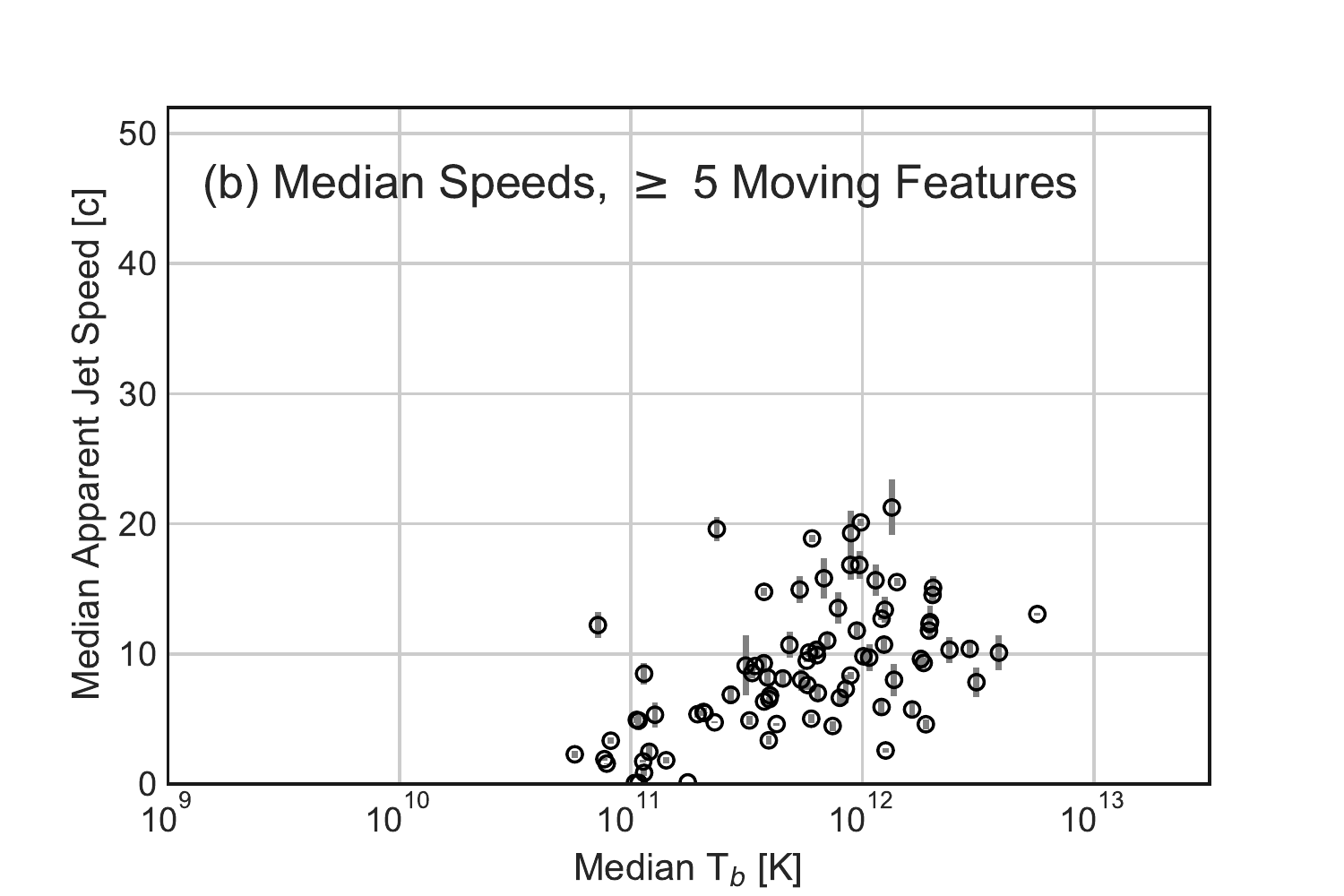}
\includegraphics[width=0.5\textwidth]{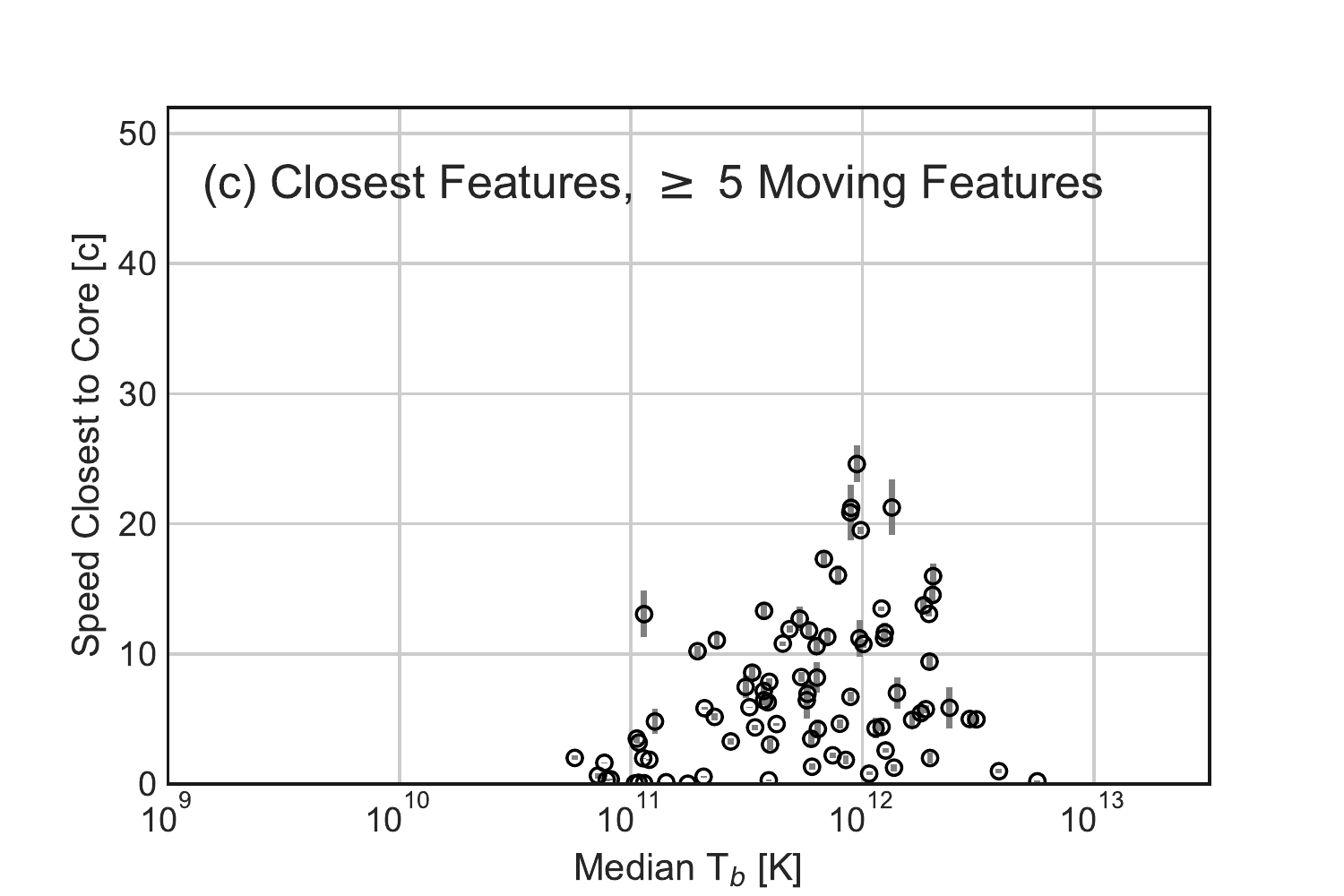}
\figcaption{\label{f:speedtest}
  Apparent Speed vs.\ Median Gaussian Brightness Temperature in the core
  for all 83 sources with $\geq 5$ moving features meeting the criteria
  described in \autoref{s:betaT_analysis}. Panels (a), (b), and (c) show respectively the
  fastest speed, median speed, and speed found closest to the core
  region. The fastest apparent speeds have the strongest correlation
  with the median brightness temperature of the core.
}
\end{figure}

\autoref{f:speedtest} compares median brightness temperature of 
the core with the fastest observed speed, $\beta_\mathrm{max}$, the median 
speed, $\beta_\mathrm{med}$, and the speed of the feature closest
to the core, $\beta_\mathrm{close}$.  The same 83 jets with at least five 
moving features are shown in each panel; the only difference is 
the speed used to represent each jet on the $y$-axis.  The strongest
correlation with median $T_b$ is for the fastest apparent speed 
(see panel (a)) with a
Spearman $\rho = 0.63$, while the median and closest features have
$\rho=0.58$ and $\rho=0.36$ respectively. It is important to note that 
even with ideal measurements, we do not expect a perfect correlation 
between the observed brightness temperature and apparent speed. At the 
``critical'' angle that maximizes apparent superluminal motion with
$\cos\theta = \beta$,
\begin{equation}
  \label{e:betaT}
  \beta_\mathrm{obs} = \beta\delta = \beta T_\mathrm{b,obs}/T_\mathrm{b,int}
\end{equation}
which would indeed suggest a strong correlation given that
$\beta$ is typically very nearly unity for powerful AGN jets; however, 
some jets may lie at
smaller or larger angles than the critical angle and consequently have
larger or smaller Doppler factors respectively.  Indeed we
will see this effect below when we look at the full data set; however, 
this subset of 83 jets includes only those that have at least five 
moving features meeting the criteria outlined above.  Jets where 
we can identify and follow several moving features may be more 
likely to be near the critical angle where we are viewing the 
jet structures from the
side in the co-moving frame, and the strong correlation seen in
panel (a) is consistent with that expectation.  In our view,
the fastest observed speed, $\beta_\mathrm{max}$, is the best speed
to use in comparing to core brightness temperatures across the
sample, and we use  $\beta_\mathrm{max}$ in the analysis that follows.
In \autoref{s:whatspeed}, we
revisit this question in the light of possible jet acceleration
and consider the effects on our results if the
median speed is used instead.

\subsubsection{Estimating the typical median intrinsic brightness temperature}
\label{s:measureTb_int}

In a complete, flux-density limited sample, jets are more likely to be
observed at a smaller angle to the line of sight than the critical angle due to 
Doppler beaming selection \citep[e.g., ][]{2007ApJ...658..232C}.  \citet{LM97} 
found that a typical beamed jet
in a flux-density limited sample like the MOJAVE 1.5\,Jy QC sample has an
angle to the line of sight about one-half of the critical angle, and
\citet{H06} used a simulation of a flux-density limited sample to estimate
that about $75$\% of the jets should lie inside the critical angle with
a Doppler beaming factor:
\begin{equation}
  \delta > \sqrt{1+\beta^2_\mathrm{obs}} \simeq \beta_\mathrm{obs}
\end{equation}
To update this estimate, we created 1000 Monte Carlo simulations
of a 174-source, flux-density limited sample based on the parameters
estimated by \citet{LHH19}, and we find that $69$\% of the simulated jets
lie within the critical angle.  While the full results of the Monte Carlo
simulation reported in that paper are based on the luminosities and
apparent speeds of the MOJAVE 1.5\,Jy QC quasars at that time, in this
work we only use the fraction of simulated jets within the critical
angle to allow us to estimate the typical median intrinsic brightness
temperature, $T_\mathrm{b,int}$, of our sample as a whole.

Following \citet{H06} we start by assuming that every source in our
sample has the same median intrinsic brightness temperature, and therefore
that any differences in observed median brightness temperatures between
sources are due to their Doppler beaming factor.  With this assumption
we can calculate the expected observed median brightness temperature
for jets at the critical angle: $T_\mathrm{b,obs} = \sqrt{1+\beta^2_\mathrm{obs}} T_\mathrm{b,int}$.
Jets with larger observed median brightness temperatures are therefore
more highly beamed and located inside the critical angle.  We vary
$T_\mathrm{b,int}$ until $69$\% of our sample lie inside the critical angle.

There are 178 sources in the MOJAVE 1.5\,Jy QC sample with both observed
median brightness temperatures and observed speeds, $149$ of which are
quasars.  Using the whole 1.5\,Jy QC sample, we find the best estimate for
the median intrinsic brightness temperature to be 
$T_\mathrm{b,int} = 10^{10.609}$\,K, and restricting the sample to only quasars
does not change this value appreciably.  We
estimate the uncertainty in this value in two ways: (1) by
creating 10,000 samples of 178 sources by randomly
drawing with replacement from the data itself to include the
effects of a limited sample size, and (2) by changing our
fraction of sources within the critical angle by $\pm 5$\%
and repeating this estimate using $64$\% and $74$\% of sources
within the critical angle. Including these uncertainties, our
best estimate for the typical median intrinsic brightness
temperature of the sample is 
$T_\mathrm{b,int} = 10^{10.609\pm0.067}$\,K $= 4.1(\pm0.6)\times10^{10}$~K.

\begin{figure}
\centering
\includegraphics[width=0.5\textwidth]{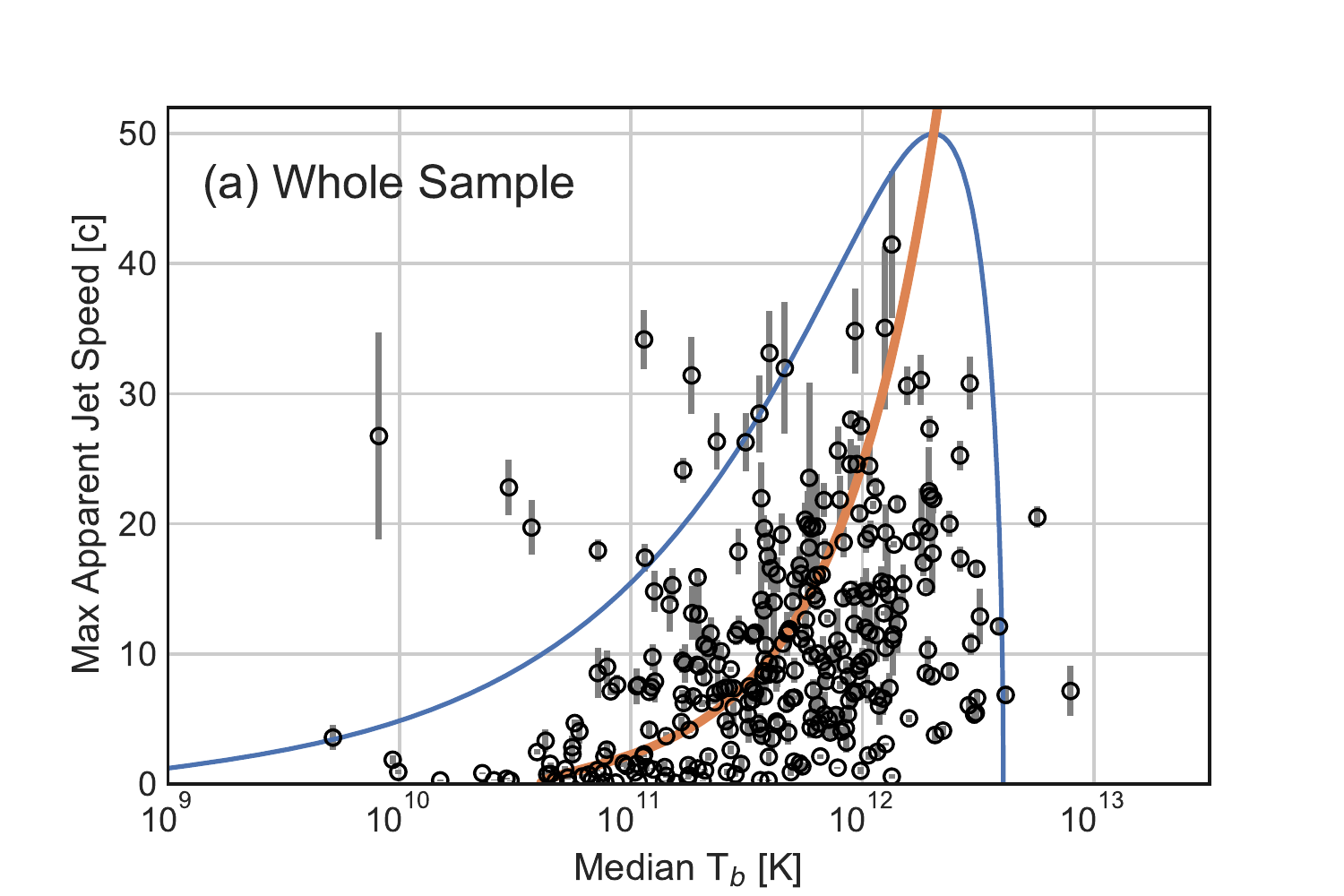}
\includegraphics[width=0.5\textwidth]{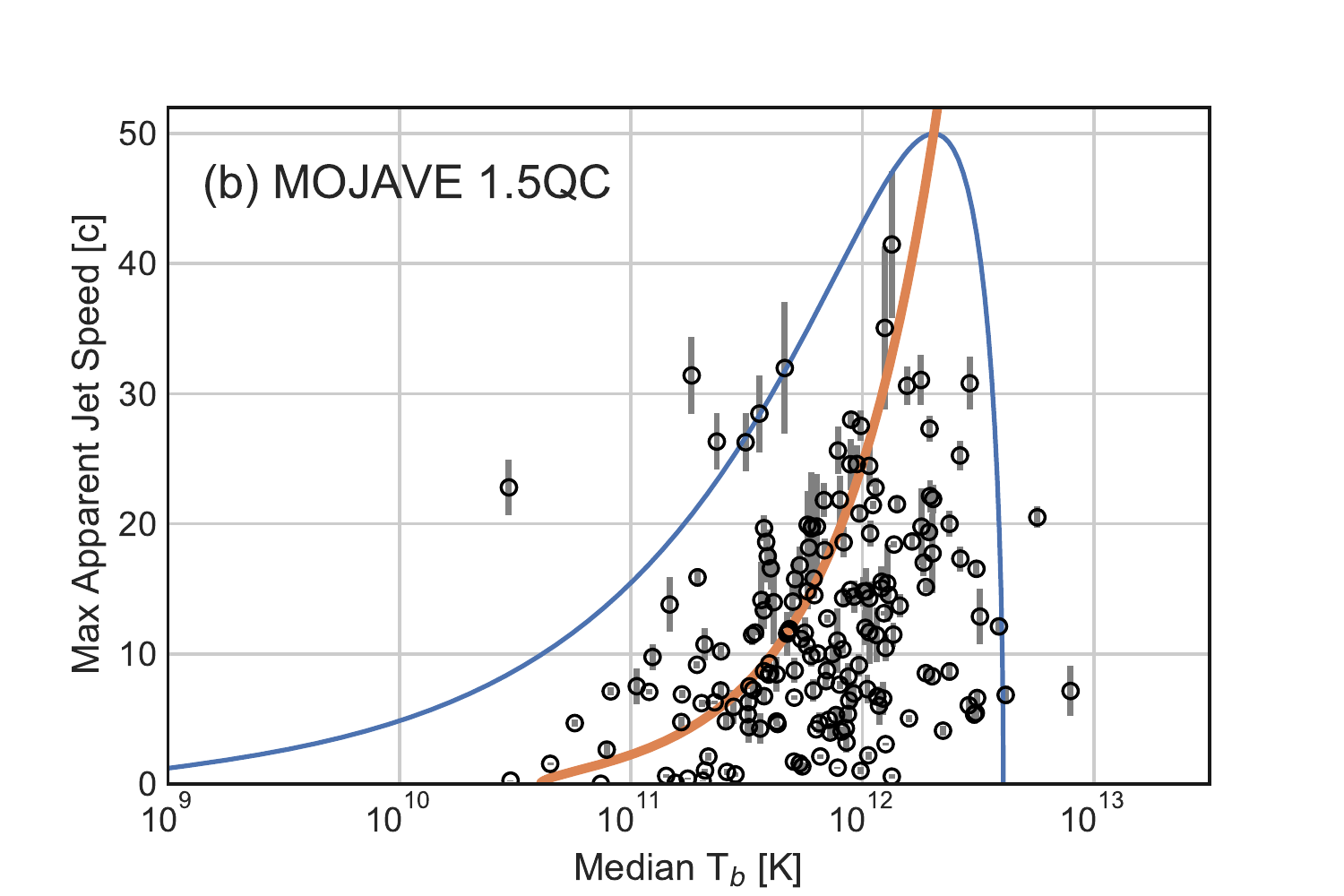}
\figcaption{\label{f:betaTb}
  Apparent Speed vs.\ Median Gaussian Brightness Temperature in the Core.
  Panel (a) includes all 309 sources with apparent speeds and 
  median brightness temperature measurements, and panel (b) 
  includes just the 178 sources from the MOJAVE 1.5\,Jy QC sample. 
  Each panel has two curves.  The first
  curve is a red-orange line through the center of the plot which 
  shows where sources with intrinsic brightness temperature = 
  $10^{10.609}$\,K, would fall if viewed at the critical angle, 
  $\cos\theta = \beta$. The second curve is a blue ``envelope'' 
  which shows where sources with a Lorentz factor of 50 would 
  fall if seen at the full range of angles to the line of sight.
}
\end{figure}

\autoref{f:betaTb} shows plots of maximum observed jet
speeds vs. observed median brightness temperature for
both our entire sample (panel a) and for the MOJAVE 1.5\,Jy QC
sample (panel b).  The superimposed lines use our estimated
value for the intrinsic median brightness temperature. The first
curve is a red-orange line through the center of the plot
which shows where jets with intrinsic brightness
temperature = $10^{10.609}$ K would fall if viewed at
the critical angle, $\cos\theta = \beta$. The second curve
is a blue ``envelope'' which shows where jets with the same
intrinsic brightness temperature and a Lorentz factor
of 50 would fall if seen at the full range of angles
to the line of sight.  If all of the jets in our sample
have this same median intrinsic brightness temperature,
jets with Lorentz factors $<50$
should fall below the blue curve, and jets viewed
inside the critical angle should fall to the right of
the red-orange curve.

\subsubsection{Finding \texorpdfstring{$\delta$}{Lg}, 
\texorpdfstring{$\Gamma$}{Lg}, and \texorpdfstring{$\theta$}{Lg}}
\label{s:calc_quant}

\begin{figure}
\centering
\includegraphics[width=0.5\textwidth]{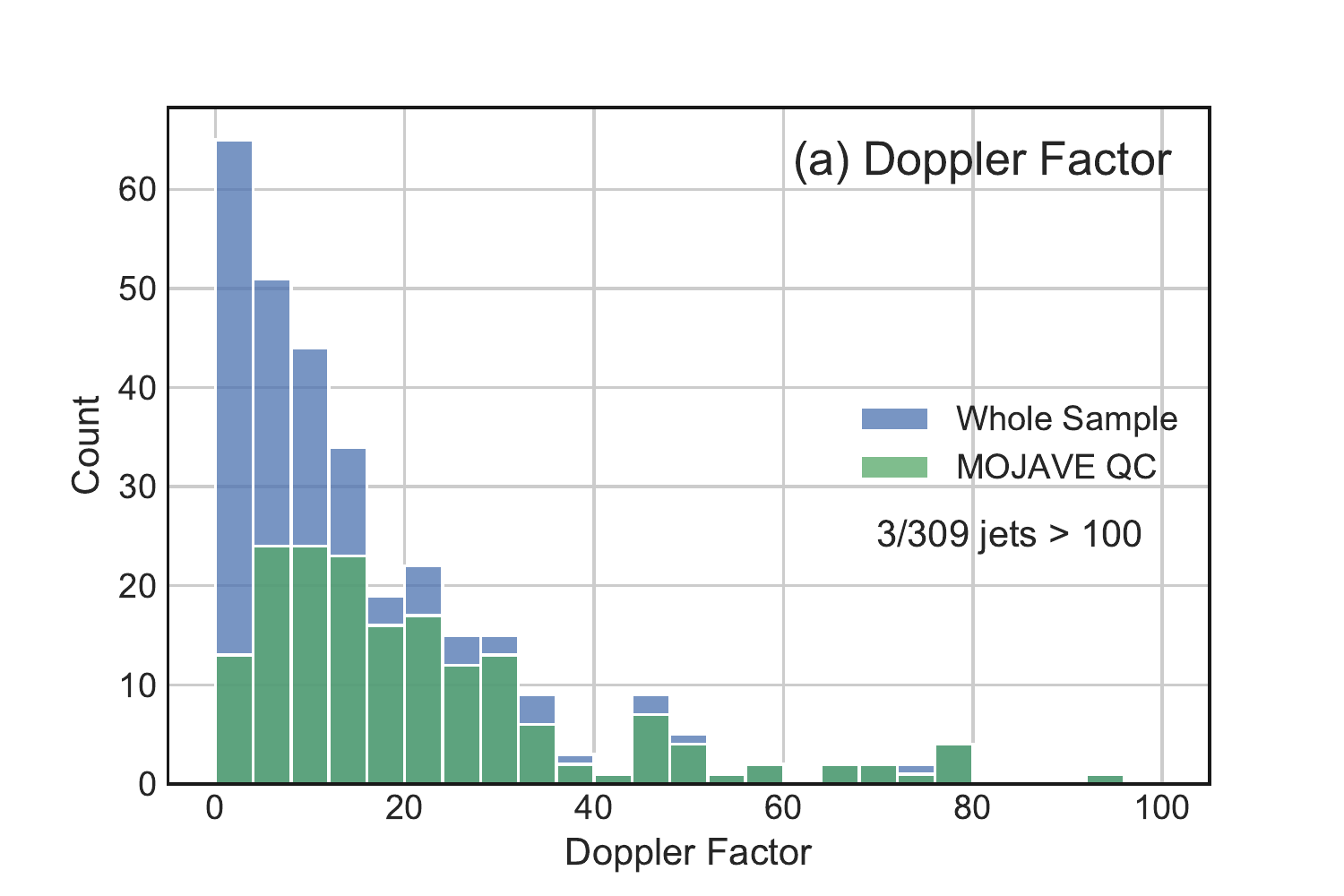}
\includegraphics[width=0.5\textwidth]{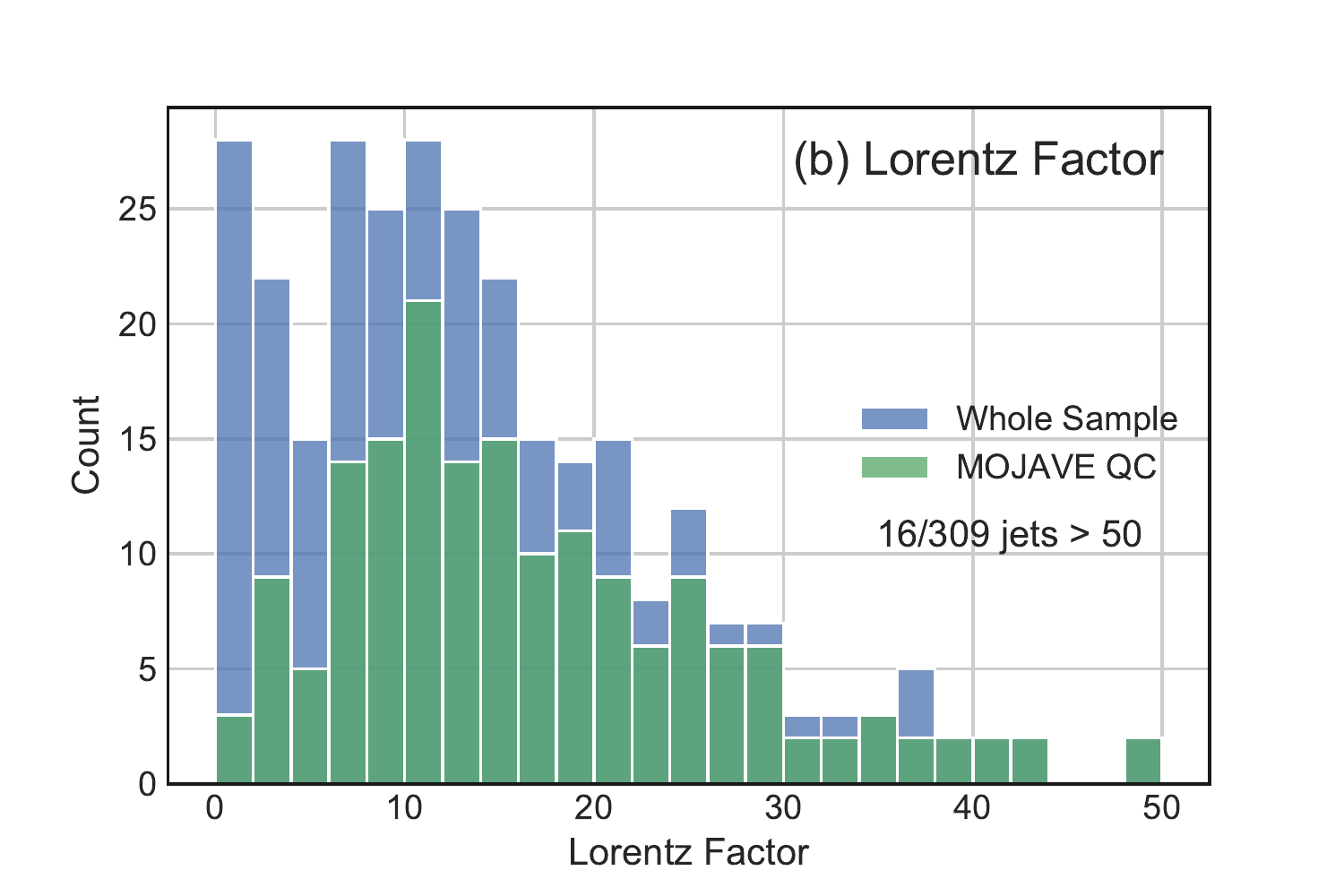}
\includegraphics[width=0.5\textwidth]{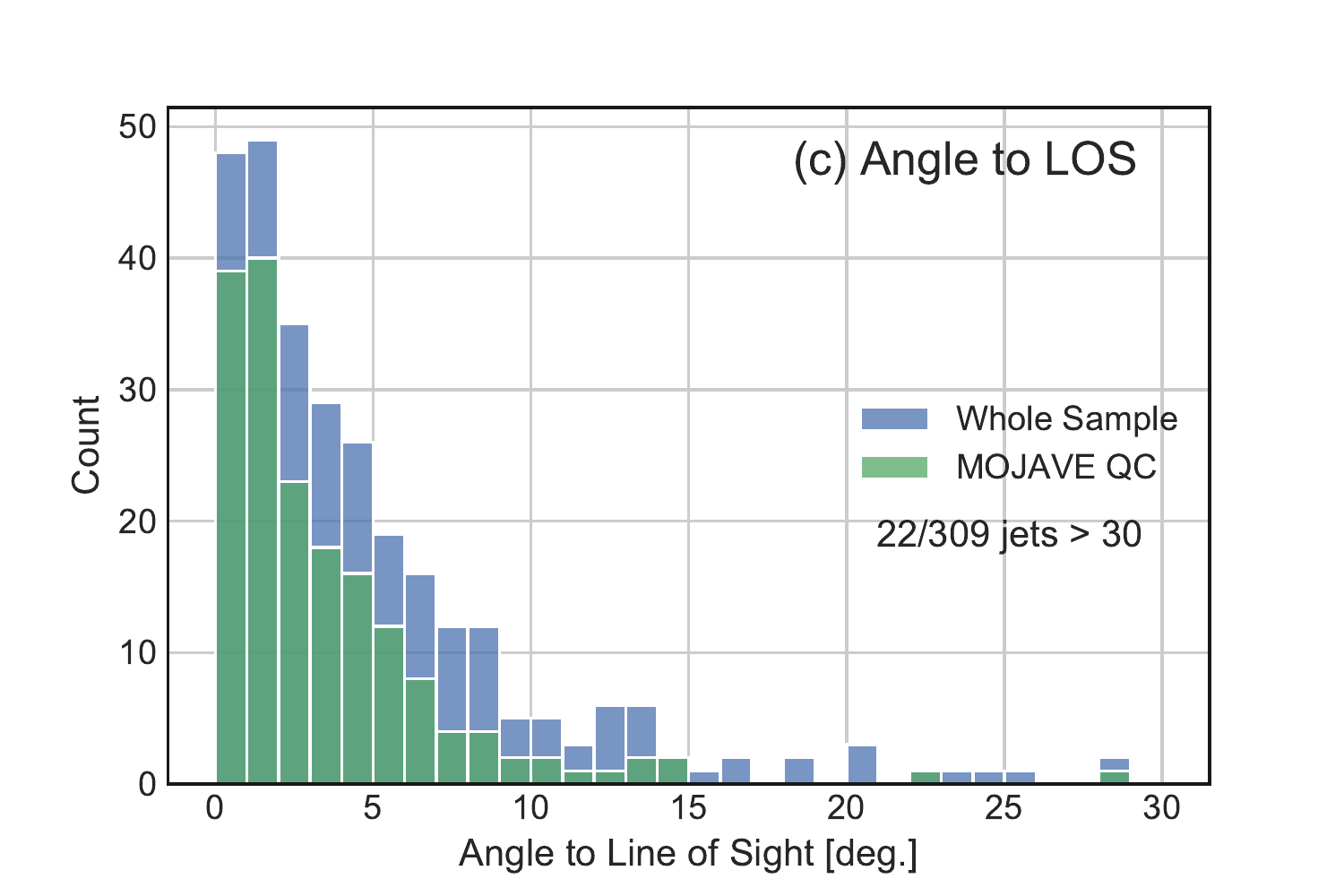}
\figcaption{\label{f:rel_hist}
  Histograms of Doppler factor, $\delta$, Lorentz Factor, $\Gamma$, and
  angle to the line of sight, $\theta$ derived from the median brightness
  temperature and apparent speeds as described in 
  \autoref{s:calc_quant}.  Note that a few
  outliers at larger values are not included on the plots for readability
  and the number of these are indicated on each panel.
}
\end{figure}

For each source in our sample, we use the assumption that
they all have the same intrinsic median brightness
temperature found above, $T_\mathrm{b,int} = 10^{10.609\pm0.067}$ K,
to estimate their Doppler factor from their median
observed brightness temperature, $\delta = T_\mathrm{b,obs}/T_\mathrm{b,int}$.
We then use their maximum observed speeds, $\beta_\mathrm{max}$,
to find their Lorentz Factor, $\Gamma$, angle to the
line of sight, $\theta$, and angle to the line of sight
in the source fluid frame, $\theta_\mathrm{src}$, as follows, e.g.,~\citet{J17}:
\begin{equation}
  \Gamma = (\beta^2_\mathrm{max} + \delta^2 + 1)/2\delta\,,
\end{equation}
\begin{equation}
  \theta = \arctan{\frac{2\beta_\mathrm{max}}{\beta^2_\mathrm{max} + \delta^2 - 1}}\,,
\end{equation}
\begin{equation}
  \theta_\mathrm{src} = \arccos{\frac{\cos\theta-\beta}{1-\beta\cos\theta}}\,.
\end{equation}
These values are listed in \autoref{t:doppler}, with distributions of $\delta$, $\Gamma$, and $\theta$ shown in \autoref{f:rel_hist}.

\begin{deluxetable}{lrrrrrr}
%\rotate
\tablenum{4}
\tablecolumns{7} 
\tabletypesize{\scriptsize} 
\tablewidth{0pt}  
\tablecaption{\label{t:doppler}Doppler Factors and Derived Properties}  
\tablehead{
\colhead{Source} & \colhead{$T_\mathrm{b,med}$} & \colhead{$\beta_\mathrm{max}$} &
\colhead{$\delta$} & \colhead{$\Gamma$} & \colhead {$\theta$} & \colhead{$\theta_\mathrm{src}$} \\
\colhead{} & \colhead{(log$_{10}$ K)} & \colhead {} &
\colhead {} & \colhead {} & \colhead {(deg)} & \colhead {(deg)} \\
\colhead{(1)} & \colhead{(2)} & \colhead{(3)} &
\colhead{(4)} &  \colhead{(5)} & \colhead{(6)} & \colhead{(7)}
}
\startdata
0003$+$380 & $ 11.550$ & $ 4.61\pm 0.36$ & $  8.7$ & $  5.6$ & $  5.5$ & $ 56.1$ \\ 
0003$-$066 & $ 11.079$ & $ 7.08\pm 0.21$ & $  3.0$ & $ 10.1$ & $ 13.8$ & $135.4$ \\ 
0006$+$061 & $>11.021$ & \ldots & $>  2.6$ &  \ldots & \ldots & \ldots \\ 
0007$+$106 & $ 11.729$ & $ 1.58\pm 0.29$ & $ 13.2$ & $  6.7$ & $  1.0$ & $ 13.7$ \\ 
0010$+$405 & $>11.425$ & $ 6.92\pm 0.64$ & $>  6.5$ &  \ldots & \ldots & \ldots \\ 
0011$+$189 & $>11.207$ & $ 4.54\pm 0.46$ & $>  4.0$ &  \ldots & \ldots & \ldots \\ 
0012$+$610 & $>10.747$\tablenotemark{a} & \ldots & $>  1.4$ &  \ldots & \ldots & \ldots \\ 
0014$+$813 & $ 11.223$ & $ 9.47\pm 0.91$ & $  4.1$ & $ 13.1$ & $ 10.2$ & $133.5$ \\ 
0015$-$054 & $ 11.246$ & \ldots & $  4.3$ &  \ldots & \ldots & \ldots \\ 
0016$+$731 & $ 11.902$ & $ 7.64\pm 0.32$ & $ 19.6$ & $ 11.3$ & $  2.0$ & $ 42.6$ \\ 
\ldots \\
\enddata
\tablenotetext{a}{Lower limit value ($z=0$) only on account of unknown source redshift.}
\tablecomments{
The complete version of this table appears in the online journal.
  Table of source properties deduced from the brightness temperature vs.\
  speed analysis. All 448 source are included in this table, but only 309
  sources have both measured apparent speeds and non-limit brightness temperatures,
  making them suitable for the full analysis as described in \autoref{s:betaT_analysis}. 
Columns are as follows: 
(1) Source name in B1950 coordinates; 
(2) Median peak Gaussian brightness temperature;
(3) Fastest apparent speed;
(4) Doppler factor assuming $T_\mathrm{b,int} = 10^{10.609}$\,K as found in 
\S{\ref{s:betaT_analysis}};
(5) Lorentz factor derived from $\delta$ and $\beta_\mathrm{max}$;
(6) Angle to the line of sight derived from $\delta$ and $\beta\mathrm{max}$;
(7) Angle to the line of sight in the co-moving jet frame;
}
\end{deluxetable}

\subsubsection{Comparing Doppler factor values to previous estimates}
\label{s:doppler_compare}

\begin{figure*}
\centering
\includegraphics[width=\textwidth]{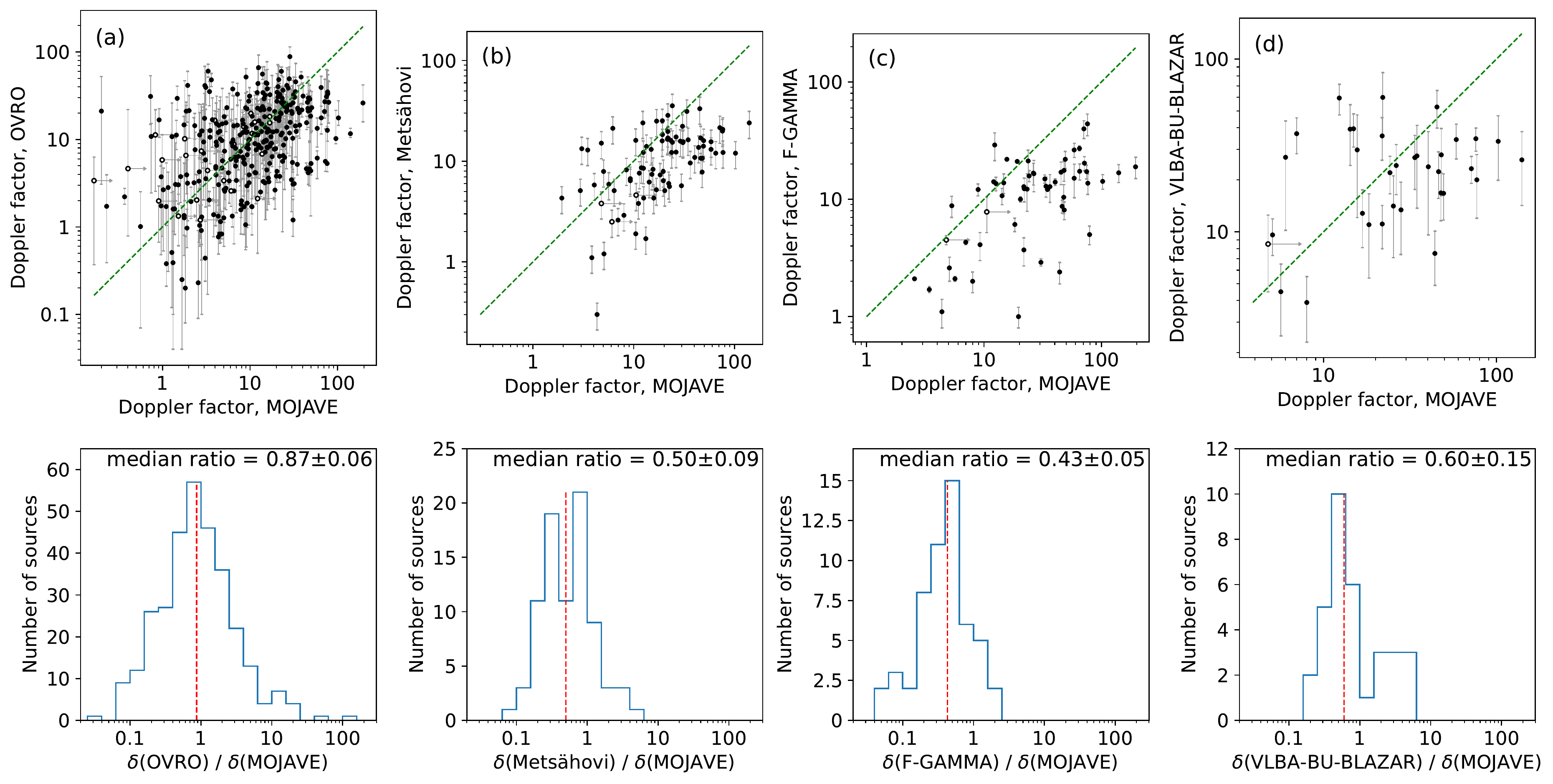}
\figcaption{\label{f:doppler_compare}
  Comparison of the Doppler factors estimated in this work with those previously estimated from different monitoring programs: (a) OVRO \citep{2018ApJ...866..137L}; (b) Mets{\"a}hovi \citep{2009A&A...494..527H}; (c) F-GAMMA \citep{2017MNRAS.466.4625L}; (d) VLBA-BU-BLAZAR \citep{J17}. {\it Upper panel:} our measured values are marked by dots, while our lower limits are marked by open circles with arrows. The dashed line marks the ideal case when Doppler factors are equal. {\it Lower panel:} distributions of the ratio of the Doppler factors. The median ratios are marked by vertical red dashed lines and are given above each histogram with their errors estimated by bootstrapping. See the discussion of the  correlations and offsets in \autoref{s:doppler_compare}.
}
\end{figure*}

It is interesting to compare Doppler factors we estimated from the median core brightness temperature to the values obtained by different methods. Doppler factors have been estimated for a large number of sources by flare modeling using the data of the single-dish monitoring programs at the OVRO 40 m radio telescope at 15~GHz \citep{2018ApJ...866..137L}, at the Mets{\"a}hovi Radio Observatory at 22 and 37~GHz \citep{2009A&A...494..527H}, and at the Effelsberg 100 m and IRAM 30 m telescopes within the F-GAMMA project at the frequencies from 2.64 to 86~GHz \citep{2017MNRAS.466.4625L}. \cite{J17} estimated Doppler factors by another method, using the flux-density decay timescale of VLBI superluminal components at 43 GHz. 
\autoref{f:doppler_compare} shows the comparison of these values with our results. There is a statistically significant correlation between our Doppler factors and those obtained from the single-dish monitoring programs (panels (a)--(c)): $p$-values determined by the Kendall partial (given redshift) correlation test, accounting also for lower limits, are no more than $10^{-3}$. 

The most significant correlation, $p\approx10^{-12}$, is with the OVRO values 
(\autoref{f:doppler_compare}a, upper panel). These values also have the smallest 
median offset, about 10\%, from our estimates 
(\autoref{f:doppler_compare}a, lower panel). The Doppler factors presented here 
and in the OVRO results are estimated by two very different methods, in different 
states of the sources with quite different corresponding estimates for 
$T_\mathrm{b,int}$ in those states. As described in \autoref{s:energy-balance}, our 
typical intrinsic core brightness temperature for the median state is at or below 
the equipartition value while the flaring state intrinsic core brightness 
temperature from \cite{2018ApJ...866..137L} is only 2 times smaller than the 
inverse-Compton limit \citep{Readhead94, KPT69}. 
The fact that the resulting Doppler factors are in such a good agreement lends 
confidence to both methods, although we note that the
two approaches are not totally independent as 
\citet{2018ApJ...866..137L} 
used population modeling of an earlier set of MOJAVE kinematics to help constrain 
their value of $T_\mathrm{b,int}$ in the flaring state. 

The values from \cite{2009A&A...494..527H} and \cite{2017MNRAS.466.4625L} also correlate with ours, but are, on average, about two times smaller 
(Figures~\ref{f:doppler_compare}b and \ref{f:doppler_compare}c). In both of these works, the authors used as intrinsic brightness temperature its equipartition value $T_\mathrm{eq}=5\times10^{10}$~K \citep{Readhead94}. 
Re-scaling their Doppler factors to the higher $T_\mathrm{b,int}=2.8\times10^{11}$~K value used by \cite{2018ApJ...866..137L} would decrease them by about a factor of two, increasing their difference from our estimates. \cite{2018ApJ...866..137L} discuss several
possible reasons for this disagreement between the otherwise similar variability approaches, including possibly 
insufficient cadence of the earlier observations. Our Doppler factors and those from \cite{J17} are 
poorly correlated, regardless of which Doppler factor values for individual jet 
components from \cite{J17} are used to represent each source: the maximum, the median, or the average value. 
For \autoref{f:doppler_compare}d, the maximum values are used. The 
Doppler factors estimated by \cite{J17} may simply have a larger scatter if the assumption 
that the observed flux density decay timescale of jet components equals to their light-crossing time 
divided by the Doppler factor is not always satisfied.

\section{Results and Discussion}
\label{s:discuss}

\subsection{Observed Brightness Temperature}
\label{s:tb-discuss}

In the frame of the host galaxy, the observed brightness temperature of
the core of an AGN jet depends on both the Doppler boosting 
factor, $\delta$,
of the jet flow and the intrinsic brightness temperature, $T_\mathrm{b,int}$ 
of the emission region: $T_\mathrm{b,obs} = \delta T_\mathrm{b,int}$.  For 
an individual jet,
observed changes in $T_\mathrm{b,obs}$ can reflect changes in either 
quantity or both.
The Doppler boosting factor can vary if there are changes in the flow
speed or direction, and the intrinsic brightness temperature can
change with optical depth (expected to be near unity in AGN jet
cores) and the balance between particle and field energy in the
emission region \citep[e.g., ][]{Readhead94}.

Our measurements of the Gaussian peak brightness temperature of the
core region of each jet, in every epoch, are reported in \autoref{t:comp_fits} and
illustrated in \autoref{f:fig2}.  From studying individual sources in \autoref{f:fig2},
it is apparent that the typical variation in $T_\mathrm{b,obs}$ over time for a given
jet is a factor of a few up to about an order of magnitude, with a few
extreme cases, like 0716$+$714, having larger variations.  However 
the differences between AGN can
be much larger, with median brightness temperature values spanning up
to three orders of magnitude across our heterogeneous 447 source sample.
The flux-density limited MOJAVE 1.5\,Jy QC sub-sample has median brightness
temperatures which span a somewhat narrower range of about two
and half orders of magnitude, see \autoref{f:fig3}.

This range of observed median brightness temperatures is consistent
with Doppler boosting being the primary difference between 
AGN jets in their median state; however, variations over time for
an individual jet may be more strongly connected to the emergence
of new features and changes in the energy balance between particles
and magnetic fields in the emission region.  In the subsections that
follow, we look first
at trends with median brightness temperature across the sample
(\autoref{s:Tb-trends}), and we then consider variability
in brightness temperature (\autoref{s:var-trends}).

\subsubsection{Trends with Median \texorpdfstring{$T_b$}{Lg}}
\label{s:Tb-trends}

\autoref{f:fig3} showed histograms of the median observed brightness
temperatures for our sample as a whole (panel a) and the MOJAVE 1.5\,Jy 
QC sub-sample (panel b), and beneath these panels we showed
box plots illustrating the range of median brightness
temperature values for different optical classes.  
Quasars ($n_\mathrm{ws} = 271$, $n_\mathrm{m15} = 158$)\footnote{The subscript 
``ws'' refers to our whole sample, while ``m15'' is the MOJAVE 1.5\,Jy 
QC flux-density limited sub-sample.},  
BL\,Lacs ($n_\mathrm{ws} = 136$, $n_\mathrm{m15} = 37$),
and 
galaxies ($n_\mathrm{ws} = 23$, $n_\mathrm{m15} = 6$)
appear to differ in their median brightness temperatures.
Because some of our median brightness temperatures are lower limits,
we use a pair-wise log-rank test from the Numerical Python ``lifelines''
distribution \citep{lifelines} to account for this censored data.
We find that galaxies are very unlikely to be drawn from
the same distribution as quasars 
($p_\mathrm{ws} < 0.001$, $p_\mathrm{m15} < 0.001$)
or BL\,Lacs ($p_\mathrm{ws} < 0.001$, $p_\mathrm{m15} < 0.001$).  BL\,Lacs
appear to differ from quasars for our whole sample ($p_\mathrm{ws} = 0.028$)
but we detect no difference in the flux-density limited 
MOJAVE 1.5\,Jy QC sub-sample ($p_\mathrm{m15} = 0.93$).

\begin{figure}
\centering
 \includegraphics[width=0.4\textwidth]{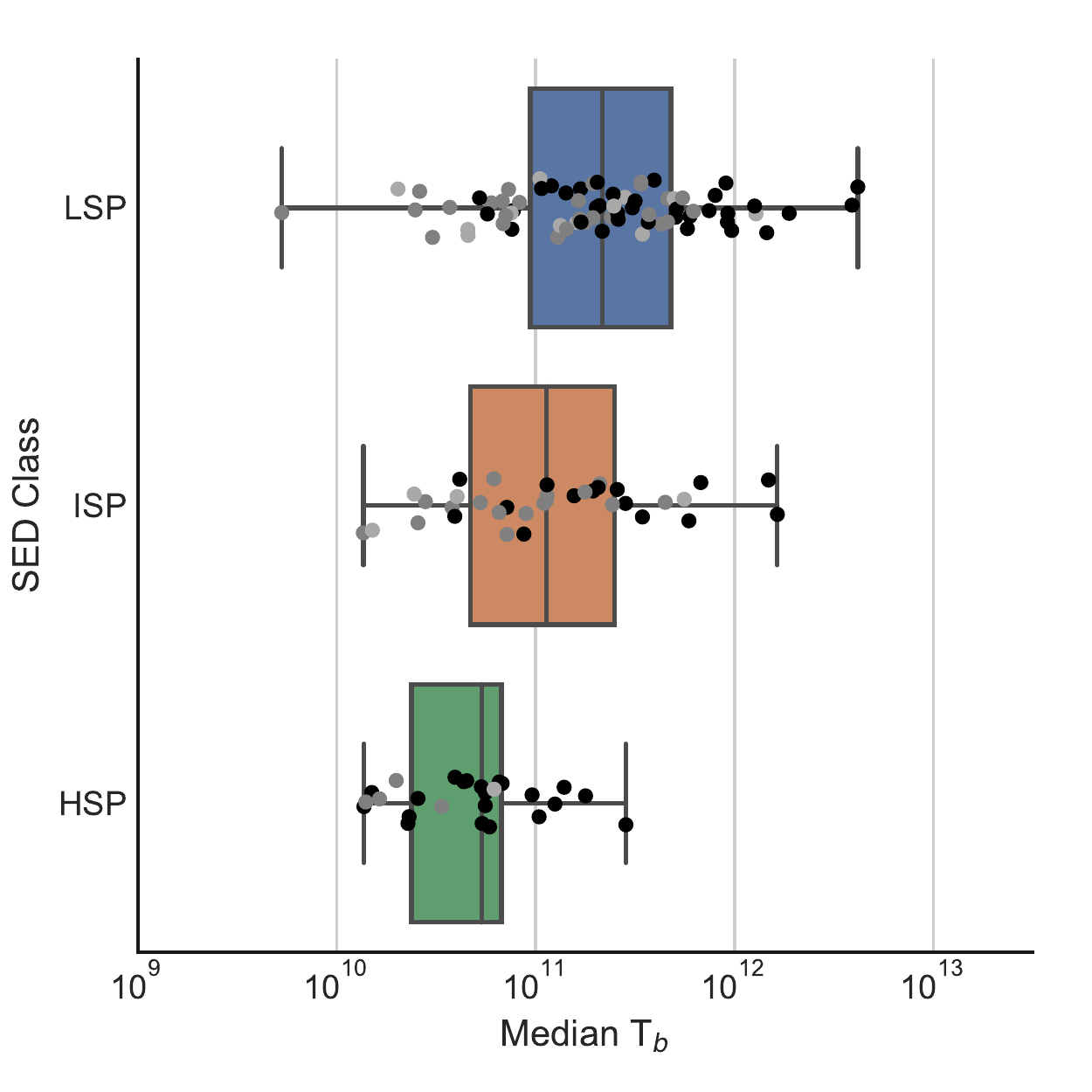}
 \includegraphics[width=0.4\textwidth]{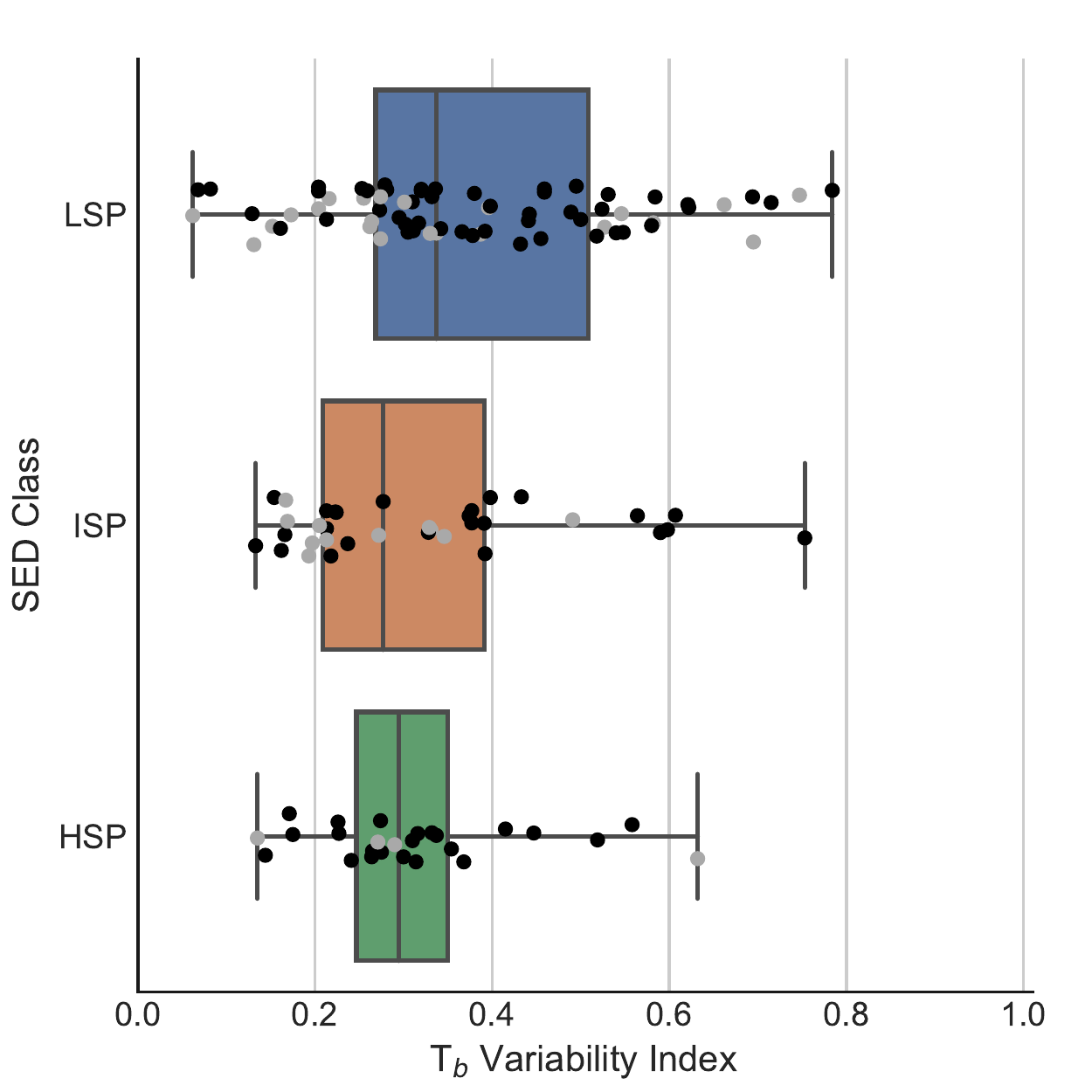}  
\figcaption{\label{f:Tb_SED_BLLacs}
  Distributions of the brightness temperature (left) and variability index
  (right) for the BL\,Lac objects in our whole sample as a function of 
  SED Class.
  The ``LSP'', ``ISP'', and ``HSP'' abbreviations indicate low, 
  intermediate, and high-synchrotron-peak sources respectively.  
  The scattered points plotted over each box plot indicates 
  the locations of the individual values for that distribution.
   Note that the inner-quartile range in each boxplot is shown without regard to limit status of the individual points; however, the overplotted points are marked as measurements or limits as described below.   In running statistical tests between distributions, we use the log-rank test, as described in the text, to properly account for the limits.
  Gray filling indicates lower limits, where the darker gray is 
  for sources where the lower limit is solely due to the missing 
  redshift.
}
\end{figure}

\begin{figure}
 \includegraphics[width=\linewidth]{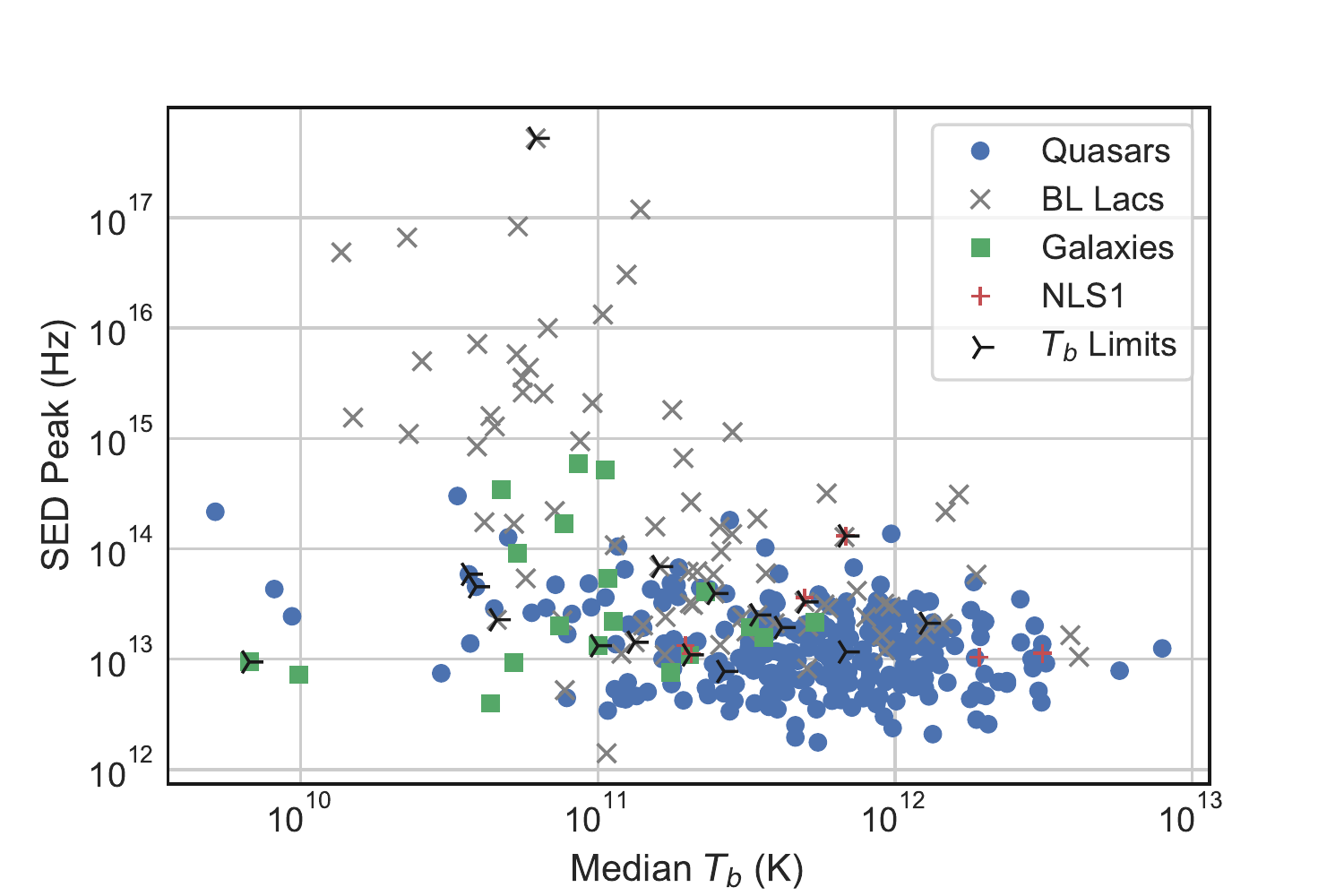} 
\figcaption{\label{f:Tb_v_SEDpeak}
Spectral energy density peak frequency in the host galaxy rest frame vs.\ 
median Gaussian brightness temperature for the whole sample.
}
\end{figure}

The BL\,Lacs in our flux-density limited, MOJAVE 1.5\,Jy QC sample are strongly
dominated by sources with a spectral energy distribution characterized
by a low synchrotron peak (LSP).  In \autoref{f:Tb_SED_BLLacs}, 
we compare the median brightness temperatures of LSP BL\,Lacs ($n=75$) to 
those with intermediate or high synchrotron peaks, ISP ($n=35$)  and 
HSP ($n=26$), which are better represented
in our whole, heterogeneous sample.  HSP BL\,Lacs have distinctly 
lower median brightness temperatures when compared to ISP or 
LSP BL\,Lacs as confirmed by a log-rank test with $p < 0.001$ for 
both comparisons; however, we detect no difference between
the median brightness temperature distributions of ISP and LSP 
BL\,Lac classes ($p = 0.14$). \autoref{f:Tb_v_SEDpeak} shows 
a plot of SED peak frequency in the galaxy rest frame 
versus median brightness temperature. BL\,Lac objects in particular 
show a strong negative correlation between SED peak frequency and median 
brightness temperature.

If the median observed brightness temperature is a good proxy 
for the Doppler beaming factor, these results mean that radio galaxies
are less beamed than BL\,Lacs and quasars as one would expect from
unification arguments \citep[e.g., ][]{1995PASP..107..803U}; however, 
we do not detect a difference between BL\,Lacs and quasars in
the flux-density limited MOJAVE 1.5\,Jy QC sample. The apparent difference
between these two classes in our larger, heterogeneous sample is likely
due to differences within the BL\,Lac optical class itself.   The 
differences in median brightness temperature between HSP and 
lower synchrotron peaked sources suggest that HSP BL\,Lacs are 
less beamed than those whose SEDs peak at lower
frequencies, consistent with earlier findings
\citep[e.g., ][]{2008AA...488..867N, 2011ApJ...742...27L}.

\begin{figure}
\includegraphics[width=\linewidth]{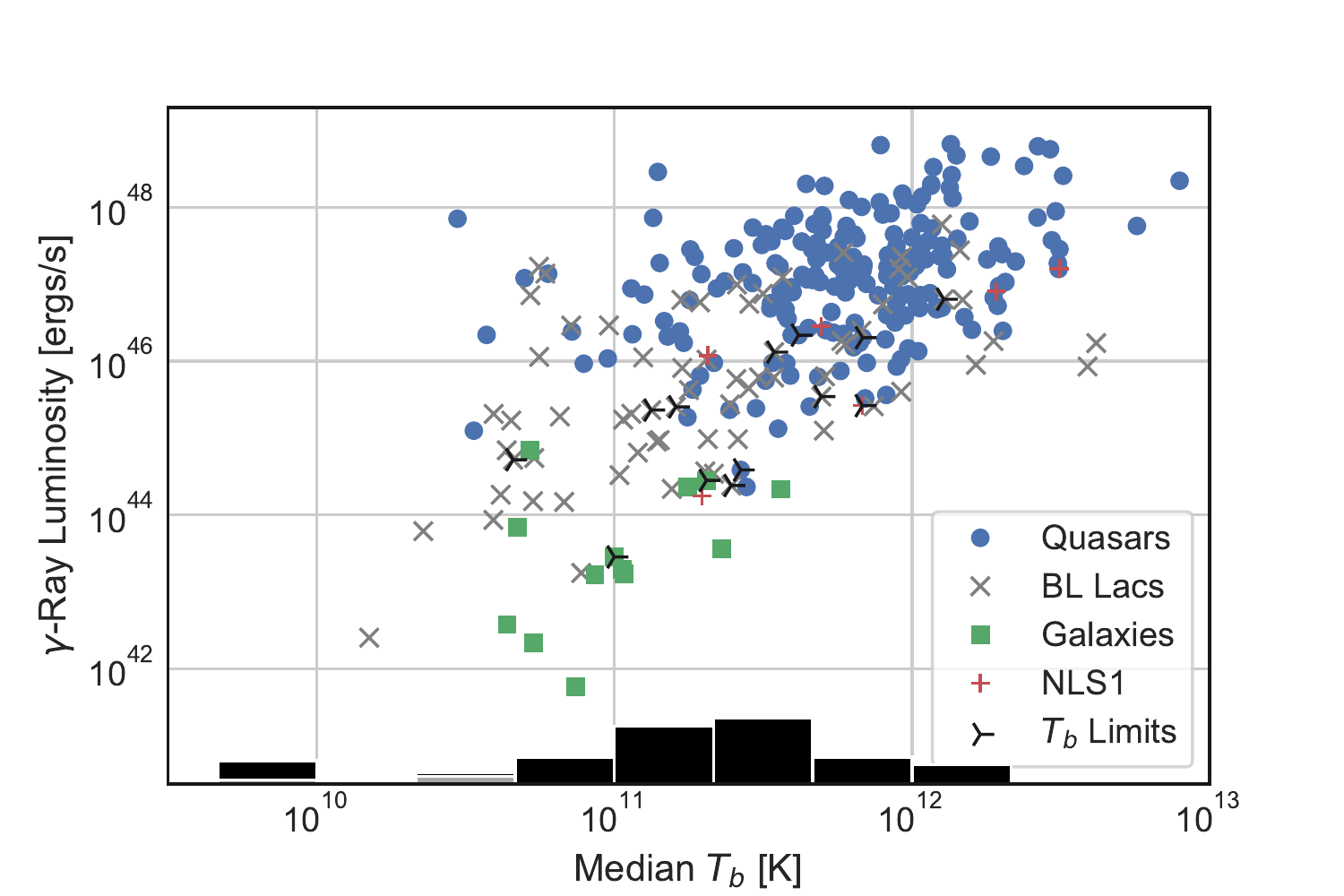}
\figcaption{\label{f:lat}
  $\gamma$-ray luminosity vs median Gaussian brightness 
  temperature for 291 \textit{Fermi}/LAT-detected AGN. The histogram at the 
  bottom of the plot shows the distribution of 60 sources in our 
  sample with measured brightness temperature but without
  \textit{Fermi}/LAT detections, gray bars in the histogram indicate lower 
  limits on the measured brightness temperature.
  Only sources with known redshifts and with a galactic 
  latitude $|b| > 10$ degrees are included in this plot.
}
\end{figure}

In \autoref{f:lat} we plot $\gamma$-ray luminosity vs median
brightness temperature for 291 \textit{Fermi}/LAT-detected AGN.
The luminosity values are computed from the
\textit{Fermi}/LAT 10-year point source catalog \citep{2020ApJ...892..105A}
using their $0.1-100$ GeV energy flux and power-law spectral index
following the approach given by \citet{2011ApJ...742...27L}, equation 3.
To allow computation of their luminosity and to avoid issues related to
galactic foreground subtraction, only sources with known redshifts and with 
a galactic latitude $|b| > 10$ degrees are included in this plot.  The 
histogram at the bottom of the plot shows the 60 sources meeting the same 
criteria which do not have \textit{Fermi}/LAT detections in the 10-year 
point source catalog.

We see a strong, positive correlation between $\gamma$-ray 
luminosity and median observed brightness temperature.  
\autoref{f:lat} includes
lower limits on the median brightness temperature of only 13/291
of our LAT detected AGN, and we measure a significant Spearman rank
correlation for the remaining 278 sources of $\rho = 0.54$ ($p < 0.001$).  
However, we must be cautious in interpreting this correlation, as selection
effects must be considered as well as common factors that affect both 
$T_\mathrm{b}$ and the $\gamma$-ray luminosity.

The observed brightness temperature in the frame of the host galaxy depends
only weakly on redshift, see %equation 
\autoref{e:Tb}, and even if we divide out
the factor of $(1+z)$, the correlation remains significant 
($\rho = 0.32$, $p < 0.001$).   Another possible confounding factor
is that many sources in our sample are selected on the basis of their
radio flux density as part of the flux-density limited MOJAVE 1.5\,Jy QC 
sub-sample, and sources at large distances are likely to be highly beamed to 
meet this criterion, creating a natural correlation between Doppler
factor and luminosity distance.  In this same group of 278 sources 
we find a correlation of $\rho = 0.44$ ($p < 0.001$) between median
brightness temperature and luminosity distance squared, $D^{2}_\mathrm{L}$.  
If we divide the $\gamma$-ray luminosity by $D^{2}_\mathrm{L}$, 
the correlation
with median brightness temperature still remains significant 
with $\rho = 0.33$ ($p < 0.001$).

We can test the relationship between median brightness temperature
and $\gamma$-ray emission 
further by comparing
these results to those of 
\citet{2009ApJ...696L..17K} and \citet{2011ApJ...742...27L} who found 
that $\gamma$-ray detected jets in earlier LAT catalogs had higher 
brightness temperatures than non-detected jets.  Here we use a
log-rank test to compare the distributions of median brightness 
temperature of the detected $\gamma$-ray sources ($n = 291$)
to the non-detected sources ($n=60$) in \autoref{f:lat}, 
and we find the two groups are very unlikely to be drawn 
from the same distribution ($p < 0.001$) with the detected 
sources having distinctly larger
median brightness temperatures on average.  The Log-Rank test
correctly accounts for the lower limits on some of our brightness 
temperature values, and by simply comparing the detected vs.
non-detected distributions we are not biased by a possible
luminosity distance correlation with median brightness 
temperature through our flux-density limited radio sample.

Taken together these results imply a common Doppler
boosting of both the $\gamma$-ray 
emission and the brightness temperature of the radio 
core and will be discussed further in \autoref{s:intrinsic_prop_trends}.

\subsubsection{\texorpdfstring{$T_b$}{Lg} Variability}
\label{s:var-trends}

As described in \autoref{s:measureTb}, we characterize the brightness
temperature variability of each jet by using a fractional measure 
of the variability between the 25\% and 75\% points in the brightness 
temperature distribution over time, see \autoref{e:var2575}.  
\autoref{f:fig4} showed histograms of this brightness temperature 
variability index for our whole sample (panel a) and the MOJAVE 1.5\,Jy 
QC sub-sample (panel b).  Box plots below each histogram showed the
distribution of variability index for different optical classes.  
Across the whole sample, quasars ($n=269$)\footnote{The number
of sources with valid variability index values may be smaller than 
the number with brightness temperature measurements due to 
ambiguous combinations of lower limits in some cases.} 
appear to have higher variability 
and a log-rank test confirms that their distribution differs 
significantly from both BL\,Lacs ($n=132$, $p = 0.006$) and radio galaxies 
($n=22$, $p = 0.011$), although we detect no difference between BL\,Lacs
and radio galaxies when compared to each other ($p=0.36$).  For the 
MOJAVE 1.5\,Jy QC flux-density limited sample, we are unable to 
detect any difference in variability index distributions between 
quasars ($n=158$), BL\,Lacs ($n=37$), and radio galaxies ($n=6$) 
with $p \geq 0.48$ for each paired comparison.

The right panel of \autoref{f:Tb_SED_BLLacs} showed box plots of the brightness 
temperature variability index of ISP ($n=71$), LSP ($n=35$), and HSP ($n=26$)
BL\,Lacs in our 
sample as a whole, and paired log-rank tests show that HSP and LSP 
BL\,Lacs differ significantly from each other ($p = 0.004$);
however, we do not detect differences from ISP BL\,Lacs for either of them
($p = 0.18$ vs LSP and $p = 0.21$ vs HSP).

\subsection{Doppler Factors and Intrinsic Jet Properties}
\label{s:dopplerfactors}

In \autoref{s:betaT_analysis} we compared median observed brightness
temperatures of jet cores in the host galaxy frame to the maximum
apparent speeds in their jets to find a single, typical intrinsic
brightness temperature,  $T_\mathrm{b,int} = 10^{10.609\pm0.067}$ K, that we
could apply to estimate Doppler factors from the median observed
brightness temperature of each source: $\delta = T_\mathrm{b,obs}/T_\mathrm{b,int}$.
Combined with our apparent speed measurements, we estimated
Lorentz factors, angles to the line of sight,
and angles to the line of sight in the source fluid frame ($\theta_\mathrm{src}$)
for 309 sources for which we had all the necessary information,
178 of which are in the MOJAVE 1.5\,Jy QC flux-density limited
sample.

Histograms of $\delta$, $\Gamma$, and $\theta$ for
those sources where we have estimates for all three quantities
were shown in \autoref{f:rel_hist}.  For the MOJAVE 1.5\,Jy
QC sample, the overall trend and shape in
these histograms is similar to the simulated Monte Carlo distribution 
discussed by \citet[][fig. 11]{LHH19}.  The latter was fit using the 
observed redshift, 15 GHz flux density, and apparent jet speed distributions
reported in that paper for the 1.5\,Jy QC sample.  Our Doppler factor
distribution peaks near $\delta = 10$ and has a long, shallow tail out
to 100 with just three jets beyond that point.  We also see that the Lorentz
factor distribution peaks near $\Gamma = 10$, with a slower fall off toward
$\Gamma = 50$ and eight sources from the flux-density limited sample at
larger values. For the angle to the line-of-sight, we do not see the
sharp decline toward $\theta = 0^\circ$ from the simulation, likely due to the
uncertainty in our Doppler factor estimates described below, but our viewing
angle distribution does peak between 1 and 2 degrees, with a sharp
decline out to 10 degrees and beyond, similar to the simulation.
It is important to note that while we did not fit to the
\citet{LHH19} simulation in a detailed way, our procedure for 
estimating the best
value for $T_\mathrm{b,int}$ did seek to match the fraction of simulated sources 
inside the critical angle for superluminal motion.

Our analysis assumes that a single value of $T_\mathrm{b,int}$ applies
to all jets in their median state, and while this assumption seems
to do a reasonable job estimating the Doppler factors of jets in our
population, there may be some natural spread in this value. Sources
with intrinsically
smaller or larger values of $T_\mathrm{b,int}$ would then appear to have
corresponding larger or smaller Doppler factors in our data, leading
to a blurring of our Doppler factor distribution.  We
estimate this effect, along with any other uncertainties
that can lead to spread in our data, by comparing the distribution of 
Doppler factors in the \citet{LHH19}
simulation with the corresponding quantity from the quasars in
our flux-density limited sample.  The distribution from the simulation is
narrower than the one that is derived from the median $T_\mathrm{b,obs}$ 
values, and by comparing the standard deviation of the logarithms of the 
two distributions, we can estimate the additional spread in the measured 
distribution.  In this way 
we estimate our Doppler factors are good to, i.e., have a $1\,\sigma$ spread of, 
a multiplicative factor of approximately $1.8$\,.\footnote{Despite the 
numerical coincidence, this factor is 
unrelated to the $1.8$ geometric conversion factor for brightness 
temperatures discussed in \autoref{s:measureTb}}

There are five sources from our whole 309 jet sample which have estimated Lorentz
factors, $\Gamma > 100$, and all 
are quasars with Doppler factors much smaller than their apparent speeds.
All five sources have multiple fast motions observed
in their jets, so the discrepancy is unlikely to be caused by
a single outlier speed.  Three of these sources: 0519$+$011, 0529$+$075, and
1420$+$326 have estimated Doppler factors $< 1.0$, making them
highly improbable to be observed at such large redshifts, and 
we note that \cite{2018ApJ...866..137L} report variability Doppler 
factors $>15$ for each of them. The
most extreme case is 0519+011 with a Doppler factor of just 0.2
and multiple features showing approximately the same $25c$ apparent
motion, leading to an estimated $\Gamma = 1790$. 0519+011 is at
a very large redshift of $z = 2.941$, and its radio
core is very dim relative to the downstream jet emission. The
jet cores in these cases may suffer from absorption or opacity or may
simply have been in an atypically low state during our observations, either 
of which could lead to a larger than expected departure 
from our assumed value for $T_\mathrm{b,int}$.

There are also five jets which have estimated viewing angles to the
line of sight, $\theta > 90^\circ$.  Three of the five are galaxies
and two are HSP BL\,Lacs, all at low redshift with $\delta < 1$ and
$\beta_\mathrm{app} < 1$.  While a $\theta > 90^\circ$ value is nonphysical
for an approaching jet, uncertainties in the 
Doppler factor consistent with our estimates given above
can bring them to more reasonable viewing angles.  For example, 1957$+$405
(Cygnus\,A), has $\theta = 127^\circ$ from this analysis, but
if its Doppler factor was $1.5\times$ higher, it would be at
$\theta = 60^\circ$, consistent with the $45^\circ < \theta < 70^\circ$
range estimated by \citet{2007ApJ...658..232C}.

Finally, there may be some jets for which the fastest apparent speed
is not a good indicator of the flow speed, and these cases will
have poor estimates of $\Gamma$ and $\theta$.  In \autoref{s:whatspeed} we 
examine the impact on our results if we had used the median instead of the fastest 
speed in our analysis; however, there may be individual sources for which the
measured speeds themselves are not reliable tracers of the flow.  One 
possible example is 1228$+$126 (M\,87), which has a Doppler
factor of $\delta = 1.8$ in our analysis, consistent with the jet to 
counter-jet ratio of 10-15 reported by \cite{2007ApJ...668L..27K}; however, its
fastest apparent speed is just 0.02$c$ as reported in \citetalias{MOJAVE_XVIII}, giving
an angle to the line of sight of $\theta=1.0^\circ$ in our analysis.
\cite{2007ApJ...668L..27K} discuss the apparent speed issue for M\,87 in 
depth including the possibility we are seeing slow pattern motions in a 
spine-sheath structure. \cite{2018ApJ...855..128W} used high cadence 43 GHz
VLBA observations to show that the apparent speed of the jet increases from 
$\lesssim 0.5$c to $\gtrsim 2$c over the first two milli-arcseconds. 
Combined with our $\delta = 1.8$, these speeds would change the 
estimated angle to the line of sight for M87 
to be in the range $22^\circ - 33^\circ$.

\subsubsection{Trends with \texorpdfstring{$\delta$}{Lg}, \texorpdfstring{$\Gamma$}{Lg}, 
\texorpdfstring{$\theta$}{Lg}, and \texorpdfstring{$\theta_\mathrm{src}$}{Lg}}
\label{s:intrinsic_prop_trends}

\begin{figure}
\centering
\includegraphics[width=0.38\textwidth]{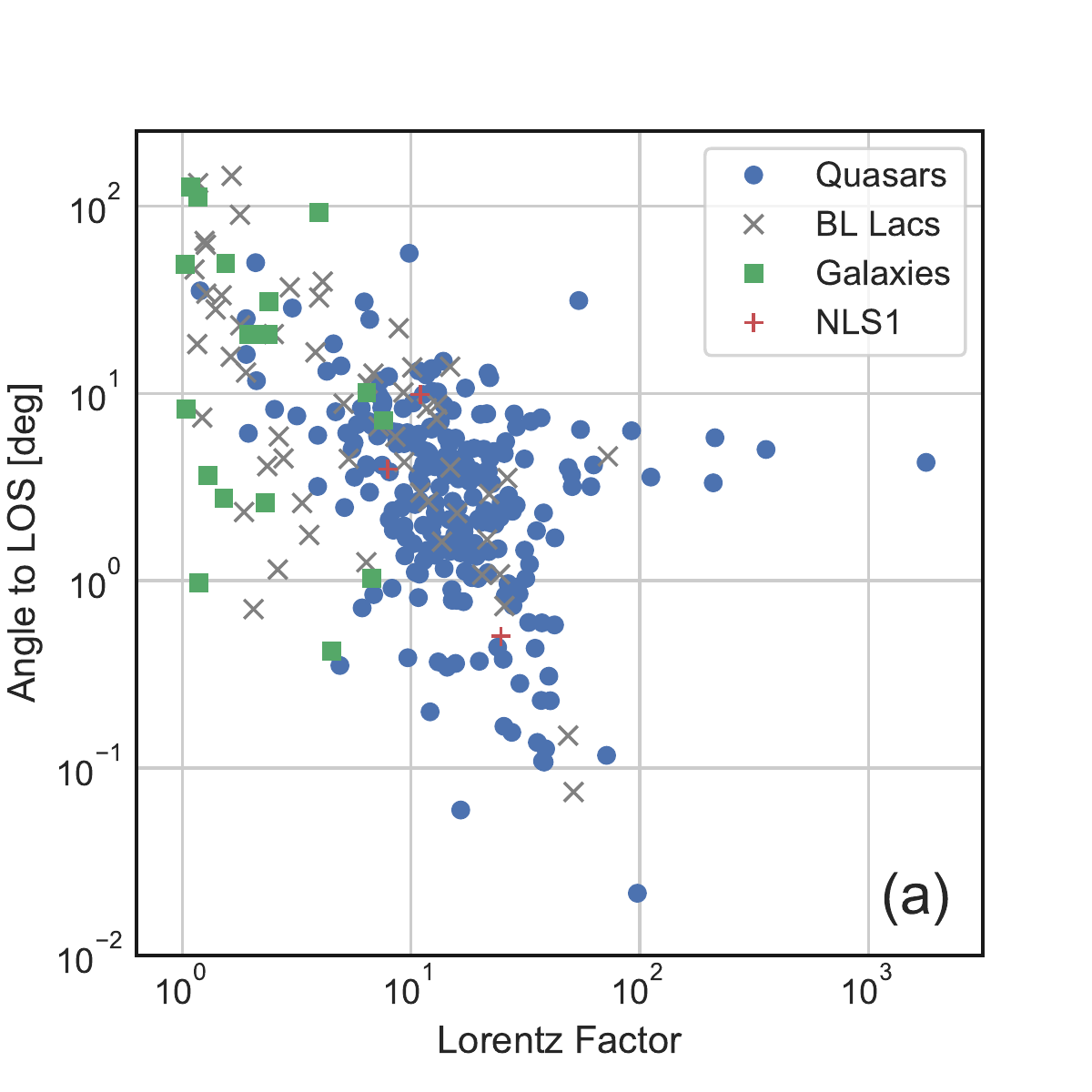}
\includegraphics[width=0.38\textwidth]{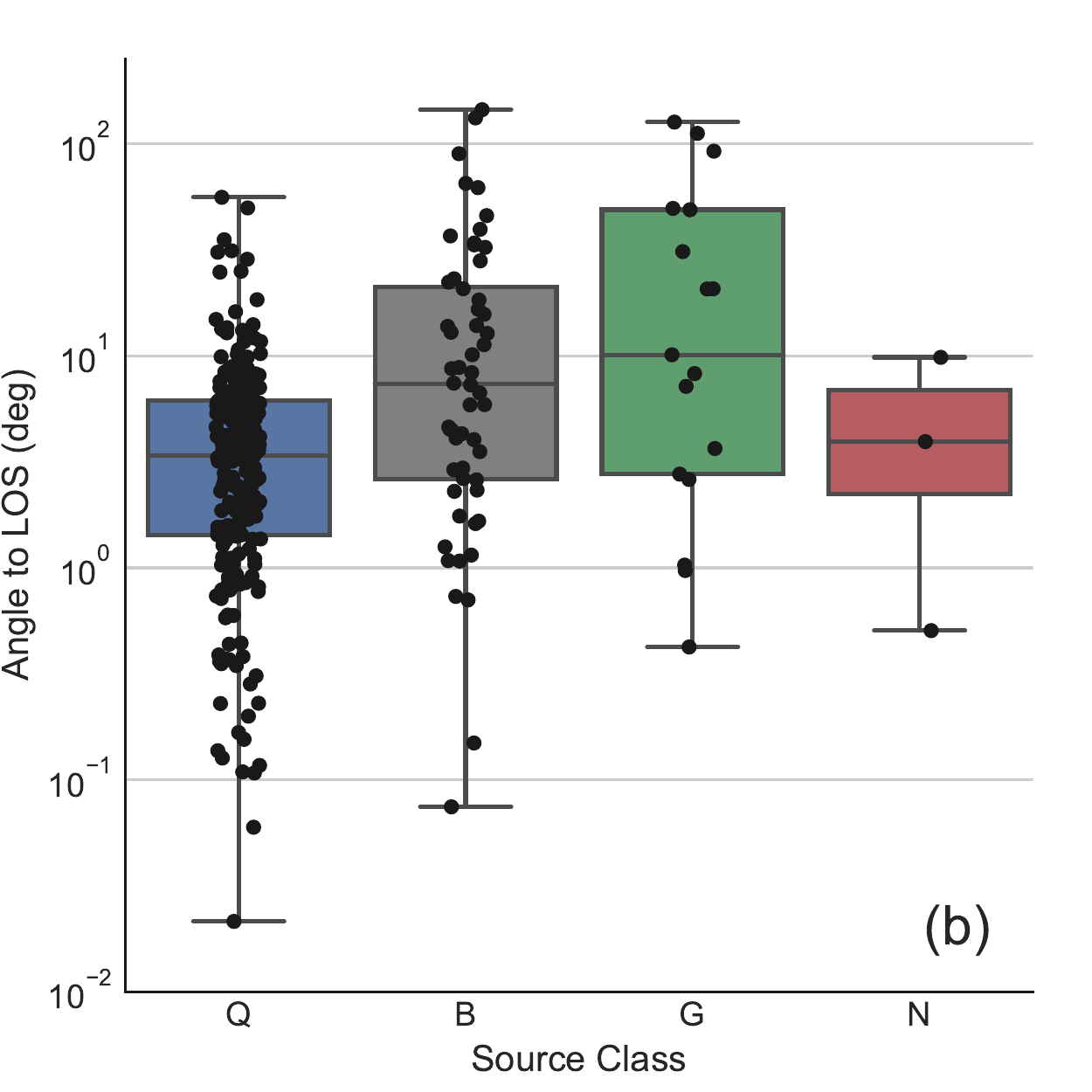}
\includegraphics[width=0.38\textwidth]{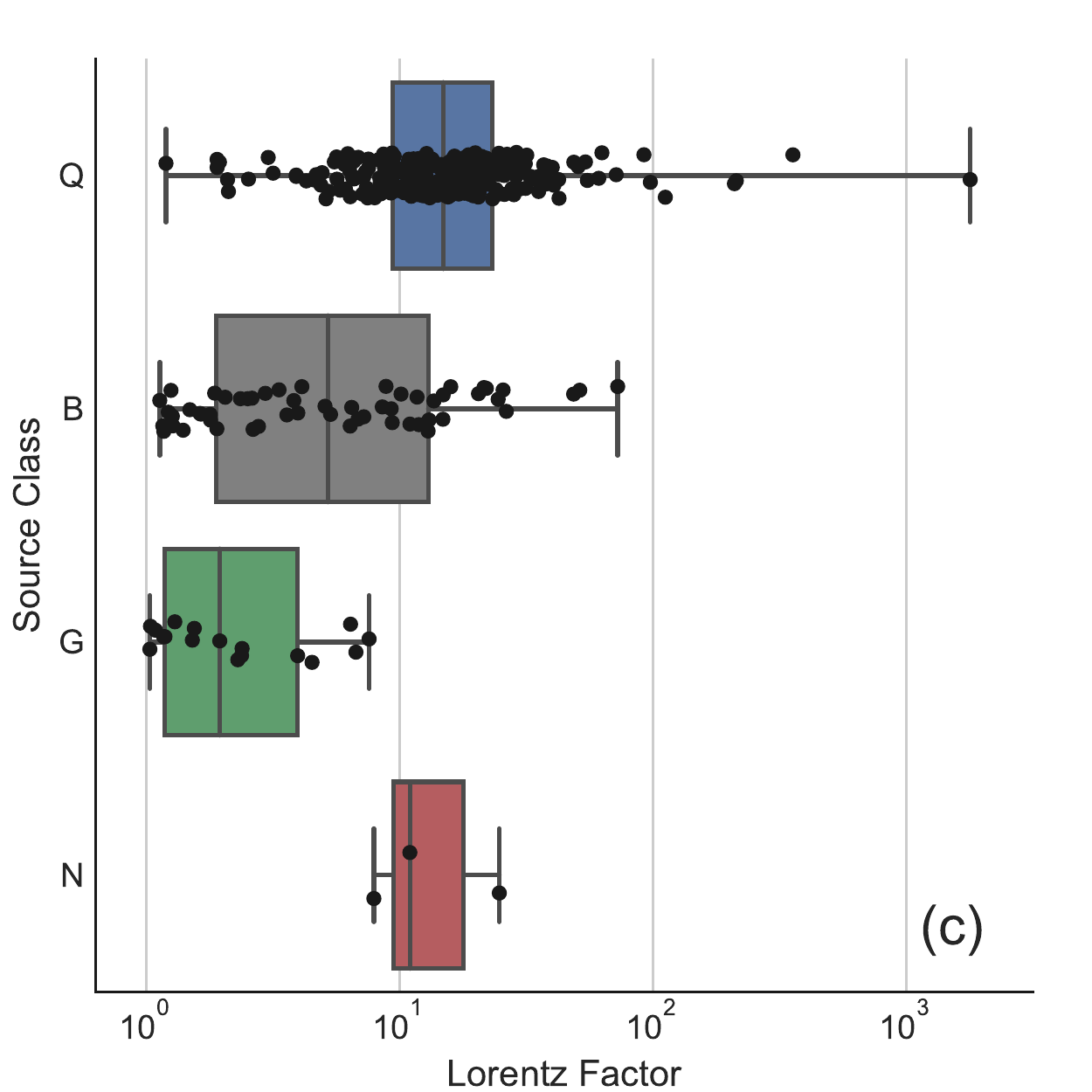}
\figcaption{\label{f:ThetaGamma_WS}
  Angle to the line of sight, $\theta$, plotted against 
  Lorentz Factor, $\Gamma$,
  (panel a) for all 309 sources with apparent speeds and median brightness
  temperature measurements. Panels (b) and (c) illustrate the distributions
  of these quantities as function of optical class, where ``Q'' = quasars,
  ``B'' = BL\,Lacs, ``G'' = radio galaxies, and ``N'' = narrow-line Seyfert Is.
  The filled regions of the box plots show in the inner-quartile range of
  each optical class, while the whiskers show the full extent of the 
  data.  Individual data points are shown as a scatter plot over the 
  box plot to better illustrate the range and density of the data.
}
\end{figure}

\begin{figure}
\centering
\includegraphics[width=0.38\textwidth]{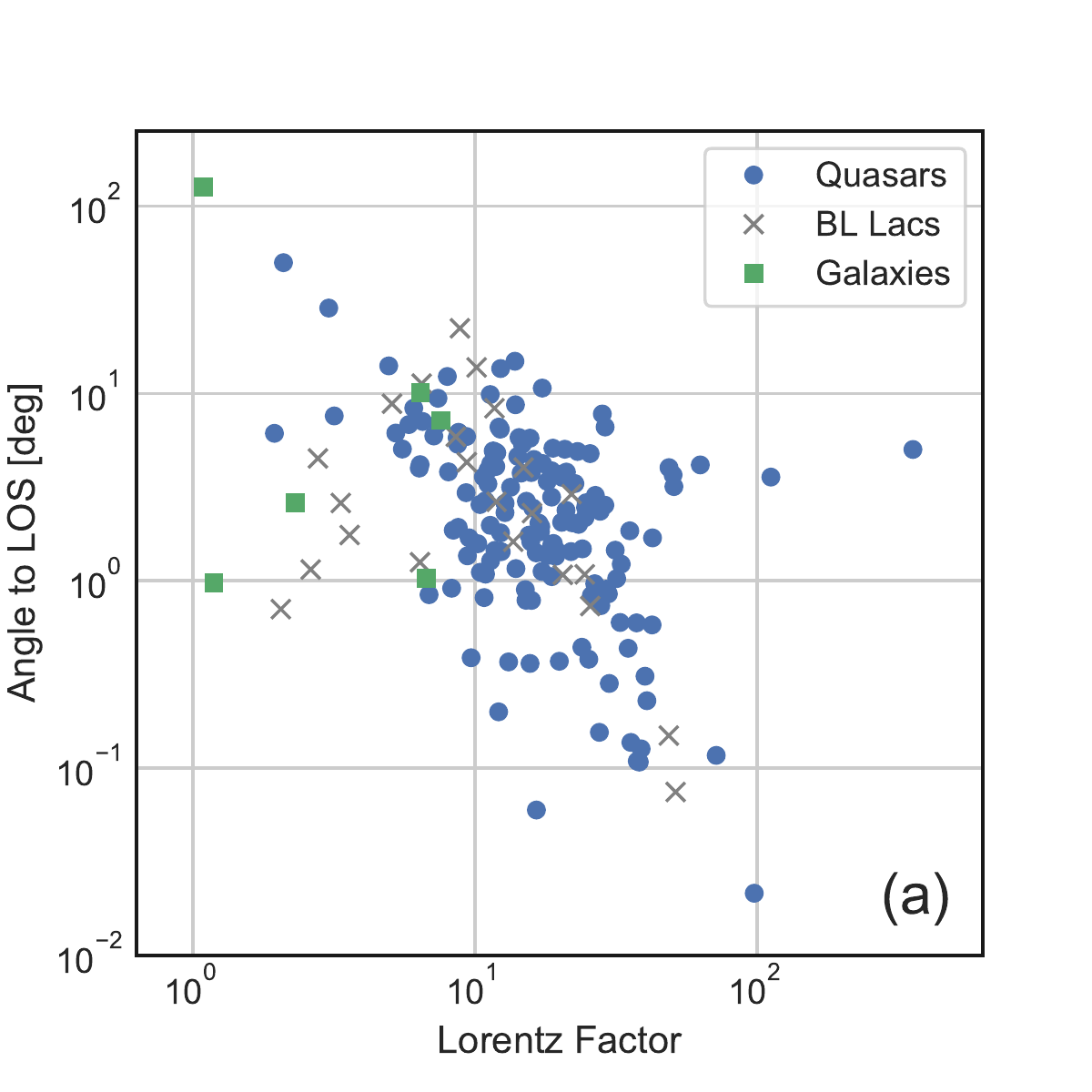}
\includegraphics[width=0.38\textwidth]{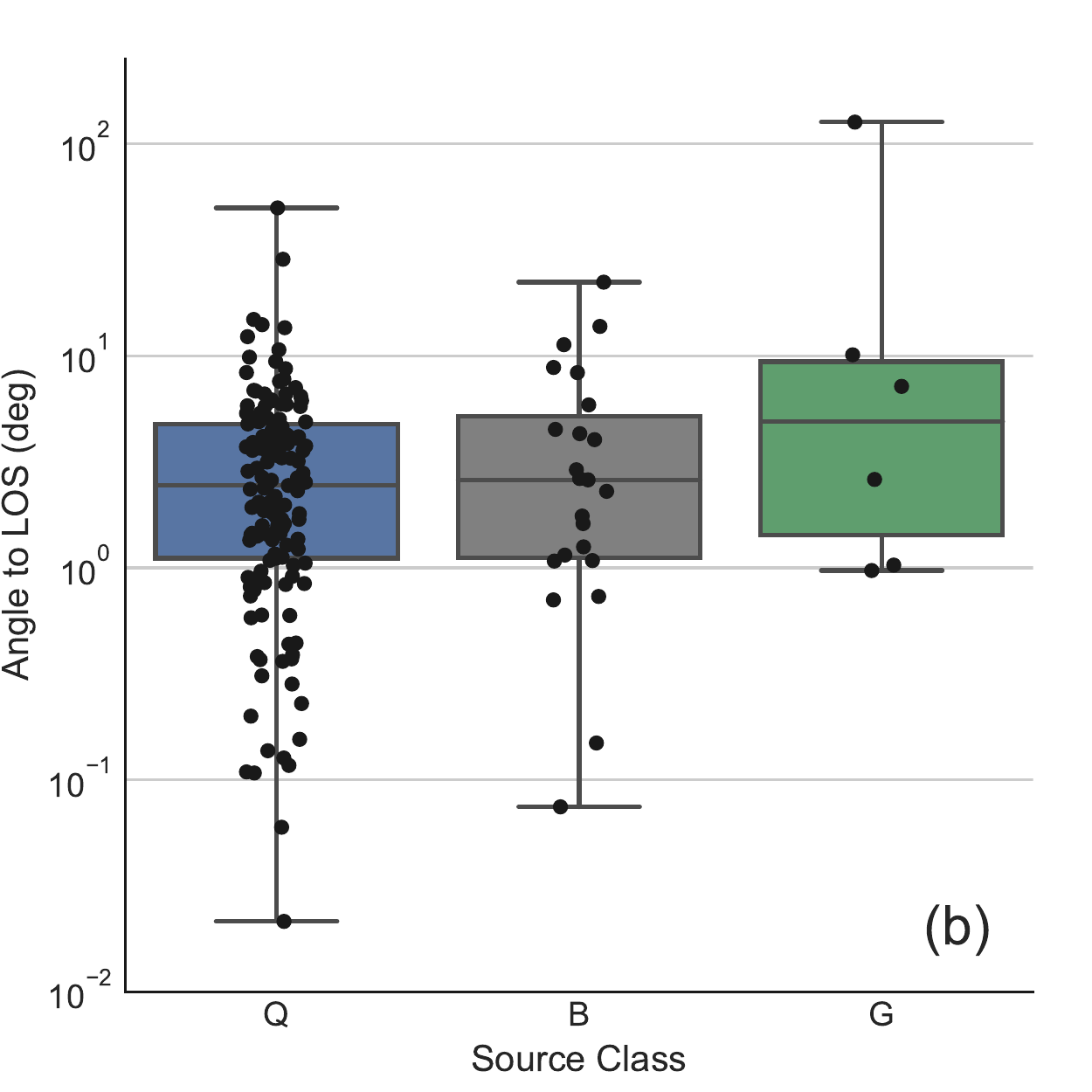}
\includegraphics[width=0.38\textwidth]{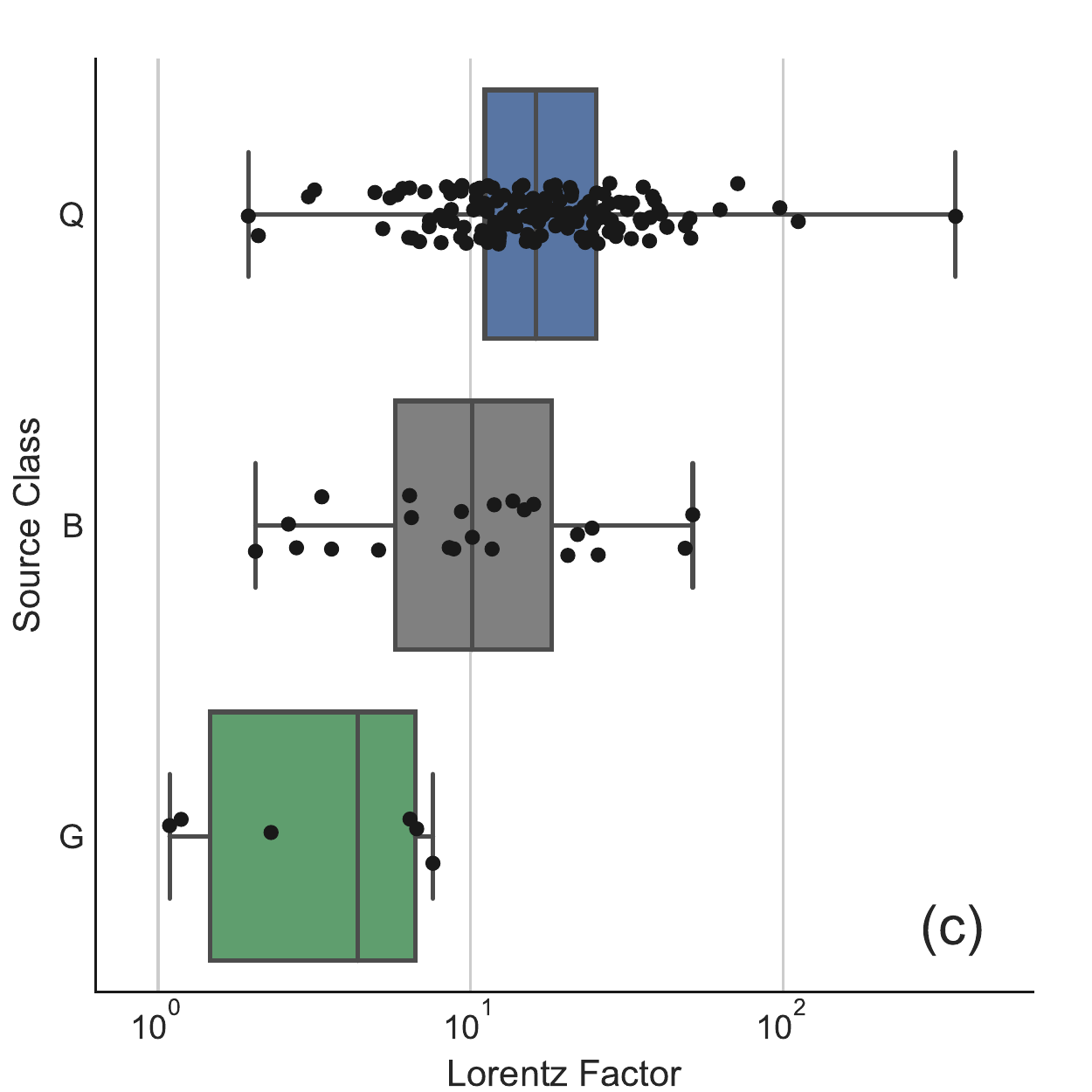}
\figcaption{\label{f:ThetaGamma_MQC}
  Angle to the line of sight, $\theta$, plotted against Lorentz Factor, 
  $\Gamma$, (panel a) for the MOJAVE 1.5\,Jy QC Sample (panel a). 
  Panels (b) and (c) illustrate the distributions of these quantities 
  as function of optical class, where ``Q'' = quasars,
  ``B'' = BL\,Lacs, and ``G'' = radio galaxies.
  The filled regions of the box plots show in the inner-quartile range of
  each optical class, while the whiskers show the full extent of the data. 
  Individual datapoints are shown as a scatter plot over the box plot to 
  better illustrate the range and density of the data.
}
\end{figure}

\begin{figure}
\centering
\includegraphics[width=0.38\textwidth]{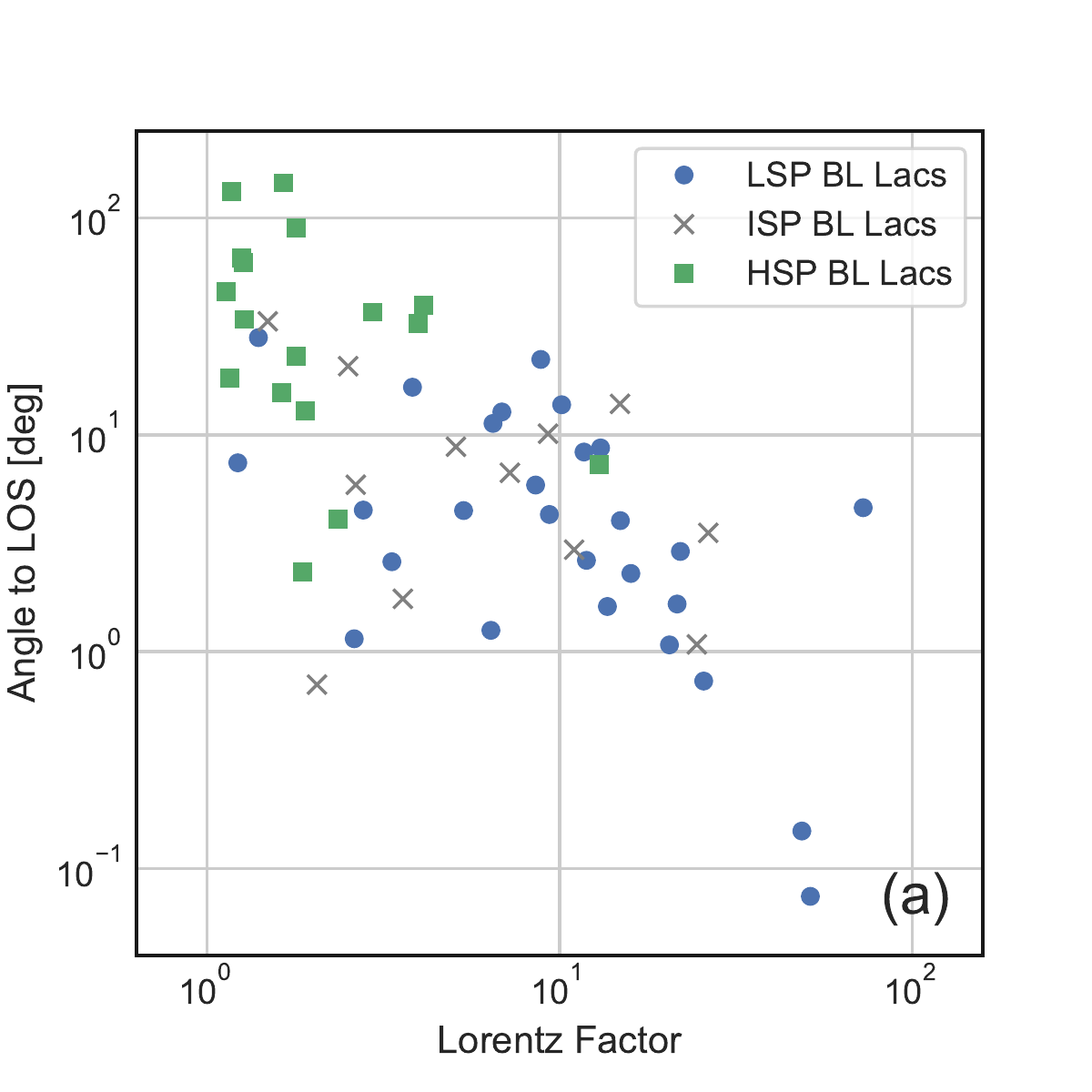}
\includegraphics[width=0.38\textwidth]{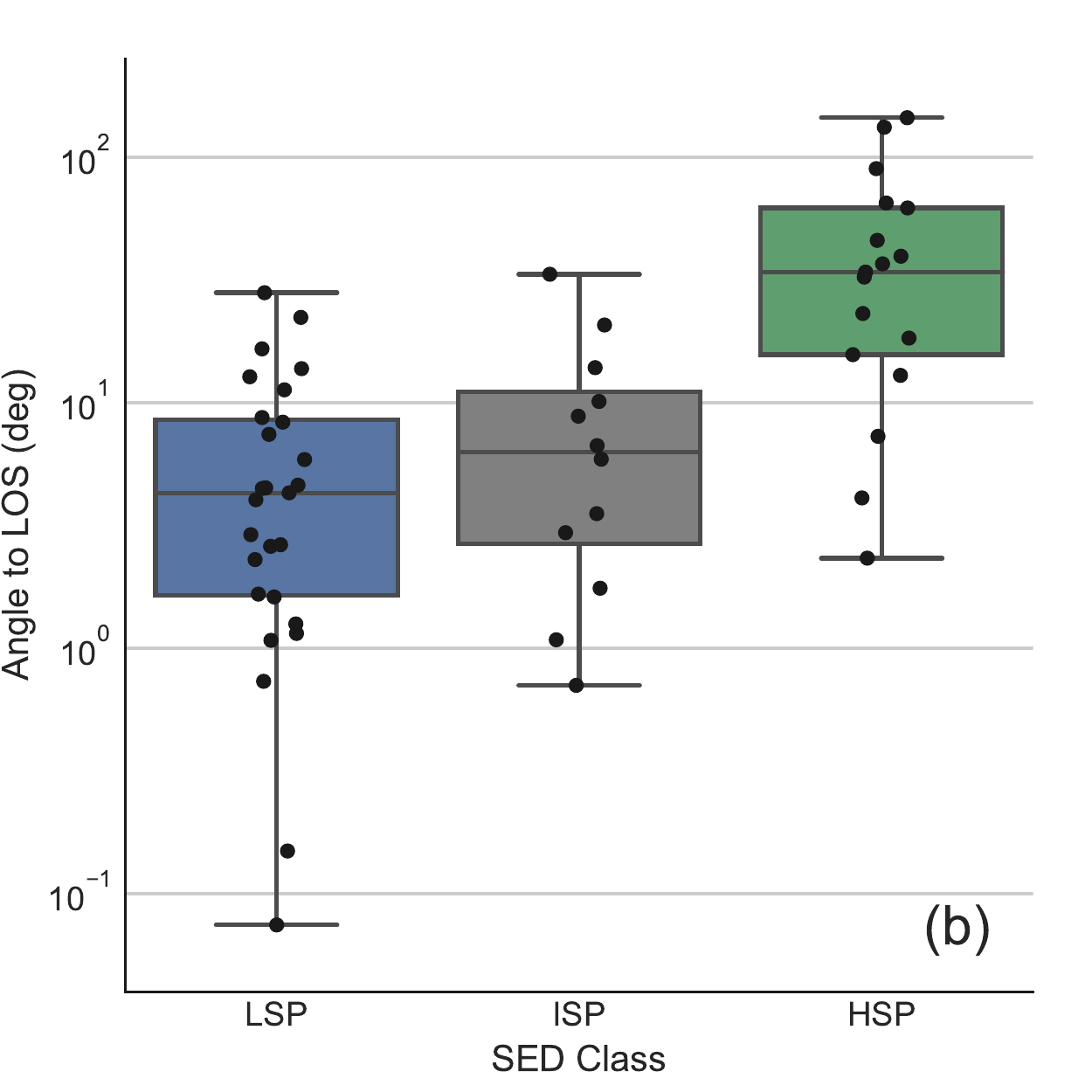}
\includegraphics[width=0.38\textwidth]{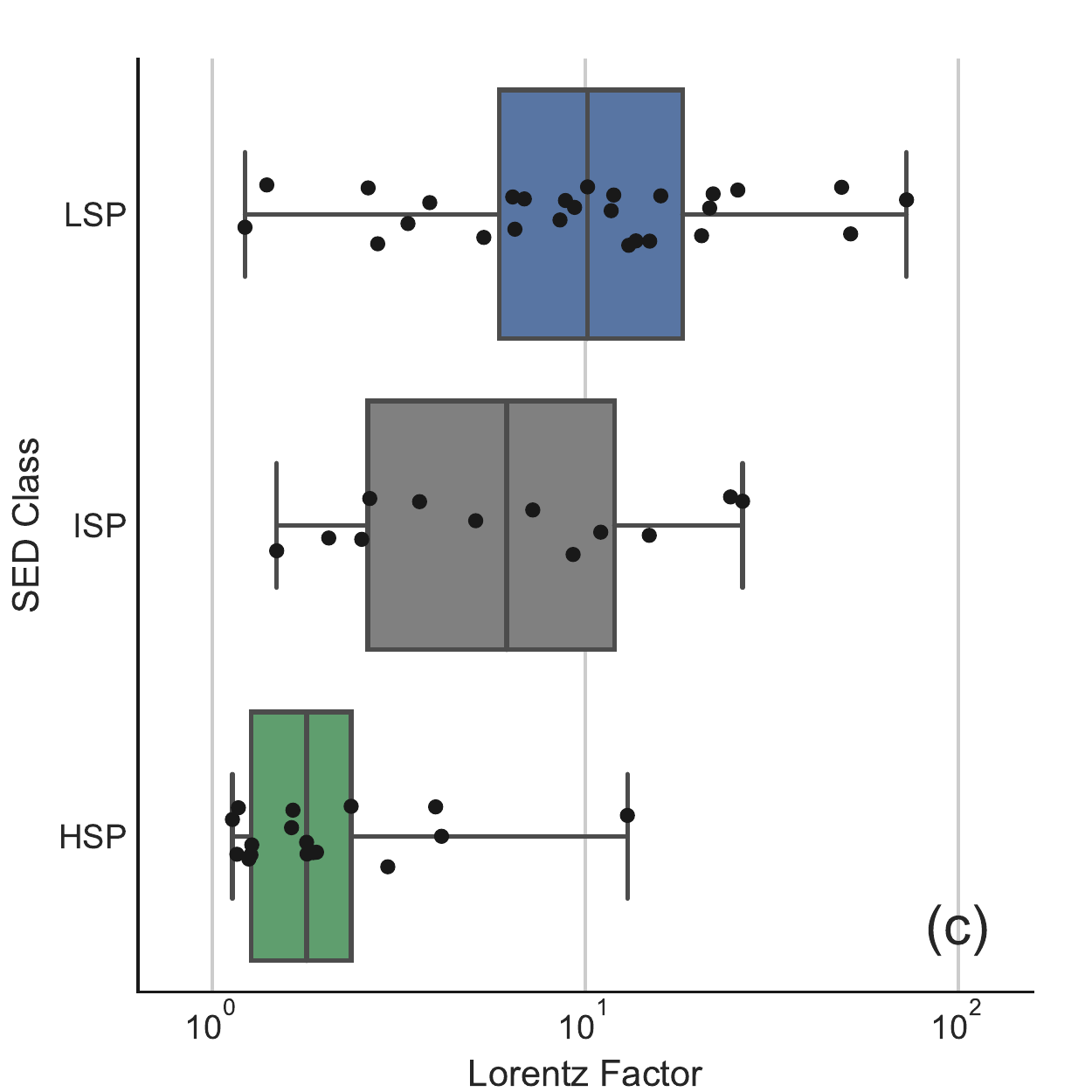}
\figcaption{\label{f:ThetaGamma_SED}
  Panel (a) plots angle to the line of sight, $\theta$, against Lorentz Factor, 
  $\Gamma$, for all BL\,Lac objects. Panels (b) and (c) illustrate 
  the distributions of these quantities as function of SED class, 
  where the ``LSP'', ``ISP'', and ``HSP'' abbreviations indicate 
  Low, Intermediate, and High Spectral Peak sources respectively.
  The filled regions of the box plots show in the inner-quartile range of
  each SED class, while the whiskers show the full extent of the data.  
  Individual data points are shown as a scatter plot over the box plot 
  to better illustrate the range and density of the data.
}
\end{figure}

\autoref{f:ThetaGamma_WS}, \ref{f:ThetaGamma_MQC}, and 
\ref{f:ThetaGamma_SED}
show scatter plots of viewing angle versus Lorentz factor
for our entire heterogeneous sample, the MOJAVE 1.5\,Jy QC 
flux-density limited sample, and BL\,Lacs divided by SED class 
respectively.  Each of these scatter plots is accompanied by two 
sets of box plots which show the distributions of these quantities as a 
function of optical or SED class.  Note that these figures and the 
following discussion are complementary to the brightness temperature
plots and discussion in section \autoref{s:Tb-trends} as we are taking
brightness temperature to be directly proportional to the Doppler factor.

\autoref{f:ThetaGamma_WS} for our whole, heterogeneous sample has 
233 quasars, 56 BL\,Lacs, 17 radio galaxies, and 3 narrow-line Seyfert I galaxies.  
Quasars have larger Lorentz factors and smaller viewing angles than 
both BL\,Lacs and galaxies as confirmed by Anderson-Darling tests 
which show the probability
they are drawn from the same distribution is $p < 0.001$ in each case.  If
we restrict the comparison to just LSP quasars ($n=227$) and BL\,Lacs ($n=27$),
the Lorentz factor difference still holds ($p = 0.010$), but we 
no longer detect a viewing angle difference ($p = 0.22$), consistent 
with the findings of \cite{2018ApJ...866..137L}.  

The MOJAVE 1.5\,Jy QC flux-density limited sample has 149 quasars, 23 
BL\,Lacs, and just 6 radio galaxies in \autoref{f:ThetaGamma_MQC}.  For 
Lorentz factor, we find all three distributions differ from one 
another ($p = 0.004$ for quasars vs. BL\,Lacs, $p < 0.001$ for 
quasars vs. galaxies, $p = 0.011$ for BL\,Lacs vs.
galaxies), with quasars having the largest Lorentz factors and galaxies the
smallest in the sequence.  For viewing angles, we can detect no difference
 between the classes with our Anderson-Darling tests, although we note the
number of galaxies is quite small ($n=6$) and includes M87 which may have
had its viewing angle underestimated as described in \autoref{s:dopplerfactors}.

We note that the Lorentz factor differences between quasars and radio galaxies 
described above are driven by our flux-density limited selection criteria 
where only nearby radio galaxies have sufficient flux-density 
without the need for large Doppler beaming factors to make it into our sample.

\begin{figure}
\centering
\includegraphics[width=0.5\textwidth]{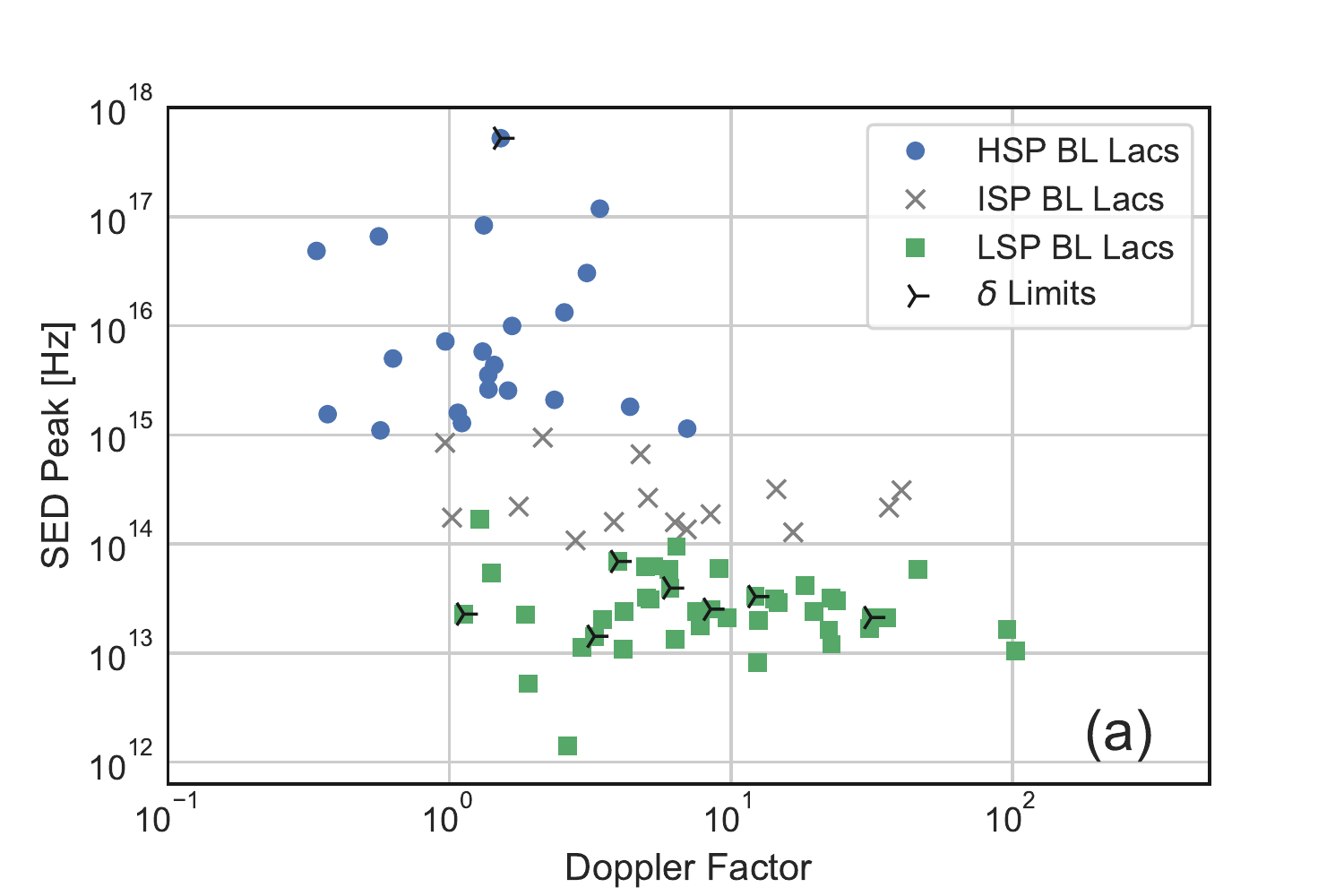}
\includegraphics[width=0.5\textwidth]{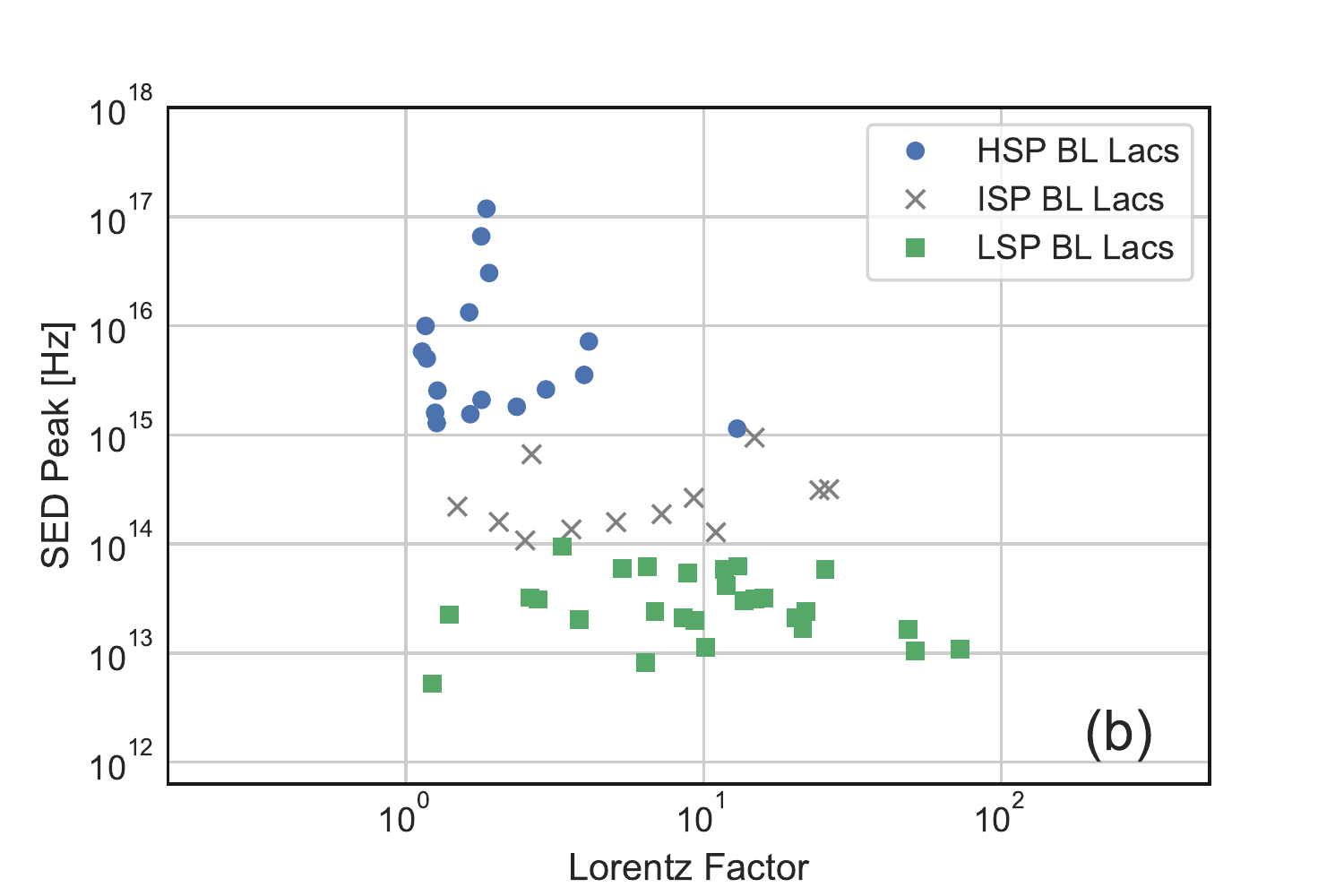}
\includegraphics[width=0.5\textwidth]{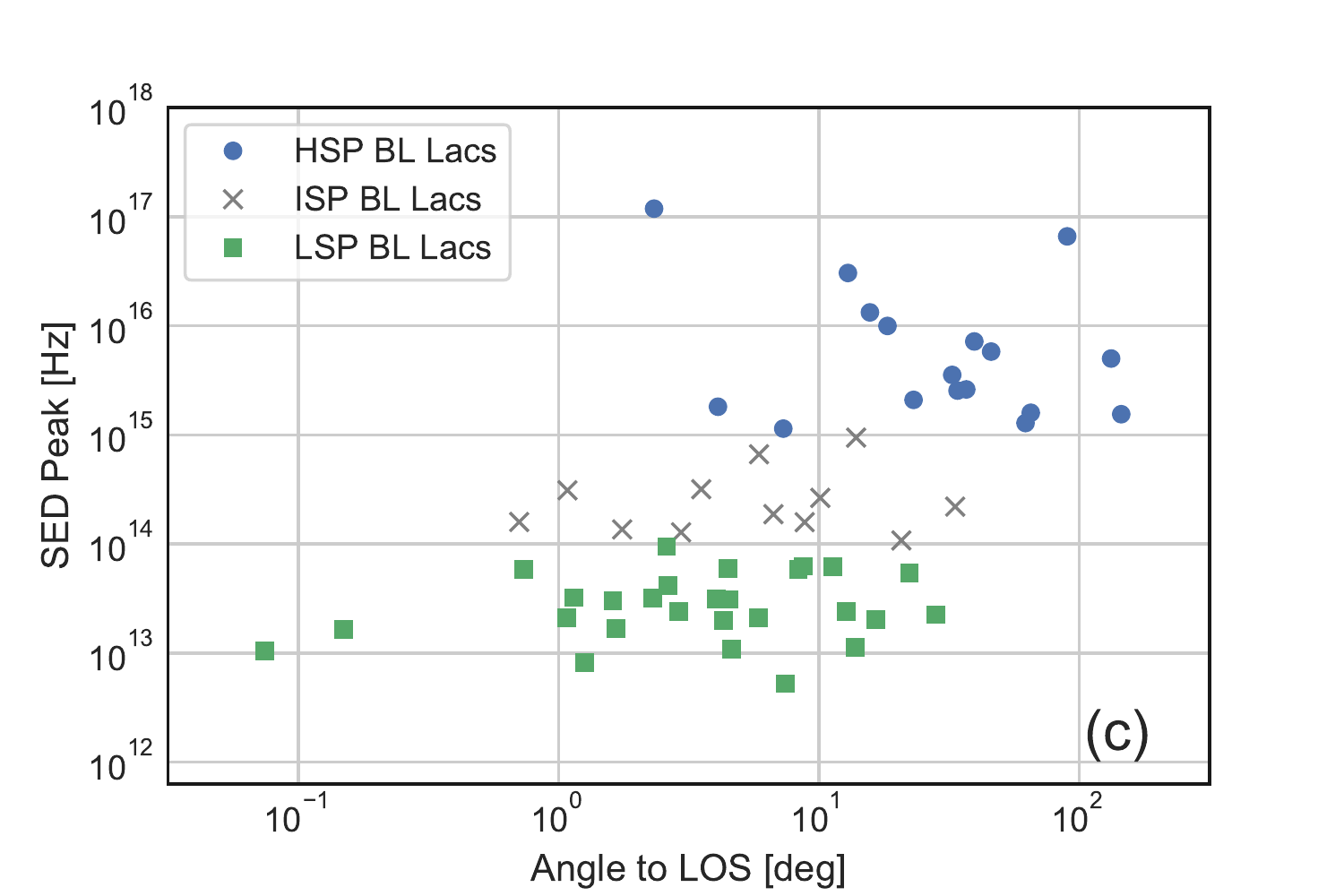}
\figcaption{\label{f:SED_plots}
  SED peak frequency in the host galaxy frame vs Doppler factor (panel a), Lorentz Factor (panel b), and Angle to the Line of Sight (panel c) 
  for BL\,Lacs identified by SED class.
  Planel (a) includes 79 BL\,Lacs 
  for which we could estimate the Doppler factor from their
  median brightness temperature.
  Panels (b) and (c)
  include just the 56 BL\,Lacs for which we could also use their measured apparent
  speeds to estimate their other properties as described in
  \autoref{s:calc_quant}.
}
\end{figure}

Finally, we look at BL\,Lacs as a function of SED class in \autoref{f:ThetaGamma_SED} which has 27 LSP, 12 ISP, and 17 HSP BL\,Lacs.
We cannot detect a difference in either Lorentz factor or viewing
angle between LSP and ISP BL\,Lacs with $p > 0.25$ for both quantities;
however, HSP BL\,Lacs have smaller Lorentz factors and larger viewing
angles than both LSPs ($p < 0.001$ for both quantities) and ISPs
($p = 0.002$ for Lorentz factor and $p = 0.001$ for viewing angle).
When combined with our finding in \autoref{s:Tb-trends} that HSP
BL\,Lacs have lower brightness temperatures, and therefore lower 
Doppler factors,
than the other classes, we get the consistent picture in
\autoref{f:SED_plots}, which shows all three quantities as a
function of SED peak frequency.  HSP BL\,Lacs appear distinct
from ISP and LSP BL\,Lacs with lower Doppler and Lorentz factors
and larger viewing angles. This is consistent with the analysis of
\citet{2018ApJ...853...68P} who estimate a maximum Lorentz factor
of about 4 for this class on the basis observed motions.

\begin{figure*}
\includegraphics[scale=0.57,angle=0]{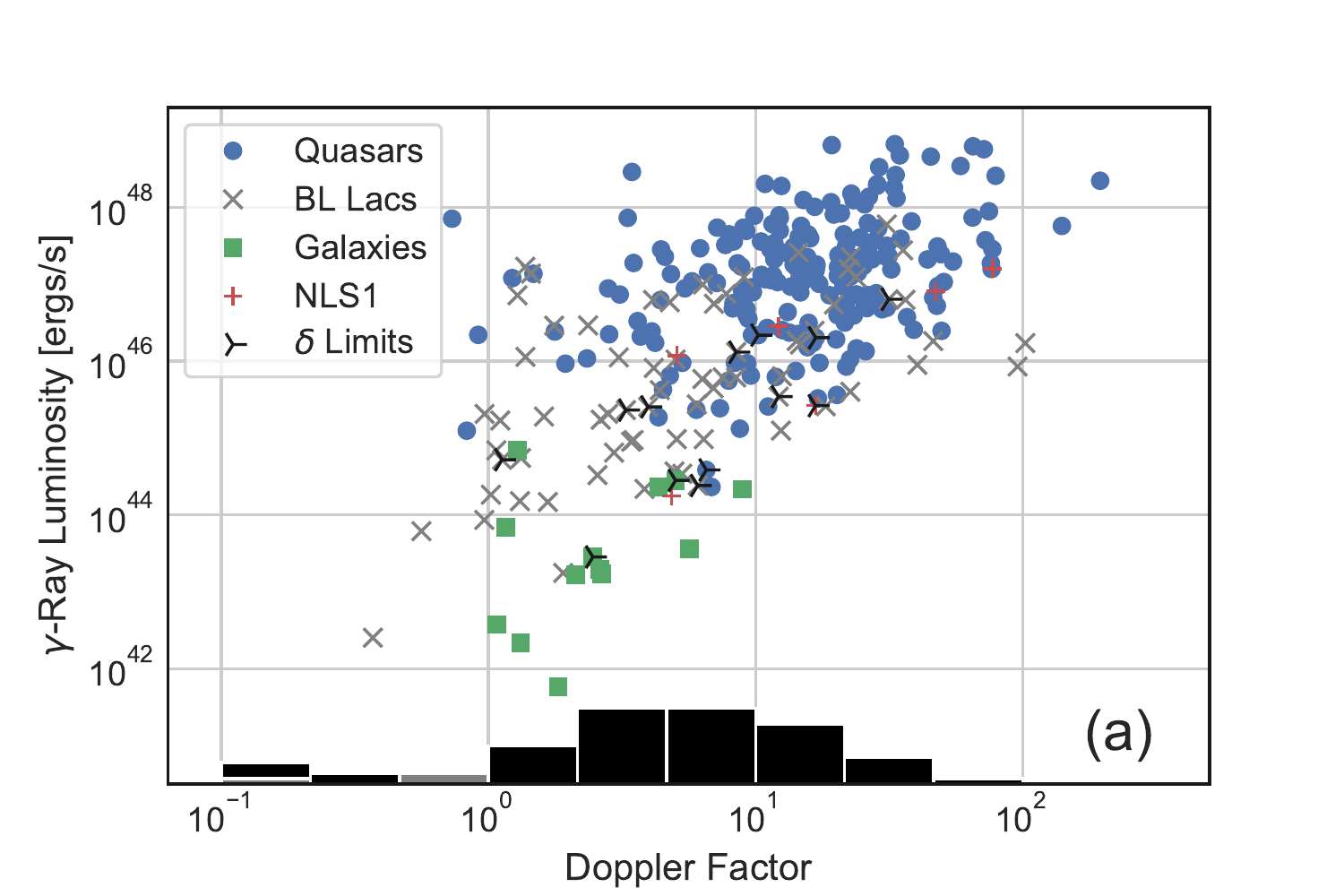}
\includegraphics[scale=0.57,angle=0]{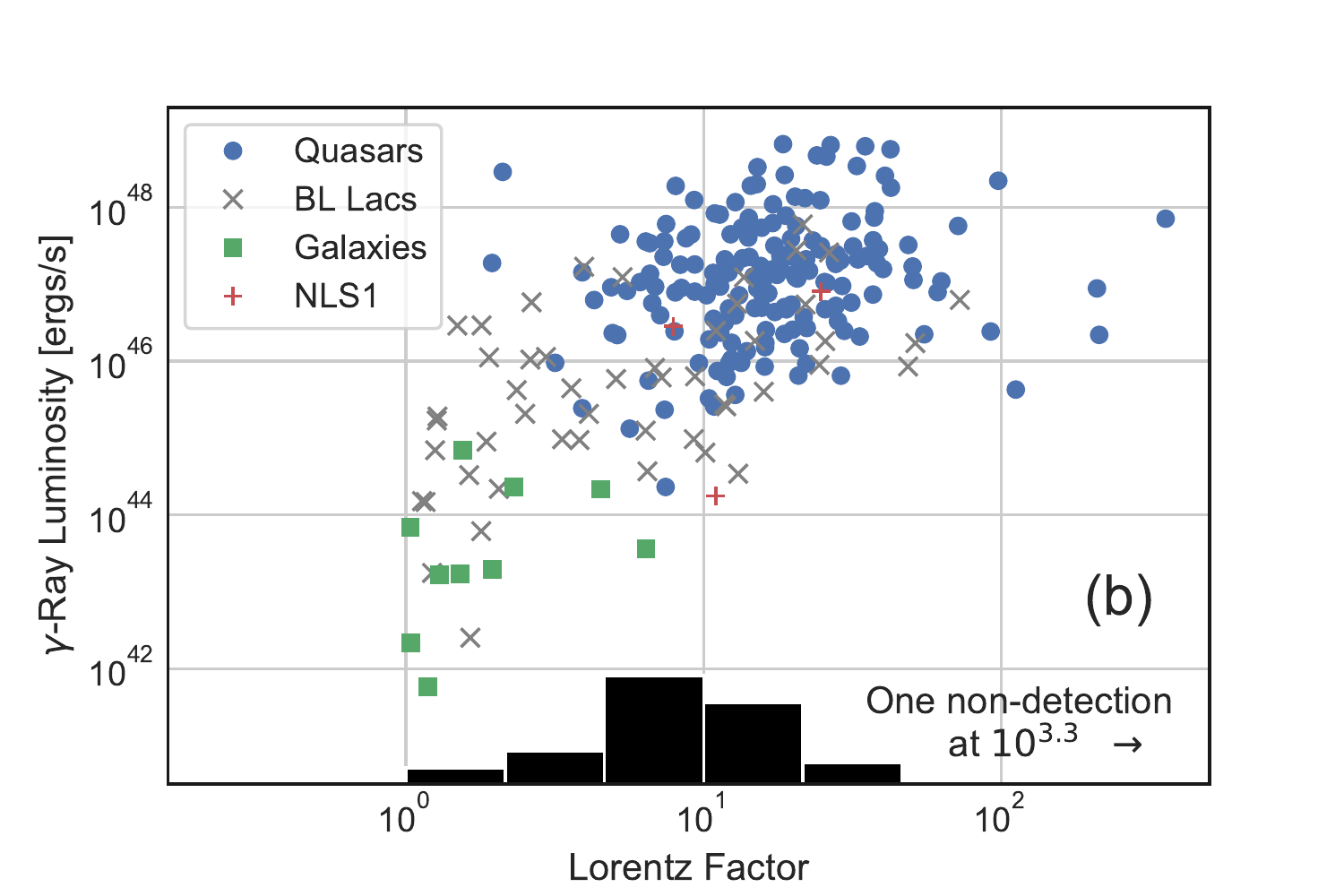}\\
\includegraphics[scale=0.57,angle=0]{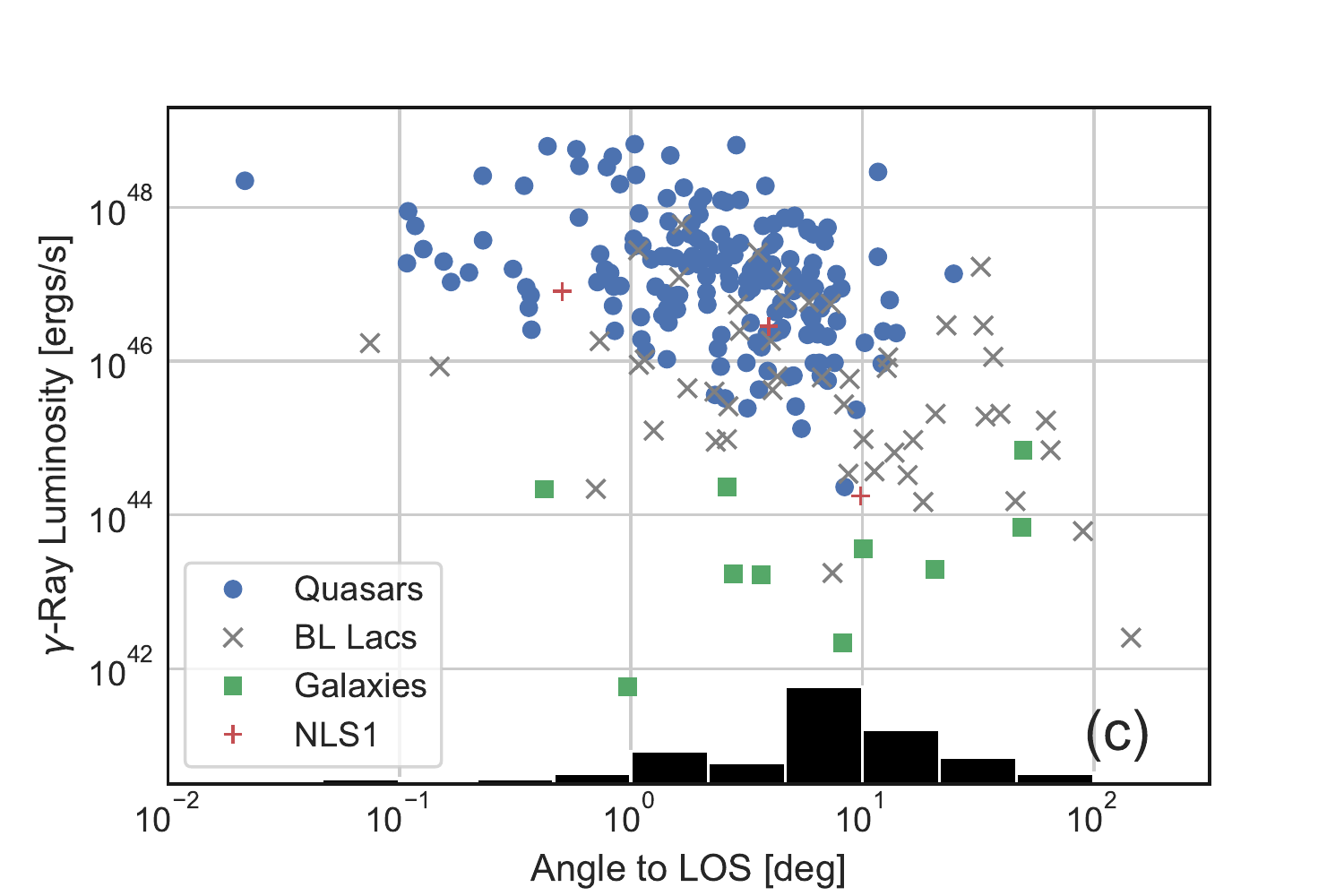}
\includegraphics[scale=0.57,angle=0]{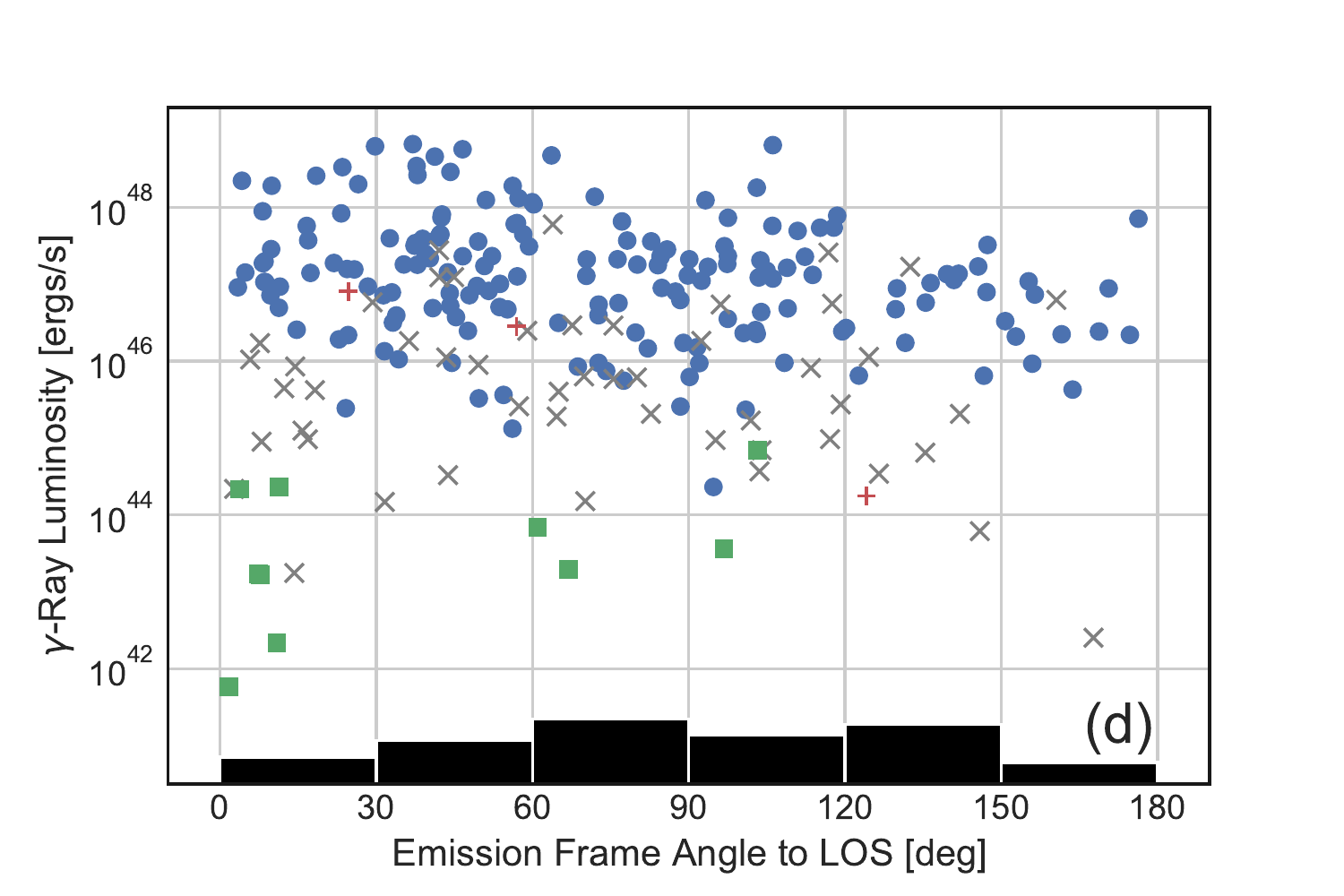}
\figcaption{\label{f:Gamma_plots}
  $\gamma$-ray luminosity vs Doppler factor (panel a), Lorentz 
  factor (panel b), angle to the line of sight (panel c), 
  and angle to the line of sight in the co-moving emission 
  frame (panel d) for \textit{Fermi}/LAT-detected AGN in our sample.
  The histogram at the bottom of each panel shows the distribution of
  sources without \textit{Fermi}/LAT-detections. Panel (a) includes 351 sources
  for which we could estimate the Doppler factor from their
  median brightness temperature,
  60 of which do not have a \textit{Fermi}/LAT detection. Panels (b) through (d)
  include 285 sources for which we could also use their measured apparent
  speeds to estimate their other properties as described in
    \autoref{s:calc_quant},
  49 of which do not have a \textit{Fermi}/LAT detection.
  Only sources with known redshifts and with a galactic latitude $|b| > 10$ degrees
  are included in this plot.
}
\end{figure*}

In \autoref{s:Tb-trends} we investigated a correlation between
$\gamma$-ray Luminosity and median brightness temperature,
most likely due to a common Doppler boosting of the radio cores
and the $\gamma$-ray emission.  \autoref{f:Gamma_plots}
examines this question further by plotting $\gamma$-ray luminosity
against each of the intrinsic quantities estimated by our analysis.
The strongest correlation is clearly with the Doppler factor, and
the somewhat weaker correlations with Lorentz factor and viewing
angle are likely a consequence of their necessary role in
producing highly Doppler boosted emission.  This is consistent
with the finding of \cite{2010A&A...512A..24S} that LAT $\gamma$-ray 
detected blazars differ significantly in their Doppler factor distribution
from non-LAT detected blazars.
We do not see a strong trend with the angle to the line of sight in
the co-moving emission frame, $\theta_\mathrm{src}$, in contradiction to
the results of \citet{2010A&A...512A..24S} from a smaller sample, but
consistent with the findings of \cite{2018ApJ...866..137L} who do
not detected a difference in source-frame viewing angle distribution between LAT
detected and non-detected sources.

\subsubsection{Fastest vs.\ Median Speeds}
\label{s:whatspeed}

In \autoref{s:betaT_analysis} we examined three possible choices
for representing the apparent jet speed in this analysis, and we chose
to use the fastest apparent speed as it correlated most strongly
with median brightness temperature and was the least likely
to be contaminated by slowly moving, ``quasi-stationary,'' features
in the jets. An additional complicating factor is that jets are
still becoming organized on these length scales and show
evidence for acceleration and collimation 
\citep[e.g., ][]{2007MNRAS.380...51K,2015ApJ...798..134H,2019MNRAS.490.2200C,2020MNRAS.495.3576K}, 
and it is possible
that choosing the fastest apparent speed may better characterize the jet
downstream from the core, rather than the core region itself
where the brightness temperature measurements are made.  When
we looked at the speed of the feature that was closest to the jet core
in its first epoch, we found it
correlated much more poorly with apparent brightness temperature,
likely due to contributions from quasi-stationary shocks near the jet
origin \citep[e.g., ][]{LCH09,J17}; however, the median jet speed correlated almost
as well with core brightness temperature as the fastest speed
and might have made a reasonable alternative for this analysis.

If we had chosen to represent jets by their
median apparent speed rather than their fastest apparent
speed, very few of our results would change.  We would conclude
the intrinsic brightness temperature was about $40\%$ larger,
$T_\mathrm{b,int} = 10^{10.762}$\,K, and would find correspondingly lower Doppler
factors for each source.  Those lower Doppler values combined with
their median speeds would lead to smaller estimated Lorentz factors
and larger estimated viewing angles for most sources by a similar factor.
However, despite these changes to $\delta$, $\Gamma$, 
and $\theta$, the relationships between these quantities 
and optical class, SED class, and $\gamma$-ray luminosity 
all remain the same without any appreciable change
to the significant statistical relationships and trends discussed
in our analysis above using the fastest speed.

\subsection{Intrinsic \texorpdfstring{$T_b$}{Lg} and Energy Balance in Jet Cores}
\label{s:energy-balance}

In \autoref{s:betaT_analysis} we find the typical intrinsic Gaussian peak 
brightness temperature for jets in their median state to be
$10^{10.609\pm0.067} = 4.1(\pm0.6)\times10^{10}$~K.  However, 
as discussed in \autoref{s:measureTb}, we found that the Gaussian
peak brightness temperature over-predicted the center brightness 
temperature of a range of homogeneous sphere models by a factor 
of $1.8$.  This factor did not depend on whether the sphere
was barely resolved and represented almost entirely by the
Gaussian, or was well-resolved with the Gaussian being fit
to the central region and the remainder of the sphere being
fit with \textsc{clean} components.  Because
this factor is constant, it cancels out and does not impact our 
analysis of Doppler factors and other derived quantities discussed above; 
however, to compare to other programs, which typically assume sphere or disk 
geometries, we take this factor of 1.8 to convert\footnote{A factor of $1.8$ 
was also estimated by \cite{2001ApJ...549L..55T}
by comparing the (u,v)-plane profile of a Gaussian to an optically thick
sphere.} our measured Gaussian brightness temperatures to 
those used or derived by variability approaches 
\citep[e.g.][]{2009A&A...494..527H, 2017MNRAS.466.4625L,J17,2018ApJ...866..137L}.  
With the
application of this factor, the typical intrinsic brightness
temperatures of jets in our program in their median state
becomes $2.3(\pm 0.3)\times10^{10}$ K.

Following \cite{Readhead94}, it has been common practice in Doppler
factor studies to assume
jets are near an equipartition balance between magnetic field and
particle energy in the emission region, even during flares, with
a canonical value of $T_\mathrm{b,int} \simeq 5.0\times10^{10}$~ K 
\citep[e.g.][]{2009A&A...494..527H, 2017MNRAS.466.4625L}; however,
as noted in \autoref{s:doppler_compare}, \cite{2018ApJ...866..137L}
found a much larger value of $T_\mathrm{b,int}=2.8\times10^{11}$~K, approaching the $\simeq 10^{11.5}$~K inverse-Compton limit \citep{KPT69, Readhead94} 
and perhaps consistent with the diamagnetic limit suggested by 
\cite{1986A&A...155..242S}.  In this paper, we have characterized
the intrinsic brightness temperatures of jets, not in their flaring
state but rather in their median state, and we find jets to be at or below
equipartition in that median state, suggesting that jet cores may even be 
magnetic field dominated in their lower brightness states.  We note
that \cite{2013JKAS...46..243L} reported even lower intrinsic brightness
temperatures at 86 GHz for compact radio jets, suggesting magnetic
field dominance closer to the central engine, although 
\cite{2016ApJ...826..135L} also concluded that the change in brightness 
temperature with frequency in VLBI jets cores indicates acceleration along the jet.

As discussed in \autoref{s:tb-discuss}, observed brightness temperatures
within individual jets can span up to an order of magnitude or more
in the most variable jets.  The typical ratio between the maximum observed
brightness temperature and its median value for the same jet is a factor 
of a few, and even if these variations are entirely due to changes in the 
intrinsic brightness temperature, we would still find intrinsic brightness
temperatures for most sources in their flaring states below the inverse-Compton 
limit of $10^{11.5}$~K \citep{KPT69, Readhead94} or even the typical 
flaring state value of $2.8\times10^{11}$~K deduced by \cite{2018ApJ...866..137L}.  
This difference between the 
maximum brightness temperatures we observe for most sources and the 
typical flaring value found by \cite{2018ApJ...866..137L}
may simply be due to the fact that we are measuring the brightness
temperature of the core region of the jet as a whole, and even during 
an outburst, the core region may not consist of just a single flaring 
component.  Indeed this suggestion is supported by the  
\textit{RadioAstron} space VLBI measurements which can detect smaller sub-components 
in the jet core \citep{2020AdSpR..65..705K}. They indicate higher peak brightness temperatures
at 22 GHz in at least two powerful AGN jets at similar epochs to those we 
observed from the VLBA alone at 15 GHz.  
For example in 3C\,273, \textit{RadioAstron} at 22 GHz 
measured an observed  
brightness temperature of $1.4\times10^{13}$~K in February 2013, an
order of magnitude larger than our $1.12\times10^{12}$~K measurement made 
eight days later \citep{2016ApJ...820L...9K}, and in BL~Lac, \textit{RadioAstron}
measured a 22 GHz brightness temperature of $>2\times10^{13}$~K a little 
more than a month before our measurement of $2.11\times10^{12}$~K 
\citep{2016ApJ...817...96G}. Note that both of these jets have estimated
Doppler factors $\delta \simeq 20$ in our analysis, so the intrinsic 
brightness temperatures implied by the \textit{RadioAstron} results are 
a couple of times larger than the flaring state value given by 
\cite{2018ApJ...866..137L}\footnote{This comparison includes the 
factor of 1.8 difference between sphere/disk model used in the 
the variability analysis and the Gaussian brightness temperatures 
used by \textit{RadioAstron}.}, confirming that compact regions in the jet
can be strongly particle dominated and approach the inverse-Compton
limit.

\section{Summary and Conclusions}
\label{s:conclude}

We have made multi-epoch, parsec-scale core brightness temperature measurements of 447 AGN 
jets from the MOJAVE VLBA program; 206 of these AGN are members of the MOJAVE 1.5 Jy QC 
flux-density limited sample.  We characterized each jet by its median core brightness 
temperature and variability over time and examined trends with optical class, SED class, 
and $\gamma$-ray luminosity computed from the \textit{Fermi}/LAT 10-year point source 
catalog \citep{2020ApJ...892..105A}.

Combined with our recently updated apparent speed measurements reported in \citetalias{MOJAVE_XVIII}, we 
followed the approach of \citet{H06} to estimate the typical intrinsic Gaussian brightness 
temperature of a jet core in its median state, 
$T_\mathrm{b,int} = 10^{10.609\pm0.067} = 4.1(\pm0.6)\times10^{10}$~K.
We used this value to derive estimates for the Doppler factor from the observed 
median brightness temperature for 447 
sources in our sample, $\delta = T_\mathrm{b,obs}/T_\mathrm{b,int}$,
and compared our results to those from other programs.
For the 309 AGN jets with both apparent speed and brightness temperature data, we 
also estimated their intrinsic Lorentz factors and viewing angles to the line of sight.  

Our main results are as follows:

1. We measured the parsec-scale core brightness temperature of each AGN jet
in every epoch by fitting a single Gaussian to the core region alone and modeling 
the remainder of the jet by \textsc{clean} components.  We find that the observed 
Gaussian brightness temperature of the jet core of a given source varies over time 
by a factor of a few 
up to about a order of magnitude, with a few extreme cases having larger 
variations; however, the differences between AGN jets in our sample can be much larger 
with median values spanning two and half to three orders of magnitude. The range 
of observed median brightness temperatures across our sample is consistent with 
Doppler boosting being the primary difference between AGN jets in their median state.

2. Median core brightness temperatures differ between AGN based on their optical classes and 
synchrotron peak classifications.  Quasars and BL\,Lacs have
larger observed brightness temperatures, and therefore Doppler beaming factors, than 
radio galaxies as one would expect according to unified models 
\citep[e.g][]{1995PASP..107..803U}, whether we consider just the MOJAVE 1.5 Jy 
QC flux-density limited sample or our entire heterogeneous sample.  If we consider 
only low synchrotron peaked (LSP) quasars and BL\,Lacs, we do not detect a 
difference between them in terms of their median core brightness temperatures, 
indicating they have similar levels of Doppler beaming. However, within the BL\,Lac 
class itself, high synchrotron peak (HSP) BL Lacs have distinctly lower median 
brightness temperatures than their intermediate and low synchrotron peaked 
counterparts, indicating they are less beamed than those whose SEDs peak at lower 
frequencies, consistent with earlier findings 
\citep[e.g.][]{2008AA...488..867N, 2011ApJ...742...27L,2018ApJ...853...68P}.

3. Combined with apparent speed measurements, the Doppler factor estimates from the 
observed median brightness temperatures allowed us to measure and compare the 
Lorentz factors and viewing angles of 309 of our AGN jets, 178 of which were
members of the MOJAVE 1.5 Jy QC sample.  The Lorentz factor distributions of 
quasars, BL\,Lacs, and radio galaxies all differ from one another with quasars 
having the largest Lorentz factors and radio galaxies the smallest.  
If we consider just LSP quasars and BL\,Lacs, we still detect a significant
Lorentz factor difference between them but do not detect a difference in 
viewing angle distribution, similar to the findings of \cite{2018ApJ...866..137L}.  
HSP BL\,Lacs appear distinct from ISP and LSP BL\,Lacs with lower Lorentz 
factors and larger viewing angles to the line of sight.

4. Median core brightness temperatures, and by extension jet Doppler factors, 
correlate strongly with $\gamma$-ray luminosity for LAT detected jets, and we 
confirm earlier findings that LAT detected jets have
larger core brightness temperatures than non-detected jets 
\citep[e.g.][]{2009ApJ...696L..17K,2011ApJ...742...27L}.  We also see clear trends
between $\gamma$-ray luminosity and Lorentz factor and viewing angle to the line of sight; however,
the strongest relationship appears to be with median core brightness temperature / Doppler factor,
and the trends with Lorentz factor and viewing angle are likely a consequence of
their necessary role in producing highly Doppler boosted emission.  We do not 
see a strong trend with angle to the line-of-sight in the co-moving emission
frame.

5. We found the typical intrinsic Gaussian peak brightness temperature for jets 
cores in their median state to be $4.1(\pm0.6)\times10^{10}$~K. 
Our Gaussian brightness temperatures are a factor of 1.8 times larger than the
spherical/disk geometries used in variability Doppler factor analyses.  The best
geometry to represent the core region is unknown; however, regardless of whether or not
we apply this geometrical factor, we find the jet cores to be at or below the
typically assumed value for equipartition between magnetic field and particle
energies of $5.0\times10^{10}$~K \citep[e.g.][]{Readhead94,1999ApJ...521..493L}
in their median state.

%\acknowledgements
%\begin{acknowledgements}
\medskip
We thank Margo Aller, Alexander Plavin, and the other members of the MOJAVE team 
for helpful conversations and their other contributions that made this work 
possible.
The MOJAVE project was supported by NASA-{\it Fermi} grants 80NSSC19K1579, NNX15AU76G and NNX12A087G.
DCH was supported by NSF grant AST-0707693. 
YYK and ABP were supported by the Russian Science Foundation grant 21-12-00241. 
AVP was supported by the Russian Foundation for Basic Research grant 19-32-90140.
TH was supported by the Academy of Finland projects 317383, 320085, and 322535.
TS was partly supported by the Academy of Finland projects 274477 and 315721.
The National Radio Astronomy Observatory is a facility of the National Science Foundation operated
under cooperative agreement by Associated Universities, Inc.
This work made use of the Swinburne University of Technology software correlator \citep{D11}, developed as part of the Australian Major
National Research Facilities Program and operated under licence.
This research has made use of data from the OVRO 40-m monitoring program \cite{2011ApJS..194...29R}, which is supported in part by NASA grants NNX08AW31G, NNX11A043G, and NNX14AQ89G and NSF grants AST-0808050 and AST-1109911. 
This research has made use of NASA's Astrophysics Data System.
This research has made use of the NASA/IPAC Extragalactic Database (NED) which is operated by the Jet Propulsion Laboratory, California Institute of Technology, under contract with the National Aeronautics and Space Administration.
\facility{VLBA, OVRO:40m, NED, ADS}

%\end{acknowledgements}

%%%%%%%%%%%%%%%%%%%%%%
%%                  %%  
%%    References    %%
%%                  %%
%%%%%%%%%%%%%%%%%%%%%%

%\bibliography{refs}{}
%\bibliographystyle{aasjournal}

\end{document}